\documentclass[10pt,letterpaper]{article}
\usepackage[top=0.85in,left=2.75in,footskip=0.75in]{geometry}

\usepackage{changepage}

\usepackage[utf8]{inputenc}

\usepackage{textcomp,marvosym}

\usepackage{fixltx2e}

\usepackage{amsmath,amssymb}

\usepackage{cite}

\usepackage{nameref,hyperref}
\newcommand{\Newnameref}[1]{``\nameref{#1}''}

\usepackage[right]{lineno}

\usepackage{microtype}
\DisableLigatures[f]{encoding = *, family = * }

\usepackage{rotating}

\usepackage{setspace} 

\newlength{\figureoffsetleft}
\newlength{\figureoffsetright}
\raggedright
\usepackage[aboveskip=1pt,labelfont=bf,labelsep=period,justification=raggedright,singlelinecheck=off]{caption}

\makeatletter
\renewcommand{\@biblabel}[1]{\quad#1.}
\makeatother

\date{}
\usepackage{color}
\usepackage[dvipsnames]{xcolor}

\usepackage{lastpage,fancyhdr,graphicx}
\usepackage{epstopdf}
\fancyheadoffset[L]{-\figureoffsetleft} \fancyfootoffset[L]{-\figureoffsetleft}
\fancyheadoffset[R]{-\figureoffsetright} \fancyfootoffset[R]{-\figureoffsetright}
\fancypagestyle{firststyle}
{
  \fancyhf{}
  
  \lhead{\includegraphics[width=2.29372in]{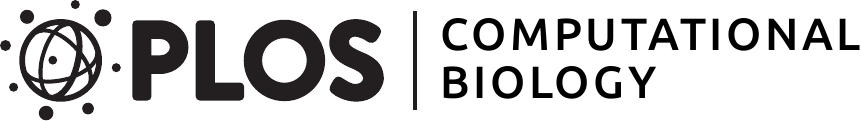}}
  \rhead{\sf Schwalger T, Deger M, Gerstner W. PLoS Comput Biol. 2017;13(4):e1005507}
  \rfoot{\thepage/\pageref{LastPage}}
  
  \fancyheadoffset[L]{-\figureoffsetleft} \fancyfootoffset[L]{-\figureoffsetleft}
  \fancyheadoffset[R]{-\figureoffsetright} \fancyfootoffset[R]{-\figureoffsetright}
  \lfoot{\sf PLOS Computational Biology $\vert$ \url{https://doi.org/10.1371/journal.pcbi.1005507} April 19, 2017 }
}

\DeclareGraphicsExtensions{%
    .pdf,.PDF,%
    .png,.PNG,%
    .jpg,.mps,.jpeg,.jbig2,.jb2,.JPG,.JPEG,.JBIG2,.JB2}

\usepackage[normalem]{ulem}

\newcommand{\rot}[1]{{\color{black} #1}}
\newcommand{\rott}[1]{{\color{black} #1}}
\newcommand{\blau}[1]{{\color{black} #1}}
\newcommand{\gruen}[1]{{\color{black} #1}}

\newcommand{\nbar}{\Delta\bar n}

\newcommand{\Xmicro}{\mathcal{X}}

\newcommand{\tl}{\hat{t}}
\newcommand{\haz}{\lambda}
\newcommand{\freehaz}{\lambda_{\text{free}}}

\newcommand{\freetheta}{\vartheta_{\text{free}}}
\newcommand{\pfree}{P_{\text{free}}}
\newcommand{\vth}{u_\text{th}}
\newcommand{\vreset}{u_\text{r}}
\newcommand{\vrest}{u_\text{rest}}

\newcommand{\tref}{t_\text{ref}}
\newcommand{\kref}{k_\text{ref}}

\newcommand{\taurel}{\tau_\text{rel}}

\newcommand{\taum}{\tau_\text{m}}

\newcommand{\taus}{\tau_\text{s}}

\newcommand{\pd}[2]{\frac{\partial#1}{\partial#2}}

\newcommand{\expect}[1]{\text{E}\!\left[#1\right]}

\newcommand{\lrk}[1]{\left\langle #1 \right\rangle}
\newcommand{\lrrund}[1]{\!\left( #1 \right)}
\newcommand{\lreckig}[1]{\!\left[ #1 \right]}

\newcommand{\od}[2]{\frac{\mathrm{d}#1}{\mathrm{d}#2}}

\begin{document}
\vspace*{0.35in}
\thispagestyle{firststyle}
\begin{flushleft}
  {\Large \textbf\newline{%
Towards a theory of cortical columns: From spiking neurons to interacting neural populations of finite size}
}
\newline
\\
Tilo Schwalger\textsuperscript{1,*},
Moritz Deger\textsuperscript{1,2},
Wulfram Gerstner\textsuperscript{1},
\\
\bigskip
\bf{1} Brain Mind Institute, School of Computer and Communication Sciences and School of Life Sciences, \'{E}cole Polytechnique F\'{e}d\'{e}rale de Lausanne (EPFL), Lausanne, Switzerland
\\
\bf{2} Institute for Zoology, Faculty of Mathematics and Natural Sciences, University of Cologne, Cologne, Germany
\\
\bigskip

%
%





* tilo.schwalger@epfl.ch

\end{flushleft}
\section*{Abstract}
Neural population equations such as neural mass or field models are
widely used to study brain activity on a large scale. However, the
relation of these models to the properties of single neurons is
unclear. Here we derive an equation for several interacting
populations at the mesoscopic scale starting from a microscopic model
of randomly connected generalized integrate-and-fire neuron
models. Each population consists of 50 -- 2000 neurons of the same
type but different populations account for different neuron types. The
stochastic population equations that we find reveal how spike-history
effects in single-neuron dynamics such as refractoriness and
adaptation interact with finite-size fluctuations on the population
level. Efficient integration of the stochastic mesoscopic equations
reproduces the statistical behavior of the population activities
obtained from microscopic simulations of a full spiking neural network
model. The theory describes nonlinear emergent dynamics such as
finite-size-induced stochastic transitions in multistable networks and
synchronization in balanced networks of excitatory and inhibitory
neurons. The mesoscopic equations are employed to rapidly integrate a
model of a cortical microcircuit consisting of eight neuron types, which
allows us to predict spontaneous population activities as well as
evoked responses to thalamic input. Our theory establishes a general
framework for modeling finite-size neural population dynamics based on
single cell and synapse parameters and offers an efficient approach to
analyzing cortical circuits and computations.

\section*{Author Summary}

Understanding the brain requires mathematical models on different
spatial scales. On the ``microscopic'' level of nerve cells, neural
spike trains can be well predicted by phenomenological spiking neuron
models. On a coarse scale, neural activity can be modeled by
phenomenological equations that summarize the total activity of many
thousands of neurons. Such population models are widely used to model
neuroimaging data such as EEG, MEG or fMRI data. However, it is
largely unknown how large-scale models are connected to an underlying
microscale model. Linking the scales is vital for a correct
description of rapid changes and fluctuations of the population
activity, and is crucial for multiscale brain models. The challenge is
to treat realistic spiking dynamics as well as fluctuations arising
from the finite number of neurons. We obtained such a link by deriving
stochastic population equations on the mesoscopic scale of 100 -- 1000
neurons from an underlying microscopic model. These equations can be
efficiently integrated and reproduce results of a microscopic
simulation while achieving a high speed-up factor.  We expect that our
novel population theory on the mesoscopic scale will be instrumental
for understanding experimental data on information processing in the
brain, and ultimately link microscopic and macroscopic activity
patterns.


\section*{Introduction}
When neuroscientists report electrophysiological, genetic, or
anatomical data from a cortical neuron, they typically refer to the
cell type, say, a layer 2/3 fast-spiking interneuron, a
parvalbumin-positive neuron in layer 5, or a Martinotti cell in layer
4, together with the area, say primary visual cortex or somatosensory
cortex \cite{WanTol04,SugHem06,LefTom09,HarShe15}.  Whatever the
specific taxonomy used, the fact that a taxonomy is plausible at all
indicates that neurons can be viewed as being organized into
populations of cells with similar properties.  In simulation studies
of cortical networks with spiking neurons, the number of different
cell types, or neuronal populations, per cortical column ranges from
eight \cite{PotDie14} to about 200 \cite{MarMul15} with 31'000 to
80'000 simulated neurons for one cortical column, but larger
simulations of several columns adding up to a million neurons (and 22
cells types) have also been performed \cite{IzhEde08}. In the
following, we will refer to a model where each neuron in each
population is simulated explicitly by a spiking neuron model as a {\em
  microscopic} model.

On a much coarser level, neural mass models \cite{Fre75,DavFri03,MorPin13},
also called field models \cite{JirHak97,Coo10,BojOos10}, population activity
equations \cite{GerKis14}, rate models \cite{DayAbb05}, or
Wilson-Cowan models \cite{WilCow72} represent the activity of a
cortical column at location $x$ by
one or at most a few variables, such as the mean activity of
excitatory and inhibitory neurons located in the region around $x$.
Computational frameworks related to neural mass models have been used
to interpret data from fMRI \cite{FriHar03,DecJir11} and EEG
\cite{DavFri03}.  Since neural mass models give a compact summary of
coarse neural activity, they can be efficiently simulated and fit to
experimental data \cite{FriHar03,DecJir11}.

However, neural mass models have several disadvantages.  While the
stationary state of neural mass activity can be matched to the
single-neuron gain function and hence to detailed neuron models
\cite{GerHem92,JirHak97,BruWan01,DecRol05,DecJir08,GerKis14},
the dynamics of neural mass models in response to a rapid change in
the input does not correctly reproduce a microscopic simulation of a
population of neurons \cite{Ger00,DecJir08,GerKis14}.  Second,
fluctuations of activity variables in neural mass models are either
absent or described by an {\it ad hoc} noise model.  Moreover, the
links of neural mass models to local field potentials are difficult to
establish \cite{EinKay13}.  Because a systematic link to microscopic
models at the level of spiking neurons is missing, existing neural
mass models must be considered as heuristic phenomenological, albeit
successful, descriptions of neural activity.

In this paper we address the question of whether equations for the
activity of populations, similar in spirit but not necessarily
identical to Wilson-Cowan equations \cite{WilCow72}, can be
systematically derived from the interactions of spiking neurons at the
level of microscopic models.  At the microscopic level, we start from
generalized integrate-and-fire (GIF) models
\cite{GerNau09,MenNau12,GerKis14} because, first, the parameters of
such GIF models can be directly, and efficiently, extracted from
experiments \cite{PozMen15} and, second, GIF models can predict
neuronal adaptation under step-current input \cite{PozNau13} as well
as neuronal firing times under in-vivo-like input \cite{MenNau12}.
In our modeling framework, the GIF neurons are organized into
different interacting populations.  The populations may correspond to
  different cell types within a cortical column with known
  statistical connectivity patterns
  \cite{LefTom09,MarMul15}.  Because of the split into different cell
types, the number of neurons per population (e.g., fast-spiking
inhibitory interneurons in layer 2/3) is finite and in the range of
50-2000 \cite{LefTom09}.  We call a model at the level of interacting
cortical populations of finite size a {\em mesoscopic} model. The
mathematical derivation of the mesoscopic model equations from the
microscopic model (i.e. network of generalized
integrate-and-fire neurons) is the main topic of this paper. The small
number of neurons per population is expected to lead to characteristic
fluctuations of the activity which should match those of the
microscopic model.

The overall aims of our approach are two-fold. As a first aim, this
study would like to develop a theoretical framework for cortical
information processing. The main advantage of a systematic link
between neuronal parameters and mesoscopic activity is that we can
quantitatively predict the effect of changes of neuronal
parameters in (or of input to) one population on the activation
pattern of this as well as other populations.  In particular, we
expect that a valid mesoscopic model of interacting cortical
populations will become useful to predict the outcome of experiments
such as optogenetic stimulation of a subgroup of neurons
\cite{BoyZha05,Dei11,LiuRam12}.  A better understanding of the
activity patterns within a cortical column may in turn, after suitable
postprocessing, provide a novel basis for models of EEG, fMRI, or LFP
\cite{DavFri03,FriHar03,DecJir11,EinKay13}.  While we cannot address
all these points in this paper, we present an example of
nontrivial activity patterns in a network model with stochastic
switching between different activity states potentially linked to
perceptual bistability \cite{MorRin07,ShpMor09,TheKov11} or resting
state activity \cite{DecJir11}.

As a second aim, this study would like to contribute to multiscale
simulation approaches \cite{ELu11} in the neurosciences by providing a
new tool for efficient and consistent coarse-grained simulation at the
mesoscopic scale.  Understanding the computations performed by the
nervous system is likely to require models on different levels of
spatial scales, ranging from pharmacological interactions at the
subcellular and cellular levels to cognitive processes at the level of
large-scale models of the brain. Ideally, a modeling framework should
be efficient and consistent across scales in the following sense.
Suppose, for example, that we are interested in neuronal membrane
potentials in one specific group of neurons which receives input from
many other groups of neurons.  In a microscopic model, all neurons
would be simulated at the same level; in a multi-scale simulation
approach, only the group of neurons where we study the membrane
potentials is simulated at the microscopic level, whereas the input
from other groups is replaced by the activity of the mesoscopic
model.  A multiscale approach is consistent, if the replacement of
parts of the microscopic simulation by a mesoscopic simulation does
not lead to any change in the observed pattern of membrane potentials
in the target population.  The approach is efficient if the change of
simulation scale leads to a significant speed-up of simulation.  While
we do not intend to present a systematic comparison of computational
performance, we provide an example and measure the speed-up factor
between mesoscopic and microscopic simulation for the case of a
cortical column consisting of eight populations \cite{PotDie14}.

Despite of its importance, a quantitative link between mesoscopic
  population models and microscopic neuronal parameters is still
largely lacking. This is mainly due to two obstacles: First, in
a cortical column the number of neurons of the same type is small
(50--2000 \cite{LefTom09}) and hence far from the $N\rightarrow\infty$
limit of classic ``macroscopic'' theories in which fluctuations vanish
\cite{GerKis14,NykTra00,MulBue07,BalFas12}.  Systematic treatments of
finite-size networks using methods from statistical physics (system
size expansion \cite{Bre09}, path integral methods
\cite{BuiCow07,BuiCow10}, neural Langevin equations
\cite{Bre10,WalBen11,TouErm11,GoyGoy15}) have so far been limited to simplified
Markov models that lack, however, a clear connection to single neuron
physiology.

Second, spikes generated by a neuron are generally correlated in time
due to refractoriness \cite{BerMei98}, adaptation and other spike
history dependencies
\cite{GeiGol66,RatNel00,ChaLon00,NawBou07,FisSch12,PozNau13}. Therefore
spike trains are often not well described by an (inhomogeneous)
Poisson process, especially during periods of high firing rates
\cite{BerMei98}. \blau{As a consequence,} the mesoscopic population
activity (i.e. the sum of spike trains) is generally not \blau{simply}
captured by a Poisson model \blau{either
  \cite{Lin06,CatRey06,DegHel12}, even in the absence of synaptic
  couplings \cite{DegSch14}.}  \blau{These non-Poissonian finite-size
  fluctuations on the mesoscopic level in turn imply temporally
  correlated synaptic} input to other neurons (colored noise) that can
drastically influence the population activity
\cite{CatRey06,DegHel12,WieBer15} but which is hard to tackle
analytically \cite{SchDro15}. Therefore, most theoretical approaches
rely on a white-noise or Poisson assumption to describe the synaptic
input \cite{BruHak99,Bru00,MatGiu02,LagRot14,GigDec15}, thereby
neglecting \blau{temporal correlations caused by spike-history
  dependencies in single neuron activity.}  Here, we will exploit
earlier approaches to treating refractoriness \cite{Ger00} and spike
frequency adaptation \cite{NauGer12,DegSch14} and combine these with a
novel treatment of finite-size fluctuations.

Our approach is novel for several reasons.  First, we use generalized
integrate-and-fire models that accurately describe neuronal data
\cite{GerNau09,MenNau12} as our microscopic reference. Second,
  going beyond earlier studies
  \cite{BruHak99,Bru00,MatGiu02,GigMat07}, we derive stochastic
  population equations that account for both strong neuronal
  refractoriness and finite population size in a consistent manner. Third, our theory has a
  non-perturbative character as it neither assumes the self-coupling
  (refractoriness and adaptation) to be weak \cite{OckJos16_arxiv} nor
  does it hinge on an expansion around the $N\rightarrow\infty$
solution for large but finite $N$
\cite{ToyRad09,BuiCho13,DegSch14}. Thus, it is also valid for
relatively small populations and non-Gaussian fluctuations. And
forth, in contrast to linear response theories
\cite{MeyVre02,LinDoi05,TroHu12,DegSch14,BarMaz14,BosDie16}, our
mesoscopic equations work far away from stationary states and
reproduce large fluctuations in multistable networks.

In the {\sc Results} section we present our mesoscopic population
equations, suggest an efficient numerical implementation, and
illustrate the main dynamical effects via a selected number of
examples.  To validate the mesoscopic theory we numerically integrate
the stochastic differential equations for the mesoscopic population
activities and compare their statistics to those of the population
activities derived from a microscopic simulation.  A detailed account
of the derivation is presented in the {\sc Methods} section. In the
discussion section we point out limitations and possible applications
of our mesoscopic theory.

\section*{Results}

\begin{figure}[p]
\begin{adjustwidth}{\figureoffsetleft}{\figureoffsetright}
  \captionsetup{font=normal}
  \centering
  \includegraphics{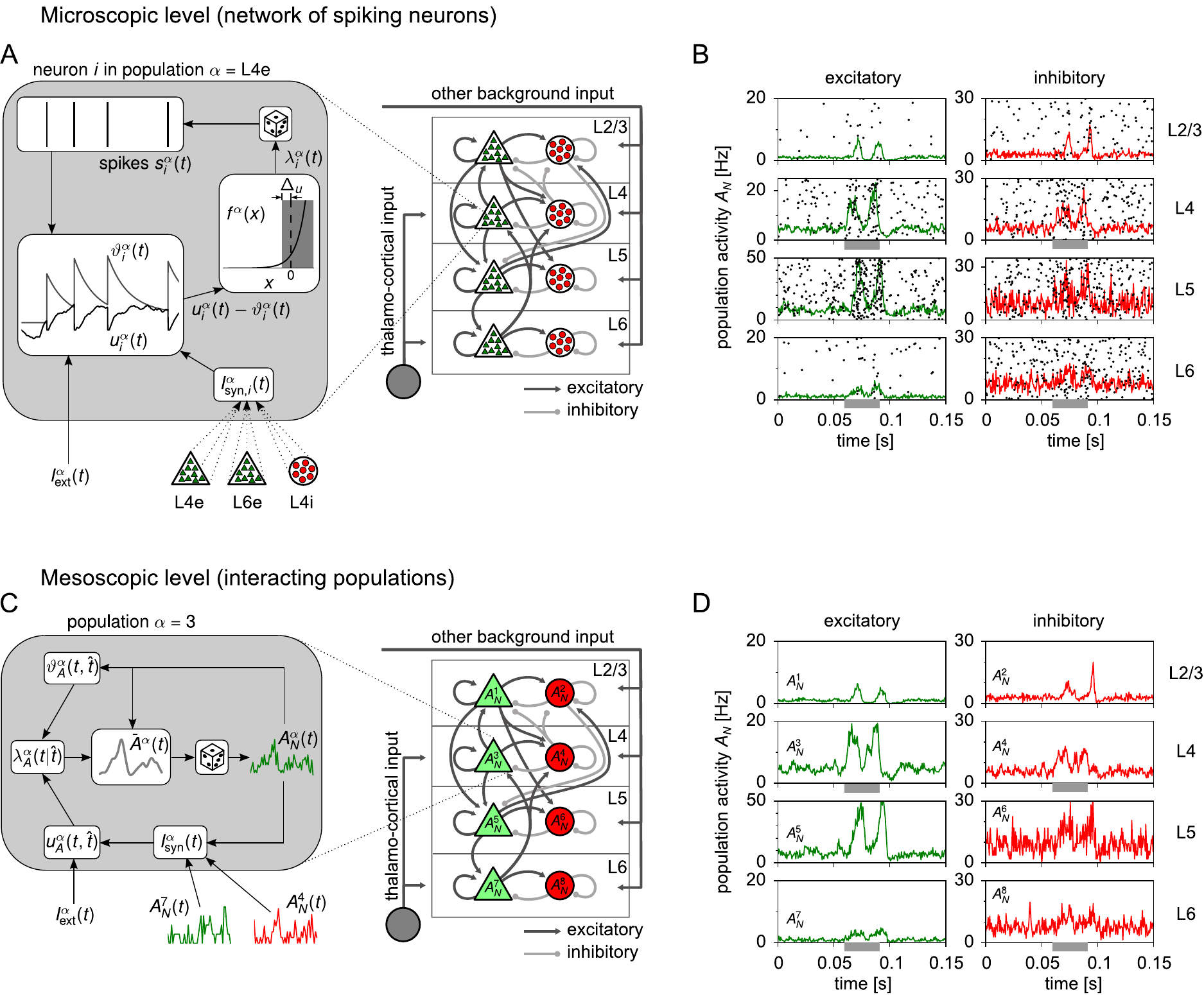}
  \caption{{\bf Illustration of population models on the microscopic and mesoscopic level.} \\
    (A) Cortical column model \cite{PotDie14} with $\sim$~80'000
    neurons organized into four layers (L2/3, L4, L5, L6) each
    consisting of an excitatory (``e'') and an inhibitory (``i'')
    population. On the microscopic level, each individual neuron is
    described by a generalized integrate-and-fire (GIF) model with
    membrane potential $u_i^\alpha(t)$, dynamic threshold
    $\vartheta_i^\alpha(t)$ and conditional intensity
    $f^\alpha\bigl(u_i^\alpha(t)-\vartheta_i^\alpha(t)\bigr)$. Inset:
    GIF dynamics for a specific neuron $i$ of population L4e
    ($\alpha=\text{L4e}$). The neuron receives spikes from neurons in
    L4e, L4i and L6e, which drive the membrane potential
    $u_i^\alpha(t)$. Spikes are elicited stochastically by a
    conditional intensity
    $\lambda_i^\alpha(t)=f^\alpha\bigl(u_i^\alpha(t)-\vartheta_i^\alpha(t)\bigr)$
    that depends on the instantaneous difference between
    $u_i^\alpha(t)$ and the dynamic threshold
    $\vartheta_i^\alpha(t)$. Spike feedback (voltage reset and
    spike-triggered threshold movement) gives rise to spike history
    effects like refractoriness and adaptation.  (B) Spike raster plot
    of the first 200 neurons of each population. The panels correspond
    to the populations in (A). Layer 4 and 6 are stimulated by a step
    current during the interval $(0.06\text{s},0.09\text{s})$
    mimicking thalamic input (gray bars). Solid lines show the
    population activities $A_N^\alpha(t)$ computed with temporal
    resolution $\Delta t=0.5$~ms, cf. Eq.~\eqref{eq:A_N-empiric}. The
    activities are stochastic due to the finite population size. (C)
    On the mesoscopic level, the model reduces to a network of 8
    populations, each represented by its population activity
    $A_N^\alpha(t)$. Inset: The mesoscopic model generates
    realizations of $A_N^\alpha(t)$ from an expected rate
    $\bar A^\alpha(t)$, which is a deterministic functional of the
    past population activities. (D) A corresponding simulation of the
    mesoscopic model yields population activities with the same
    temporal statistics as in (B).
    }
  \label{fig:potjans-scheme}
\end{adjustwidth}
\end{figure}

We consider a structured network of interacting homogeneous
populations. Homogeneous here means that each population consists of
spiking neurons with similar intrinsic properties and random
connectivity within and between populations. To define such
populations, one may think of grouping neurons into
genetically-defined cell classes of excitatory and inhibitory neurons
\cite{HarShe15}, or, more traditionally, into layers and cell types
(Fig.~\ref{fig:potjans-scheme}A). For example, pyramidal cells in
layer 2/3 of rodent somatosensory cortex corresponding to whisker C2
form a population of about 1700 neurons \cite{LefTom09}. Pyramidal
cells in layer 5 of the same cortical column form another one
($\sim 1200$ neurons \cite{LefTom09}), fast-spiking inhibitory cells
in layer 2/3 a third one ($\sim 100$ neurons \cite{LefTom09}) and
non-fast-spiking inhibitory cells in layer 2/3 a fourth one
($\sim 130$ neurons \cite{LefTom09}), and so on
\cite{LefTom09,AveTom12,MarMul15}. We suppose that the parameters
of typical neurons from each population
\cite{AveTom12,GenKre12,PozMen15}, 
the number of neurons per population \cite{LefTom09,AveTom12} and the
typical connection probabilities \cite{PotDie14} and strengths within
and between populations
\cite{AveTom12,CruLew07,PacYus11,PfeXue13,LiJi14,KarJac16} are known
from experimental data. The resulting spiking neural network can be
simulated on a cellular level by numerical integration of the spiking
dynamics of each individual neuron
(Fig.~\ref{fig:potjans-scheme}B). In the following, we will refer to
this level of description as the {\em microscopic} level. Apart from
being computationally expensive, the full microscopic network dynamics
is highly complex and hence difficult to understand. To overcome these
shortcomings, we have developed a new mean-field description for the
mesoscopic dynamics of interacting populations.

\subsection*{Mesoscopic population equations.}
\label{sec:mesosc-popul-equat}

A population $\alpha$ of size $N^\alpha$ is represented by its
population activity $A_N^\alpha(t)$ (Greek superscripts label the
populations, Fig.~\ref{fig:potjans-scheme}C) defined as
\begin{equation}
  \label{eq:pop-activ-def}
  A_N^\alpha(t)=\frac{1}{N^\alpha}\sum_{i=1}^{N^\alpha}s_i^\alpha(t).
\end{equation}
Here, $s_i^\alpha(t)=\sum_k\delta(t-t_{i,k}^\alpha)$ with the Dirac
$\delta$-function denotes the spike train of an individual neuron $i$
in population $\alpha$ with spike times $t_{i,k}^\alpha$. Empirically,
the population activity is measured with a finite temporal resolution
$\Delta t$.  In this case, we define the coarse-grained population activity as 
\begin{equation}
  \label{eq:A_N-empiric}
  A_N^\alpha(t)=\frac{\Delta n^\alpha(t)}{N^\alpha \Delta t},
\end{equation}
where $\Delta n^\alpha(t)$ is the number of neurons in population $\alpha$
that have fired in a time bin of size $\Delta t$ starting at time
$t$. The two definitions converge in the limit
$\Delta t\rightarrow 0$.

An example of population activities derived from spiking activity in a
cortical circuit model under a step current stimulation is shown in
Fig.~\ref{fig:potjans-scheme}B. To bridge the scales between neurons
and populations, the corresponding mean-field model should ideally
result in the same population activities as obtained from the full
microscopic model (Fig.~\ref{fig:potjans-scheme}D).  Because of the
stochastic nature of the population activities, however, the qualifier
``same'' has to be interpreted in a statistical sense. The random
fluctuations apparent in Fig.~\ref{fig:potjans-scheme}B,D are a
consequence of the finite number of neurons because microscopic
stochasticity is not averaged out in the finite sum in
Eq.~\eqref{eq:pop-activ-def}. This observation is important because
estimated neuron numbers reported in experiments on local cortical
circuits are relatively small \cite{LefTom09,AveTom12}. Therefore, a
quantitatively valid population model needs to account for finite-size
fluctuations. As mentioned above, we will refer to the
population-level with finite size populations ($N\sim 50$ to $2000$
per population) as the {\em mesoscopic} level.  In summary, we face
the following question: is it possible to derive a closed set of
evolution equations for the mesoscopic variables $A_N^\alpha(t)$ that
follow the same statistics as the original microscopic model?

To address this question, we describe neurons by generalized
integrate-and-fire (GIF) neuron models (Fig.~\ref{fig:potjans-scheme}A
(inset) and {\sc Methods}, Sec.~\Newnameref{sec:glm-model}) with
escape noise \cite{GerKis14}. In particular, neuron $i$ of population
$\alpha$ is modeled by a leaky integrate-and-fire model with dynamic
threshold \cite{ChaLon00,LiuWan01}. The variables of this model are
the membrane potential $u_i^\alpha(t)$ and the dynamic threshold
\blau{$\vartheta_i^\alpha(t)=\vth+\int_{-\infty}^t\theta^\alpha(t-t')s_i^\alpha(t')\,\mathrm{d}t'$
(Fig.~\ref{fig:potjans-scheme}A, inset), where $\vth$ is a baseline
threshold and $\theta^\alpha(t)$ is a spike-triggered adaptation
kernel or filter function that accounts for adaptation
\cite{GeiGol66,BibIva85,SchFis10,MenNau12,SchLin13,WebPil16} and other
spike-history effects \cite{GerKis14,WebPil16} via a convolution with
the neurons spike train $s_i^\alpha(t)$. In other words, the dynamic threshold depends on earlier spikes
$t_{i,k}^\alpha$ of neuron $i$:
$\vartheta_i^\alpha(t)\equiv
\vartheta^\alpha(t,t_{i,k}^\alpha<t)$.}
Spikes are elicited stochastically depending on the present state of
the neuron (Fig.~\ref{fig:potjans-scheme}A, inset). Specifically, the
probability that neuron $i$ fires a spike in the next time step
$[t,t+\Delta t]$ is given by $\lambda_i(t)\Delta t$, where
$\lambda_i^\alpha(t)$ is the conditional intensity of neuron $i$ (also
called conditional rate or hazard rate):
\begin{equation}
  \label{eq:hazard-def-results}
  \lambda_i^\alpha(t)=f^\alpha\left(u_i^\alpha(t)-\vartheta^\alpha(t,t_{i,k}^\alpha<t)\right)
\end{equation}
with an exponential function
$f^\alpha(x)=c^\alpha \exp(x/\Delta_u^\alpha)$. Analysis of
experimental data has shown that the ``softness'' parameter
$\Delta_u^\alpha$ of the threshold is in the range of 1 to 5 mV
\cite{JolRau06}. The parameter $c^\alpha$ can be interpreted as the
instantaneous rate at threshold. 

Besides the effect of a spike on the threshold as mediated by the
filter function $\theta^\alpha(t)$, a spike also triggers a change of
the membrane potential. In the GIF model ({\sc Methods},
Sec.~\Newnameref{sec:glm-model}), the membrane potential $u_i^\alpha(t)$ is
reset after spiking to a reset potential $\vreset$, to which
$u_i^\alpha(t)$ is clamped for an absolute refractory period
$\tref$. Absolute refractoriness is followed by a period of relative
refractoriness, where the conditional intensity
Eq.~\eqref{eq:hazard-def-results} is reduced. This period is
determined by the relaxation of the membrane potential from the reset
potential to the unperturbed or ``free'' potential, denoted $h(t)$,
which corresponds to the membrane potential dynamics in the absence of
resets.

The GIF model accurately predicts spikes of cortical neurons under
noisy current stimulation mimicking in-vivo like input
\cite{GerNau09,MenNau12} and its parameters can be efficiently
extracted from single neuron recordings
\cite{MenNau12,PozMen15}. \blau{Variants of this model have also been
suggested that explicitly incorporate biophysical properties such as
fast sodium inactivation \cite{PlaBre11,MenHag16}, conductance-based currents
\cite{ChiGra07} and synaptically-filtered background noise
\cite{ChiGra08}.}

\paragraph{Mean-field \gruen{approximations}.}

In order to derive a mesoscopic mean-field theory for populations of
GIF neurons, we first approximate the conditional intensity
$\lambda_i^\alpha(t)$ of an individual neuron by an effective rate
$\lambda_A^\alpha(t|\tl_i^\alpha)$ that only depends on its last spike
time $\tl_i^\alpha$ and on the history of the population activity
$A_N^\alpha(t')$, $t'<t$ (as expressed by the subscript ``$A$''). This is
called the quasi-renewal approximation \cite{NauGer12}. Taking into
account the dependence on the last spike is particularly important
because of neuronal refractoriness.

To obtain such a quasi-renewal description we make two
approximations. Firstly, we approximate the random connectivity by an
effective full connectivity with proportionally scaled down synaptic
weights (``mean-field approximation''). As a result, all neurons of
the same population are driven by identical synaptic input (see {\sc
  Methods}). This implies that for all neurons that had the same last
spike time, the time course of the membrane potential is identical,
$u_i^\alpha(t)\approx u_A(t,\tl_i^\alpha)$. Secondly, we make the
quasi-renewal approximation for GIF neurons \cite{NauGer12}, which
replaces the threshold $\vartheta_i(t)$ by an effective threshold
$\vartheta_A^\alpha(t,\tl_i^\alpha)$. Again, the effective threshold
only depends on the last spike time and the history of the population
activity. As a final result we obtain the conditional intensity for
all neurons with a given last spike time $\tl$ as
\begin{align}
  \label{eq:lambda-short}
  \lambda_A^\alpha(t|\tl)=f^\alpha\left(u_A^\alpha(t,\tl)-\vartheta_A^\alpha(t,\tl)\right)
\end{align}
(Fig.~\ref{fig:potjans-scheme}C, inset). A comparison of Eq.~(\ref{eq:lambda-short}) with
Eq.~(\ref{eq:hazard-def-results}) shows that the explicit dependence
on {\em all} past spike times $t_{i,k}^\alpha<\tl$ of a given neuron $i$
has disappeared. Instead, the conditional
intensity now only depends on the {\em last} firing time $\tl$ and the
past population activity $A_N^\alpha(t')$, $t'<t$. To keep the
notation simple, we omit in the following the population label
$\alpha$ at all quantities.

\paragraph{Finite-size mean field theory.}

\begin{figure}[tp]
\begin{adjustwidth}{\figureoffsetleft}{\figureoffsetright}
  \centering
  \includegraphics{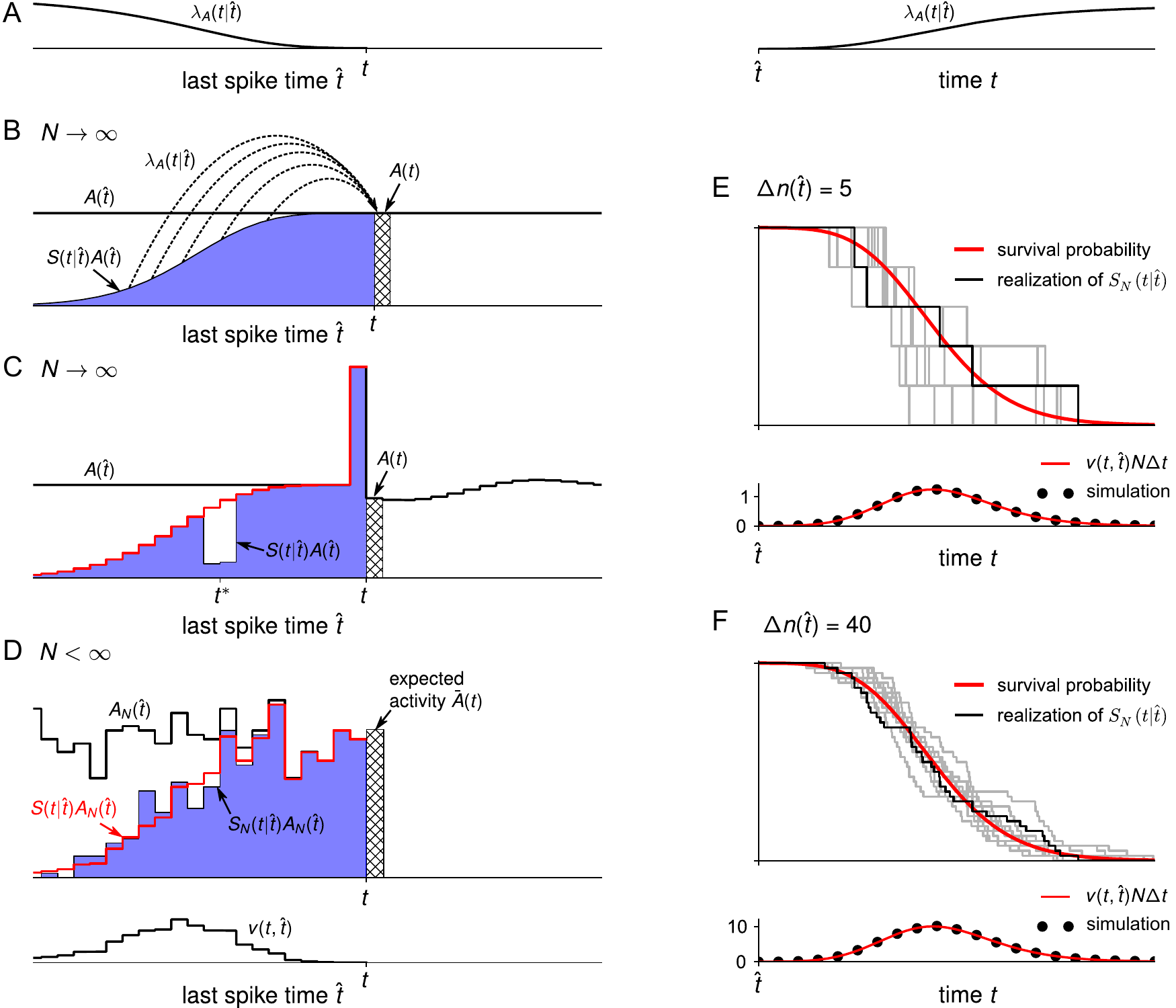}
  \caption{{\bf How fluctuations of the refractory density effect the population activity.}\\
    (A) The conditional intensity $\lambda_A(t|\tl)$ shown as a
    function of $\tl$ (left) and $t$ (right). Typically, the
    conditional intensity increases as the time since the last spike
    grows and the neuron recovers from refractoriness. (B) For
    $N\rightarrow\infty$, the population activity $A(t)$ (hatched bin)
    results from $\lambda_A(t|\tl)$ averaged over the \blau{last spike times $\tl$ with a weighting factor 
    $S(t|\tl)A(\tl)$ (blue) corresponding to the density of last spike times. Here,} $S(t|\tl)$ is the
    survival probability. (C) The \blau{last spike times} $\tl$ are
    discretized into time bins. In the bin immediately before $t$, a
    large fluctuation (blue peak) was induced by forcing some of the
    neurons with last spike time around $t^*$ to fire. At time $t$,
    the density of last spike times (blue) has a hole and a peak
    of equal probability mass. The red line shows the pseudo-density
    $S_0(t|\tl)A(\tl)$ that would be obtained if we had used the
    survival probability $S_0(t|\tl)$ of the unperturbed system. \blau{The peak at $\tl=t-\Delta t$ does not contribute to the activity $A(t)$ because of refractoriness, but the hole at $\tl=t^*$ contributes with a non-vanishing rate} (A), implying a reduction of $A(t)$ (hatched
    bin). (D) For a finite population size (here $N=400$), the
    refractory density $S_N(t|\tl)A_N(\tl)$ (blue), determines the
    expectation $\bar{A}(t)$ (hatched bin) of the fluctuating activity
    $A_N(t)$.  Analogous to the forced fluctuation in (C), the
    finite-size fluctuations are associated with negative and positive
    deviations in the refractory density (holes and overshoots)
    compared to the non-normalized density $S(t|\tl)A_N(\tl)$ (red
    line) that would be obtained if only the mean $S(t|\tl)$ and not
    the exact survival fraction $S_N(t|\tl)$ had been taken into
    account. The variance of the deviations is proportional to
    $v(t,\tl)$ given by Eq.~(\ref{eq:pt}). As a function of $\tl$,
    $v(t,\tl)$ shows the range of $\tl$ where deviations are most
    prominent (bottom).  (E,F) Given the number of neurons firing in
    the bin $[\tl,\tl+\Delta t)$, $\Delta n(\tl)$, the fraction of neurons
    that survive until time $t$ is shown for ten realizations (gray
    lines, one highlighted in black for clarity). The mean fraction
    equals the survival probability $S(t|\tl)$ (red line, top
    panel). The variance of the number of survived neurons at time
    $t$, $v(t,\hat{t})N\Delta t$, is shown at the bottom (red line:
    semi-analytic theory, Eq.~(\ref{eq:pt}; circles: simulation). (E)
    $\Delta n(\tl)=5$, (F) $\Delta n(\tl)=40$.}
  \label{fig:scheme-theory}
\end{adjustwidth}
\end{figure}

In this section, we present the main theoretical results with a focus
on the finite-size effects arising from neuronal
refractoriness. So far, we have effectively reduced the firing
probability of a neuron to a function $\lambda_A(t|\tl)$ that only
depends on its last spike time $\tl$
(Fig.~\ref{fig:scheme-theory}A). This allows us to describe the
evolution of the system by the density of the last spike time
\cite{Ger00,MeyVre02,ChiGra07,ChiGra08}. Because the last spike time
characterizes the refractory state of the neuron, this density will
also be referred to as the refractory density. Before we describe the
novel finite-$N$ theory, it is instructive to first recall the
population equations for the case of infinite $N$
(Fig.~\ref{fig:scheme-theory}B,C). Let us look at the population of
neurons at time $t$ and ask the following question: What fraction of
these neurons has fired their last spike between $\tl$ and
$\tl+\mathrm{d}\tl$? This fraction is given by the number of neurons
$A(\tl)\mathrm{d}\tl$ that have fired in this interval multiplied by
the survival probability $S(t|\tl)$ that such a neuron has not fired
again until time $t$. In other words, the product $Q_\infty(t,\tl)=S(t|\tl)A(\tl)$
evaluated at time $t$ represents the density of last spike times
$\tl$. Because a neuron with last spike time $\tl$ emits a spike with
rate $\lambda_A(t|\tl)$, the total population activity at time $t$ is
given by the integral \cite{Ger00}
\begin{equation}
  \label{eq:pop-eq-infty}
   A(t)=\int_{-\infty}^t\lambda_A(t|\tl)S(t|\tl)A(\tl)\,\mathrm{d}\tl.
\end{equation}
This situation is depicted in Fig.~\ref{fig:scheme-theory}B. At the
same time, the survival probability $S(t|\tl)$ of neurons that fired
their last spike at $\tl$ decays according to
\begin{equation}
  \label{eq:Q-dyn}
  \pd{S(t|\tl)}{t}=-\lambda_A(t|\tl)S(t|\tl)
\end{equation}
with initial condition $S(\tl|\tl)=1$
(Fig.~\ref{fig:scheme-theory}E,F red line). Equations \eqref{eq:pop-eq-infty} and
\eqref{eq:Q-dyn} define the population dynamics for
$N\rightarrow\infty$ \cite{WilCow72,Ger00,ChiGra08}.

In the limit $N\rightarrow\infty$, the dynamics of $A_N(t)=A(t)$ is
deterministic because microscopic noise averages out. Nevertheless,
the infinite-$N$ case is useful to understand the main effect of
fluctuations in the finite-$N$ case. To this end, let us perform the
following thought experiment: in a small time interval of length
$\Delta t$ immediately before time $t$, we induce a large, positive
fluctuation in the activity by forcing many of the neurons with last
spike close to a given time $\tl=t^*$ to fire a spike
(Fig.~\ref{fig:scheme-theory}C). As a result, the density of last
spike times at time $t$ exhibits a large peak just prior to time $t$
corresponding to the large number of neurons that have been forced to
fire in the time interval $[t-\Delta t,t)$. At the same time, these
neurons leave behind a ``hole'' in the density around
$\tl=t^*$. Because the number of neurons is conserved, this hole
exactly balances the area under the peak, and hence, the density of
last spike times remains normalized.  However, the two fluctuations
(the hole and the peak) have two different effects on the population
activity after time $t$. Specifically, the hole implies that some of
the neurons which normally would have fired at time $t$ with a nonzero
rate $\lambda_A(t|t^*)>0$ are no longer available. Moreover, neural
refractoriness implies that neurons which fired in the peak have a
small or even vanishing rate $\lambda_A(t|t-\Delta t)\approx 0$ at
time $t$. As a result, the population activity is reduced shortly
after the perturbation (Fig.~\ref{fig:scheme-theory}C). This example
highlights the importance of the normalization of the refractory
density as well as the state-dependence of the firing probability for
understanding the effect of fluctuations. In particular, the
normalization condition and neuronal refractoriness imply that a
positive fluctuation of the population activity is succeeded by a
negative fluctuation, and vice versa. This behavior is characteristic
for spiking neurons, which are known to exhibit negative
auto-correlations of their mean-centered spike trains at short time
lags (see e.g. \cite{PerGer67,CatRey06,GerKis14}).

We now turn to the finite-size case. In this case, it is advantageous
to discretize the last spike times into small bins of size $\Delta t$
that begin at the grid points $\tl_k=k\Delta t$,
$k\in\mathbb{Z}$. Furthermore, we adopt the definition of the
coarse-grained population activity, Eq.~\eqref{eq:A_N-empiric},
i.e. we consider the number of
spikes $\Delta n(\tl_k)$ in the time bin $[\tl_k,\tl_k+\Delta t)$. Instead of
the survival probability, we introduce the fraction of survived
neurons $S_N(t|\tl_k)$, $t>\tl_k$, such that $S_N(t|\tl_k)\Delta n(\tl_k)$ is
the number of neurons from bin $k$ that have not fired again until
time $t$ (Fig.~\ref{fig:scheme-theory}D,E). Dividing this number by
$N\Delta t$ and taking the continuum limit $\Delta t\rightarrow 0$,
yields the density of last spike times $Q_N(t,\tl)= S_N(t|\tl)A_N(\tl)$ in the
case of finite $N$. Since all neurons are uniquely identified by their
last spike time, this density is normalized \cite{Ger00}
\begin{equation}
    \label{eq:normalization-gen-contin}
    1=\int_{-\infty}^tS_N(t|\tl)A_N(\tl)\,\mathrm{d}\tl.
\end{equation}
We note that differentiating this equation with respect to time
  $t$ yields the population activity
  $A_N(t)=-\int_{-\infty}^t\partial_tS_N(t|\tl)A_N(\tl)\,\mathrm{d}\tl$
  as the formal finite-size analog of Eq.~\eqref{eq:pop-eq-infty}. The
  change of the survival fraction $\partial_tS_N(t|\tl)$, however, is
  not deterministic anymore as in Eq.~\eqref{eq:Q-dyn} but follows a
  stochastic jump process
(Fig.~\ref{fig:scheme-theory}E and F): In \rott{the time step
  $[t,t+\Delta t)$}, the number of survived neurons for a given bin $\tl_k$
in the past, \rott{$S_N(t|\tl_k)\Delta n(\tl_k)$,} makes a downstep $X(t,\tl_k)$ that corresponds to
  the number of neurons that fire in the group with last spike time
  $\tl_k$. \rott{For sufficiently small $\Delta t$, this} number is Poisson-distributed with mean
$\lambda_A(t|\tl_k)S_N(t|\tl_k)\Delta n(\tl_k)\Delta t$.  Hence, the
fraction of survived neurons $S_N(t|\tl_k)$ evolves in time like a
random stair case according to the update rule
$S_N(t+\Delta t|\tl_k)=S_N(t|\tl_k)-X(t,\tl_k)/\Delta n(\tl_k)$.
\blau{The activity $A_N(t)=\Delta n(t)/(N\Delta t)$ in the time bin
  starting at $t$ is given by the sum of all the downsteps,
  $\Delta n(t)=\sum_{\tl_k<t}X(t,\tl_k)$, where the sum runs over all
  possible last spike times. This updating procedure represents the evolution
  of the density of last spike times, $Q_N(t,\tl)=S_N(t|\tl)A_N(\tl)$,
  for finite $N$ under the quasi-renewal approximation (cf. {\sc
    Methods}, Eq.~\eqref{eq:QN-microdyn} and
  \eqref{eq:popact-micro}). Although it is possible to simulate such a
  finite-$N$ stochastic process using the downsteps $X(t,\tl_k)$, this
  process will not yield the reduced mesoscopic dynamics that we are
  looking for. The reason is that the variable $S_N(t|\tl_k)$ refers
  to the subpopulation of neurons that fired in the small time
  bin at $\tl_k$. For small $\Delta t$, the size of the subpopulation,
  $\Delta n(\tl_k)$, is a {\em small} number, much smaller than $N$. In
  particular, in the limit $\Delta t \rightarrow 0$, the simulation of
  $S_N(t|\tl_k) \Delta n(\tl_k)$ for all $\tl_k$ in the past would be
  as complex as the original microscopic dynamics of $N$
  neurons. Therefore we must consider such a simulation as a
  microscopic simulation. To see the difference to a mescoscopic
  simulation, we note that the population activity $A_N(t)$ involves
  the summation of many random processes (the downsteps $X(t,\tl_k)$)
  over many small bins. If we succeed to simulate directly $A_N(t)$
  from an underlying rate dynamics that depends deterministically on
  the past activities $A_N(t')$, $t'<t$, we will have a truely
  mescoscopic simulation. How to arrive at a formulation directly at
  the level of mescocopic quantities will be the topic of the rest of
  this section.}

\blau{The crucial question is} whether the stochasticity of the many different
random processes $\{S_N(t|\tl_k)\}_{\tl_k<t}$ can be reduced to a
single effective noise process that drives the dynamics on the
mesoscopic level. \rot{To this end, we note that for small $\Delta t$ and given history $\Delta n(\tl_k)$, $\tl_k<t$,}
each bin $\tl_k$ contributes with rate
$\lambda_A(t|\tl_k)S_N(t|\tl_k)\Delta n(\tl_k)$ a Poisson random
number of spikes to the total activity at time $t$
(Fig.~\ref{fig:scheme-theory}D). \rot{Therefore, the total number of
  spikes $\Delta n(t)$ is Poisson distributed with mean
  $N\bar A(t)\Delta t$, where $\bar A(t)$ is} the {\em expected}
population rate
\begin{equation}
  \label{eq:Abar-def}
    \bar A(t)=\int_{-\infty}^t\lambda_A(t|\tl)S_N(t|\tl)A_N(\tl)\,\mathrm{d}\tl.
\end{equation}
Here, the integral extends over all last spike times $\tl$ up to but
not including time $t$. \rot{Equation \eqref{eq:Abar-def} still depends on
the stochastic variables $\{S_N(t|\tl_k)\}_{\tl_k<t}$. The main
  strategy to remove this microscopic stochasticity is to
  \rott{use} the evolution of the survival probability $S(t|\tl)$, given by Eq.~\eqref{eq:Q-dyn}, and the normalization condition
  Eq.~\eqref{eq:normalization-gen-contin}. \rott{For finite $N$, the quantity $S(t|\tl_k)$ is formally defined as the solution of Eq.~\eqref{eq:Q-dyn} and}
  can be interpreted as the mean of the survival fraction \rott{$S_N(t|\tl_k)$} (Fig.~\ref{fig:scheme-theory}E,F, see {\sc Methods}). Importantly,
  $S(t|\tl_k)$ is a valid mesoscopic quantity since it only depends on
  the mesoscopic population activity $A_N$ (through
  $\lambda_A(t|\tl_k)$, cf. Eq.~(\ref{eq:Q-dyn}))}, and not on a
specific microscopic realization. \rot{Combining the survival
  probability $S(t|\tl)$ with the actual history of the mesoscopic
  activity $A_N(\tl)$ for $\tl<t$ yields the pseudo-density
  $Q(t,\tl_k)=S(t|\tl_k)A_N(\tl_k)$.}  In contrast to the macroscopic
density $Q_\infty(t,\tl)=S(t|\tl)A(\tl)$ in
Eq.~\eqref{eq:pop-eq-infty} or the microscopic density
$Q_N(t,\tl_k)=S_N(t|\tl_k)A_N(\tl_k)$, the pseudo-density is not
normalized. However, the pseudo-density $S(t|\tl_k)A_N(\tl_k)$ has the
advantage that it is based on mesoscopic quantities only.  

\blau{Let us} split the survival fraction into \rott{the mesoscopic quantity $S(t|\tl_k)$} and a
microscopic deviation, $S_N(t|\tl_k)=S(t|\tl_k)+\delta S(t|\tl_k)$.
\rot{In analogy to the artificial fluctuation in our thought
  experiment,} endogenously generated fluctuations in the finite-size
system are accompanied by deviations of the microscopic density
$S_N(t|\tl_k)A_N(\tl_k)$ from the pseudo-density
$S(t|\tl_k)A_N(\tl_k)$ (Fig.~\ref{fig:scheme-theory}C and D, red
line). A negative deviation (\rot{$\delta S(t|\tl_k)<0$}) can be
interpreted as a hole and a positive deviation
(\rot{$\delta S(t|\tl_k)>0$}) as an overshoot (compare red curve and
blue histogram in Fig.~\ref{fig:scheme-theory}D).  \blau{Similar to the thought experiment, the effect of these deviations needs to be weighted by the conditional intensity $\lambda_A(t|\tl_k)$ in order to arrive at the population activity. Equation (\ref{eq:Abar-def}) yields}
\begin{equation}
  \label{eq:abar-split}
    \bar A(t)=\int_{-\infty}^t\lambda_A(t|\tl)S(t|\tl)A_N(\tl)\,\mathrm{d}\tl+\int_{-\infty}^t\lambda_A(t|\tl)\delta S(t|\tl)A_N(\tl)\,\mathrm{d}\tl.  
\end{equation}
Analogously, the normalization of the refractory density, Eq.~\eqref{eq:normalization-gen-contin}, can be written as
\begin{equation}
  \label{eq:norm-cond-split}
    1=\int_{-\infty}^tS(t|\tl)A_N(\tl)\,\mathrm{d}\tl+\int_{-\infty}^t\delta S(t|\tl)A_N(\tl)\,\mathrm{d}\tl.
\end{equation}
We refer to the second integral in Eq.~\eqref{eq:abar-split} as a
correction term because it corrects for the error that one would make
if one sets $S_N=S$ in Eq.~(\ref{eq:Abar-def}). This correction term
represents the overall contribution of the holes ($\delta S<0$) and
overshoots ($\delta S>0$) to the expected activity.

\rot{To eliminate the microscopic deviations $\delta S(t|\tl)$ in
  Eq.~\eqref{eq:abar-split} we use the normalization condition,
  Eq.~\eqref{eq:norm-cond-split}.} This is possible because the
correction term is tightly constrained by the sum of all holes and
overshoots, $\int_{-\infty}^t\delta S(t|\tl)A_N(\tl)\,\mathrm{d}\tl$,
which by Eq.~\eqref{eq:norm-cond-split}, is completely determined by
the past mesoscopic activities. \blau{Equations \eqref{eq:abar-split} and
\eqref{eq:norm-cond-split}} suggest to make the deterministic ansatz
$\int_{-\infty}^t\lambda_A\delta
S(t|\tl)A_N(\tl)\,\mathrm{d}\tl\approx\Lambda(t)\int_{-\infty}^t\delta
S(t|\tl)A_N(\tl)\,\mathrm{d}\tl$
for the correction term.  As shown in {\sc Methods} (``Mesoscopic
population equations''), the optimal rate $\Lambda(t)$ that minimizes
the mean squared error of this approximation is given by
\begin{equation}
  \label{eq:def-func}  
  \Lambda(t)=\frac{\int_{-\infty}^t\lambda_A(t|\tl)v(t,\tl)\,\mathrm{d}\tl}{\int_{-\infty}^tv(t,\tl)\,\mathrm{d}\tl}.
\end{equation}
\rot{Here, the quantity $v(t,\tl)$, called variance function, obeys} the differential equation
\begin{equation}
  \label{eq:pt}
  \pd{v}{t}=-2\lambda_A(t|\tl)v+\lambda_A(t|\tl)S(t|\tl)A_N(\tl)
\end{equation}
with initial condition $v(\tl,\tl)=0$ (see {\sc Methods},
Eq.~\eqref{eq:dvar}). \rot{Importantly, the dynamics of $v$ involves
  mesoscopic quantities only, and hence $v$ is mesoscopic.}  \blau{As
  shown in {\sc Methods} and illustrated in Fig.~\ref{fig:scheme-theory}D
  (bottom),} we can interpret $v(t,\tl_k)N\Delta t$ as the variance of
the number of survived neurons, $S_N(t|\tl_k)\Delta n(\tl_k)$.  \blau{
  To provide an interpretation of the effective rate $\Lambda(t)$ we
  note that, for fixed $t$, the normalized variance
  $v(t,\tl)/\int_{-\infty}^tv(t,\tl)\,\mathrm{d}\tl$ is a probability
  density over $\tl$.} Thus, the effective rate $\Lambda(t)$ can be
regarded as a weighted average of the conditional intensity
$\lambda_A(t|\tl)$ that accounts for the expected amplitude of the
holes and overshoots.

Using the effective rate $\Lambda(t)$ in Eq.~(\ref{eq:abar-split}) results in the
expected activity
\begin{equation}
  \label{eq:master-cont-a-inf-main}
  \bar{A}(t)=\int_{-\infty}^t \lambda_A(t|\tl)S(t|\tl)A_N(\tl)\,\mathrm{d}\tl+\Lambda(t)\left(1-\int_{-\infty}^t S(t|\tl)A_N(\tl)\,\mathrm{d}\tl\right).
\end{equation}
Looking at the structure of Eq.~\eqref{eq:master-cont-a-inf-main}, we
find that the first term is the familiar population integral known
from the infinite-$N$ case, Eq.~\eqref{eq:pop-eq-infty}. The second
term is a correction that is only present in the finite-$N$ case. In
fact, in the limit $N\rightarrow\infty$, the pseudo-density
$S(t|\tl)A_N(\tl)$ converges to the macroscopic density
$S(t|\tl)A(\tl)$, which is normalized to unity. Hence the correction
term vanishes and we recover the population equation
\eqref{eq:pop-eq-infty} for the infinite system.

To obtain the population activity we consider \rot{an infinitesimally small}
time scale $\mathrm{d}t$ such that the probability of a neuron to fire
during an interval $[t,t+\mathrm{d}t)$ is much smaller than one,
i.e. $\bar A(t)\mathrm{d}t\ll 1$. On this time scale, \rot{the total number
of spikes $\mathrm{d} n(t)$} is an independent, Poisson distributed
random number with mean $\bar A(t)N\mathrm{d}t$, where
$\bar A(t)$ is given by Eq.~\eqref{eq:master-cont-a-inf-main}. \rot{From
Eq.~\eqref{eq:A_N-empiric} thus follows} the population activity
\begin{equation}
  \label{eq:pop-act-coarse}
  A_N(t)=\frac{1}{N}\frac{\mathrm{d}n(t)}{\mathrm{d}t},\qquad \mathrm{d}n(t)\sim\text{Pois}[\bar A(t)N\mathrm{d}t].
\end{equation}
\rot{Alternatively, the population
activity can be represented as a $\delta$-spike train, or ``shot
noise'', $A_N(t)=\frac{1}{N}\sum_{k}\delta(t-t_{\text{pop},k})$, where $\{t_{\text{pop},k}\}_{k\in\mathbb{Z}}$ is a random point
process with a conditional intensity function
$\lambda_{\text{pop}}\bigl(t|\mathcal{H}_t\bigr)=N\bar A(t)$.  Here, the condition $\mathcal{H}_t$ denotes the
history of the point process $\{t_{\text{pop},k}\}$ up to (but not
including) time $t$, or equivalently the history of the population
activity $A_N(\tl)$ for $\tl<t$. The conditional intensity means that the conditional expectation of
the population activity is given by
$\lrk{A_N(t)}|_{\mathcal{H}_t}=\bar A(t)$, which according to
Eq.~\eqref{eq:master-cont-a-inf-main} is indeed a deterministic
functional of the past activities.   Finally, we note that the
case of large but finite populations permits a Gaussian approximation, which yields the more
explicit form}
\begin{equation}
  \label{eq:AN-Gaussian}
  A_N(t)=\bar A(t)+\sqrt{\frac{\bar A(t)}{N}}\xi(t).
\end{equation}
Here, $\xi(t)$ is a Gaussian white noise with correlation function
$\langle\xi(t)\xi(t')\rangle=\delta(t-t')$.  \blau{The correlations of
  $\xi(t)$ are white because spikes at $t'$ and $t>t'$ are independent
  {\em given} the expected population activity $\bar A(t)$ at time
  $t$. However, we emphasize that the expected population activity
  $\bar A(t)$ does include information on the past fluctuations
  $\xi(t')$ at time $t'$. Therefore the fluctuations of the total
  activity $A_N(t)$ are not white but a sum of a colored process
  $\bar A(t)$ and a white-noise process $\xi(t)$ \cite{DegSch14}. The
  white noise gives rise to the delta peak of the auto-correlation
  function at zero time lag which is a standard feature of any spike
  train, and hence also of $A_N(t)$. The colored process $\bar A(t)$,
  on the other hand, arises from Eq.~\eqref{eq:master-cont-a-inf-main}
  via a filtering of the {\em actual} population activity $A_N(\tl)$
  which includes the past fluctuations $\xi(\tl)$. For neurons with
  refractoriness, $\bar A(t)$ is negatively correlated with recent
  fluctuations $\xi(\tl)$ (cf. the thought experiment of
  Fig.~\ref{fig:scheme-theory}B) leading to a trough at small time
  lags in the spike auto-correlation function
  \cite{PerGer67,CatRey06,GerKis14}.}

The set of coupled equations \eqref{eq:Q-dyn},~(\ref{eq:pt}),
\eqref{eq:def-func}~--~\eqref{eq:pop-act-coarse} constitute the desired mesoscopic
population dynamics and is the main result of the paper. The dynamics
is fully determined by the history of the mesoscopic population
activity $A_N$. The Gaussian white noise in
Eq.~\eqref{eq:AN-Gaussian} or the independent random number involved
in the generation of the population activity via
Eq.~\eqref{eq:pop-act-coarse} is the only
source of stochasticity and summarizes the effect of microscopic noise
on the mesoscopic level. Microscopic detail such as the knowledge of
how many neurons occupy a certain microstate $\tl$ has been
removed. 

One may wonder where the neuronal interactions enter in the population
equation. Synaptic interactions are contained in the conditional
intensity $\lambda_A(t|\tl)$ which depends on the membrane potential
$u_A(t,\tl)$, which in turn is driven by the synaptic current that
depends on the population activity via Eq.~(\ref{eq:I_syn-fully}) in
{\sc Methods}. An illustration of the derived mesoscopic model is
shown in Fig.~\ref{fig:potjans-scheme}C (inset). In this section, we
considered a single population to keep the notation simple. However,
it is straightforward to formulate the corresponding equations for the
case of several equations as shown in {\sc Methods},
Sec.~\Newnameref{sec:several-pop}.
\rot{
\paragraph*{Stochastic population dynamics can be efficiently
  simulated forward in time. }
\label{sec:sim-algo-results}

The stochastic population equations provide a rapid means to integrate
the population dynamics on a mesoscopic level. To this end, we devised
an efficient integration algorithm based on approximating the infinite
integrals in the population equation
Eq.~\eqref{eq:master-cont-a-inf-main} by discrete sums over a finite
number of refractory states $\tl$ ({\sc Methods},
Sec.~\Newnameref{sec:sim-algo}). The algorithm involves the generation
of only one random number per time step and population, because the
activity is sampled from the mesoscopic rate $\bar A^\alpha(t)$. In
contrast, the microscopic simulation requires in each time step to
draw a random number for each neuron. Furthermore, because the
population equations do not \blau{depend on} the number of neurons, we
expect a significant speed-up factor for large neural networks
compared to a corresponding microscopic simulation. For example, the
microscopic simulation of the cortical column in
Fig.~\ref{fig:potjans-scheme}B took 13.5 minutes to simulate 10
seconds of biological time, whereas the corresponding forward
integration of the stochastic population dynamics
(Fig.~\ref{fig:potjans-scheme}D) took only 6.6 seconds on the same
machine (see Sec.~\Newnameref{sec:micro-vs-micro}).

A pseudocode of the algorithm to simulate neural population dynamics
is provided in {\sc Methods} (Sec.~\Newnameref{sec:sim-algo}). In
addition to that, a reference implementation of this algorithm \blau{is}
publicly available under \url{https://github.com/schwalger/mesopopdyn_gif}, and
\blau{will be integrated} in the Neural Simulation Tool (NEST)
\cite{GewDie07}, \url{https://github.com/nest/nest-simulator}, as a module \blau{presumably}
named {\tt gif\_pop\_psc\_exp}.
}

\subsection*{Two different consequences of finite $N$}
\label{sec:poiss-abs-refrac}

For a first analysis of the finite-size effects, we consider the
special case of a fully-connected network of Poisson neurons with
absolute refractory period \cite{GerKis14}. In this case,
the conditional intensity can be represented as
$\lambda_A(t|\tl)=f\bigl(h(t)\bigr)\Theta(t-\tl-\tref)$, where $\tref$
is the absolute refractory period, $\Theta(\cdot)$ is the Heaviside
step function and $h(t)$ is the free membrane potential, which obeys the
passive membrane dynamics
\begin{equation}
  \label{eq:rate-eq}
  \taum \od{h}{t}=-h+\mu(t) +\taum J\, (\epsilon*A_N)(t),
\end{equation}
where $\taum$ is the membrane time constant,
\blau{$\mu(t)=\vrest+RI(t)$ (where $\vrest$ is the resting potential
  and $R$ is the membrane resistance)} accounts for all currents
\blau{$I(t)$} that are independent of the population activities, $J$
is the synaptic strength and $\epsilon(t)$ is a synaptic filter kernel
(see {\sc Methods}, Eq.~(\ref{eq:h-eta}) for details).  \blau{For the
mathematical analysis, we assume that the activity $A_N(t)$ and input
$\mu(t)$ have started at $t = -\infty$ so that we do not need to worry
about initial conditions.  In a simulation, we could for example start
at $t=0$ with initial conditions $A_N(t)=\delta(t)$ for $t\le 0$ and
$h(0) =0$.}

For the conditional intensity \rot{given above}, the effective rate
$\Lambda(t)$, Eq.~\eqref{eq:def-func}, is given by
$\Lambda(t)=f\bigl(h(t)\bigr)$ because the variance $v(t,\tl)$ is zero
during the absolute refractory period $t-\tref\le\tl<t$. As a result,
the mesoscopic population equation (\ref{eq:master-cont-a-inf-main})
reduces to the simple form
\begin{equation}
 \label{eq:wilson-cowan-abs-refract}
 \bar A(t)=f(h(t))\left[1-\int_{t-\tref}^tA_N(\tl)\,\mathrm{d}\tl\right].
\end{equation}
This mesoscopic equation is exact and could have been constructed
directly in this simple case.  For $N\rightarrow\infty$, where
$A_N(t)$ becomes identical to $\bar A(t)$, this equation has been
derived by Wilson and Cowan \cite{WilCow72}, see also
\cite{Ger00,DegHel10,GerKis14}. The intuitive interpretation of
Eq.~(\ref{eq:wilson-cowan-abs-refract}) is that the activity at time
$t$ consists of two factors, the ``free'' rate
$\freehaz(t)=f\bigl( h(t)\bigr)$ that would be expected in the absence
of refractoriness and the fraction of actually available (``free'')
neurons that are not in the refractory period. For finite-size
populations, these two factors explicitly reveal two distinct
finite-size effects: firstly, the free rate is driven by the
fluctuating population activity $A_N(t)$ via Eq.~(\ref{eq:rate-eq})
and hence the free rate exhibits finite-size fluctuations. This effect
originates from the transmission of the fluctuations through the
recurrent synaptic connectivity. Secondly, the fluctuations of the
population activity impacts the refractory state of the population,
i.e. the fraction of free neurons, as revealed by the second factor in
Eq.~(\ref{eq:wilson-cowan-abs-refract}). In particular, a large
positive fluctuations of $A_N$ in the recent past reduces the fraction
of free neurons, which causes a negative fluctuation of the number
$N\bar A(t)\Delta t$ of expected firings in the next time
step. Therefore, refractoriness generates negative correlations of the
fluctuations $\langle\Delta A(t)\Delta A(t')\rangle$ for small
$|t-t'|$. \gruen{We note that such a decrease of the expected rate would
  not have been possible if the correction term in
  Eq.~\eqref{eq:master-cont-a-inf-main} was absent. However,
  incorporating the effect of recent fluctuations (i.e. fluctuations
  in the number of refractory neurons) on the number of free
  neurons by adding the correction term, and thereby balancing the total number of neurons,  recovers
  the correct equation~\eqref{eq:wilson-cowan-abs-refract}.}

The same arguments can be repeated in the general setting of
Eq.~\eqref{eq:master-cont-a-inf-main}. Firstly, the conditional
intensity $\lambda_A(t|\tl)$ depends on the past fluctuations of the
population activity because of network feedback. Secondly, the
fluctuations lead to an imbalance in the number of
neurons across different states of relative refractoriness
(i.e. fluctuations do not add up to zero)
which gives rise to the ``correction term'', i.e. the second term on the
r.h.s. of Eq.~\eqref{eq:master-cont-a-inf-main}.

\subsection*{Comparison of microscopic and mesoscopic simulations}
\label{sec:micro-vs-micro}

We wondered how well the statistics of the population activities
obtained from the integration of the mesoscopic equations compare with
the corresponding activities generated by a microscopic simulation. As
we deal with a finite-size system, not only to the first-order
statistics (mean activity) but also higher-order statistics needs to
be considered. Because there are several approximations involved
(e.g. full connectivity, quasi-renewal approximation and effective
rate of fluctuations in the refractory density), we do not expect a
perfect match.  To compare first- and second-order statistics, we will
mainly use the power spectrum of the population activities in the
stationary state (see {\sc Methods}, Sec.~\Newnameref{sec:power-spectrum}).

\paragraph{Mesoscopic equations capture refractoriness.}

\begin{figure}[t]
\begin{adjustwidth}{\figureoffsetleft}{\figureoffsetright}
  \centering
  \includegraphics{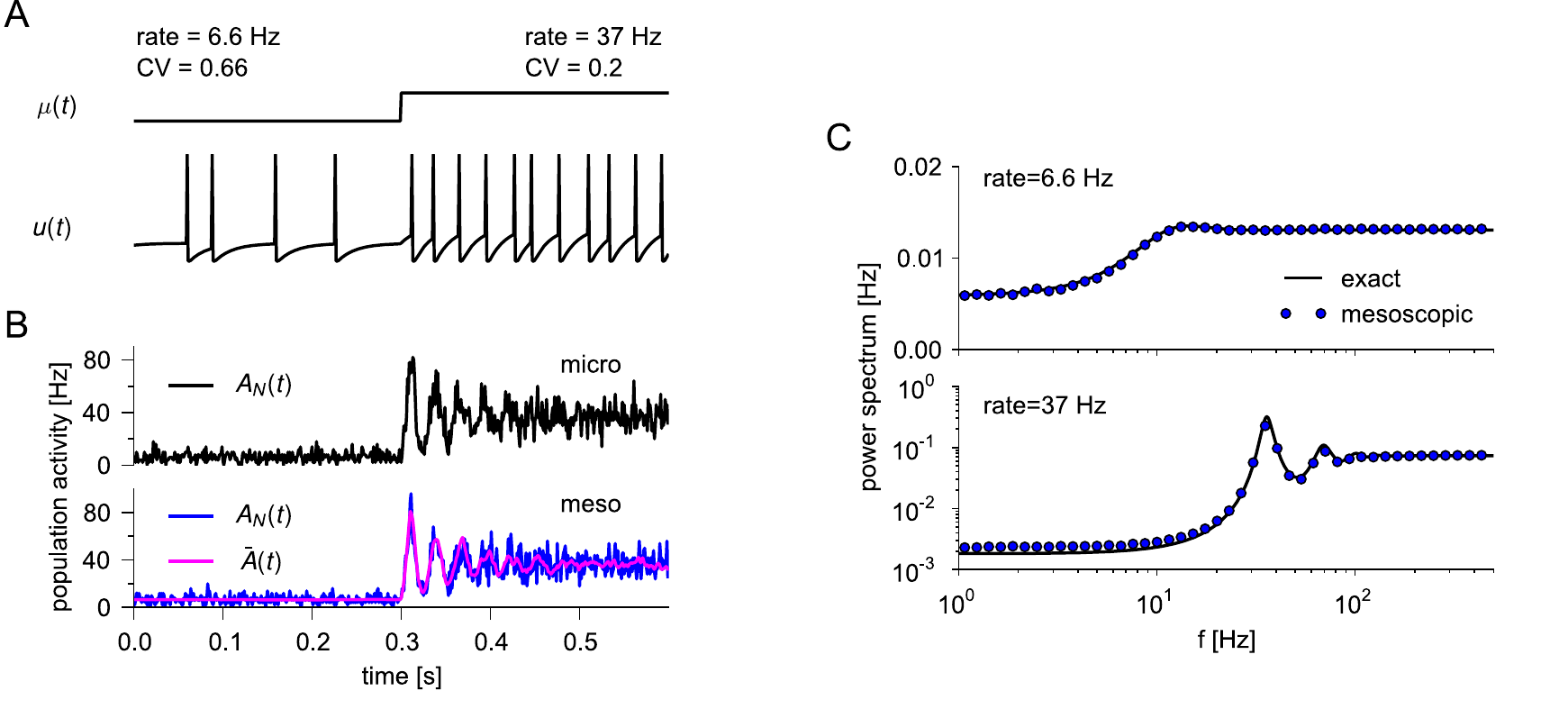}
  \caption{{\bf Population activity of uncoupled leaky integrate-and-fire neurons without adaptation.} \\
    (A) Neurons were stimulated by a step current
    $I_\text{ext}(t)$ such that $\mu=\vrest+RI_\text{ext}(t)$ increased
    from $\mu=15$~mV to $\mu=30$~mV (top). Voltage trace of one of 500
    neurons (bottom). Stationary firing statistics (rate and
    coefficient of variation (CV) of the interspike intervals)
    corresponding to the two stimuli are indicated above the step
    current. (B) Realizations of the population activity $A_N(t)$ for
    the microscopic (top) and mesoscopic (bottom, blue line)
    simulation. The magenta line shows the expected population rate
    $\bar A(t)$ given the past actual realization $A_N(t')$ for
    $t'<t$. (C) The power spectrum of the stationary activity $A_N(t)$
    obtained from renewal theory, Eq.~\eqref{eq:renewal-psd}, (black
    solid line) and from the mesoscopic simulation (blue circles). The
    top and bottom panel corresponds to weak ($\mu=15$~mV) and strong
    ($\mu=30$~mV) constant stimulation (transient removed).}
  \label{fig:lif}
\end{adjustwidth}
\end{figure}

Our theory describes the interplay between finite-size fluctuations
and spike-history effects.  The most prominent spike-history effect is
refractoriness, i.e. the strong effect of the last spike on the
current probability to spike. To study this effect, we first focus on
a population of uncoupled neurons with a constant threshold
corresponding to leaky integrate-and-fire (LIF) models without
adaptation  (Fig~\ref{fig:lif}). The reset of the membrane
potential after each spike introduces a period of relative
refractoriness, where spiking is less likely due to a hyper-polarized
membrane potential (Fig.~\ref{fig:lif}A). Because of the reset to a
fixed value, the spike trains of the LIF neurons are renewal
processes. In the stationary state, the fluctuation statistics as
characterized by the power spectrum is known analytically for the case
of a population of independent renewal spike trains
(Eq.~\eqref{eq:renewal-psd} in {\sc Methods}).

\rot{A single realization of the} population activity $A_N(t)$
fluctuates around the expected activity $\bar A(t)$ that exhibits a
typical ringing in response to a step current stimulation
\cite{Kni72,Ger00}. The time course of the expected activity as well
as the size of fluctuations are \rot{roughly} similar for microscopic simulation
(Fig.~\ref{fig:lif}B, top) and the numerical integration of the
population equations (Fig.~\ref{fig:lif}B, bottom). We also note that
the expected activity $\bar A(t)$ is not constant in the stationary
regime but shows weak fluctuations. This is because of the feedback of
\blau{the actual realization of $A_N(t')$ for $t'<t$} onto the dynamics of $\bar A(t)$,
Eq.~\eqref{eq:master-cont-a-inf-main}.

A closer inspection confirms that the fluctuations generated by the
mesoscopic population dynamics in the stationary state exhibit the
same power spectrum \blau{(Fig.~\ref{fig:lif}C)} as the theoretically
predicted one, which is given by Eq.~\eqref{eq:renewal-psd}. In
particular, the mesoscopic equations capture the fluctuation
statistics even at high firing rates, where the power spectrum
strongly deviates from the white (flat) power spectrum of a Poisson
process (Fig.~\ref{fig:lif}C bottom). The pronounced dip at
low-frequencies is \blau{a well-known signature of} neuronal refractoriness
\cite{FraBai95}.

\paragraph{Mesoscopic equations capture adaptation and burstiness.}

\begin{figure}[p]
\begin{adjustwidth}{\figureoffsetleft}{\figureoffsetright}
  \centering
  \includegraphics{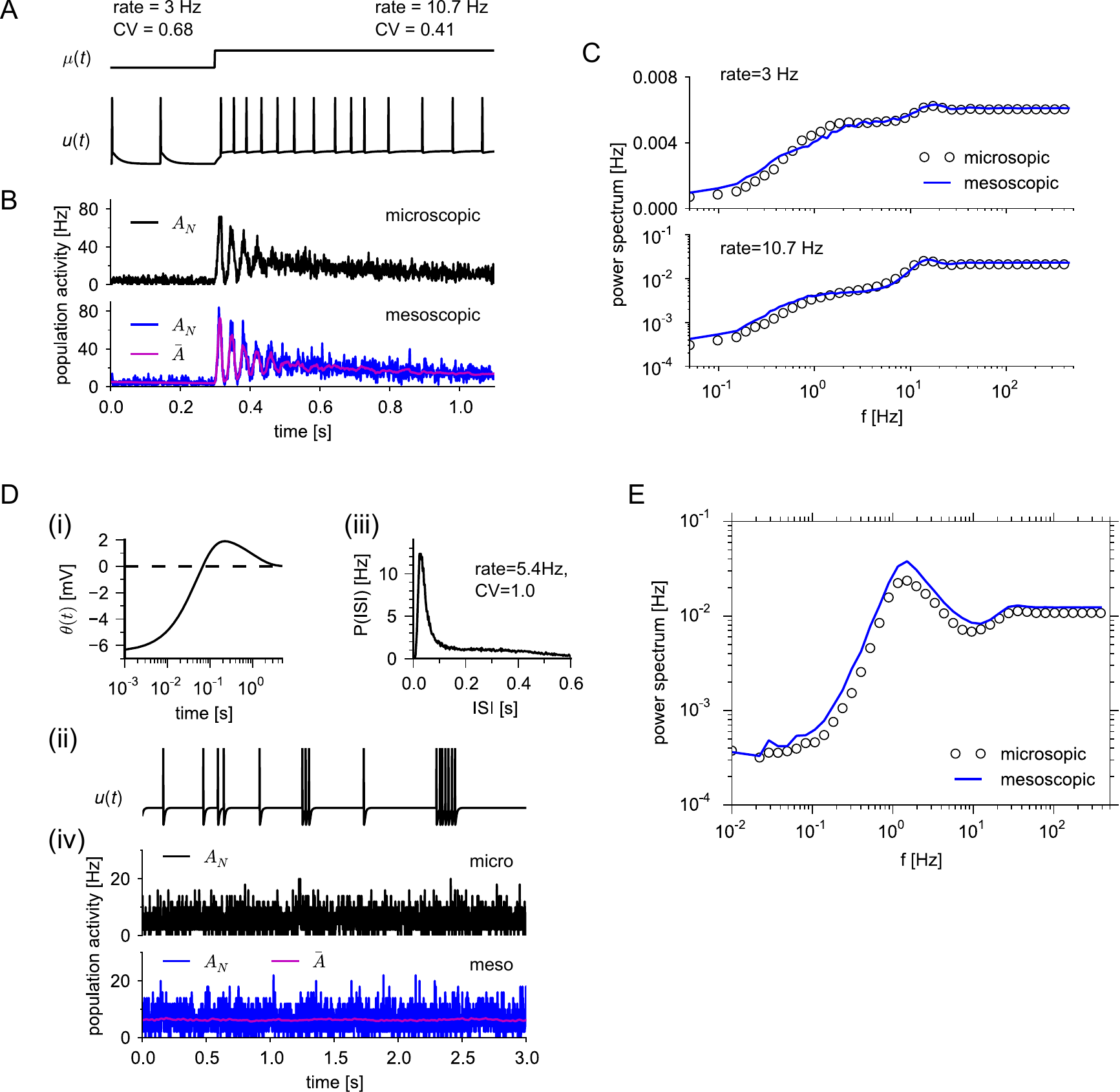}
  \caption{{\bf Population dynamics captures adaptation and burstiness.}\\
    (A) 500 adapting leaky integrate-and-fire neurons were stimulated
    by a step current $I_\text{ext}$ that increased $\mu=\vrest+RI_\text{ext}(t)$
    from $\mu=12$~mV to $\mu=27$~mV (top). Voltage trace of one neuron
    (bottom). Stationary firing statistics (rate and coefficient of
    variation (CV)) corresponding to the two stimuli are indicated
    above the step current. (B) Realizations of the population
    activity obtained from microscopic simulation (black) and
    mesoscopic population equation (blue) as well as $\bar A(t)$
    (magenta). (C) Power spectra corresponding to the stationary
    activity (averaged over 1024 trials each of 20~s length) at
    low and high firing rates as in (A), circles and lines depict
    microscopic and mesoscopic case, respectively. Parameters in
    (A)--(C): $\vth=10$~mV, $\vreset=25$~mV, threshold kernel
    $\theta(t)=\sum_{\ell=1,2}(J_{\theta,\ell}/\tau_{\theta,\ell})e^{-t/\tau_{\theta,\ell}}$ for
    $t\ge\tref$ with $J_{\theta,1}=1.5$~mV$\cdot$s,
    $\tau_{\theta,1}=0.01$~s, $J_{\theta,2}=1.5$~mV$\cdot$s,
    $\tau_{\theta,2}=1$~s.  (D) Bursty neuron model.  (i) Biphasic
    threshold kernel $\theta(t)$, where a combination of a negative
    part (facilitation) and a positive part (adaptation) yields a
    bursty spike pattern, (ii) sample firing pattern of one neuron.
    (iii) The interspike interval distribution with values of rate and
    CV. (E) Power spectrum of the population activity $A_N(t)$ shown
    in (D)-(iv). Parameters in (D) and (E): $\mu=20$, $\vth=10$~mV,
    $\vreset=0$~mV, $\taum=0.01$~s, facilitation:
    $J_{\theta,1}=-0.45$~mV$\cdot$s, $\tau_{\theta,1}=0.05$~s;
    adaptation: $J_{\theta,2}=2.5$~mV$\cdot$s,
    $\tau_{\theta,2}=1$~s }
  \label{fig:bursting}
\end{adjustwidth}  
\end{figure}

Further important spike-history effects can be realized by a dynamic
threshold. For instance, spike-frequency adaptation, where a neuron
adapts its firing rate in response to a step current after an initial
strong response (Fig.~\ref{fig:bursting}A,B), can be modeled by an
accumulating threshold that slowly decays between spikes
\cite{GeiGol66,ChaLon00}. In \rot{single realizations}, the mean
population rate as well as the size of fluctuations appear to be
\rot{similar} for microscopic and mesoscopic case
(Fig.~\ref{fig:bursting}B, top and bottom, respectively). \blau{For a
  more quantitative comparison we compared the ensemble statistics as
  quantified by the power spectrum. This comparison reveals that the
  fluctuation statistics is well captured by the mesoscopic model
  (Fig.~\ref{fig:bursting}C). The main} effect of adaptation is a
marked reduction in the power spectrum at low frequencies \blau{compared to
the non-adaptive neurons of Fig.~\ref{fig:lif}.} The small
discrepancies compared to the microscopic simulation originate from
the quasi-renewal approximation, which does not account for the
individual spike history of a neuron before the last spike but only
uses the population averaged history. This approximation is expected
to work well if the threshold kernel changes slowly, effectively
averaging the spike history locally in time \cite{NauGer12}.

In the case of fast changes of the threshold kernel, we do not expect
that the quasi-renewal approximation holds. For example, a biphasic
kernel \cite{GerHem96} with a facilitating part at short interspike
intervals (ISI) and an adaptation part for large ISIs
(Fig.~\ref{fig:bursting}D-(i)) can realize bursty spike patterns
(Fig.~\ref{fig:bursting}D-(ii)). The burstiness is reflected in the
ISI density \blau{by} a peak at small ISIs, corresponding to ISIs
within a burst, and a tail that extends to large ISIs representing
interburst intervals (Fig.~\ref{fig:bursting}D-(iii)). Remarkably, the
mesoscopic equations with the quasi-renewal approximation
qualitatively capture the burstiness, as can be seen by the strong
low-frequency power at about $1$~Hz (Fig.~\ref{fig:bursting}E). At the
same time, the effect of adaptation manifests itself in a reduced
power at even lower frequencies. \rot{The systematic overestimation of
  the power across frequencies implies a larger variance of the
  empirical population activity obtained from the mesoscopic
  simulation, which is indeed visible by looking at the single realizations
  (Fig.~\ref{fig:bursting}D-(iv)).}  As an aside, we note that
facilitation which is strong compared to adaptation can lead to
unstable neuron dynamics even for isolated neurons \cite{GerDeg17}.

\paragraph{Recurrent network of randomly connected neurons.}
 
\begin{figure}[p]
\begin{adjustwidth}{\figureoffsetleft}{\figureoffsetright}
  \centering
  \includegraphics[width=7.5in]{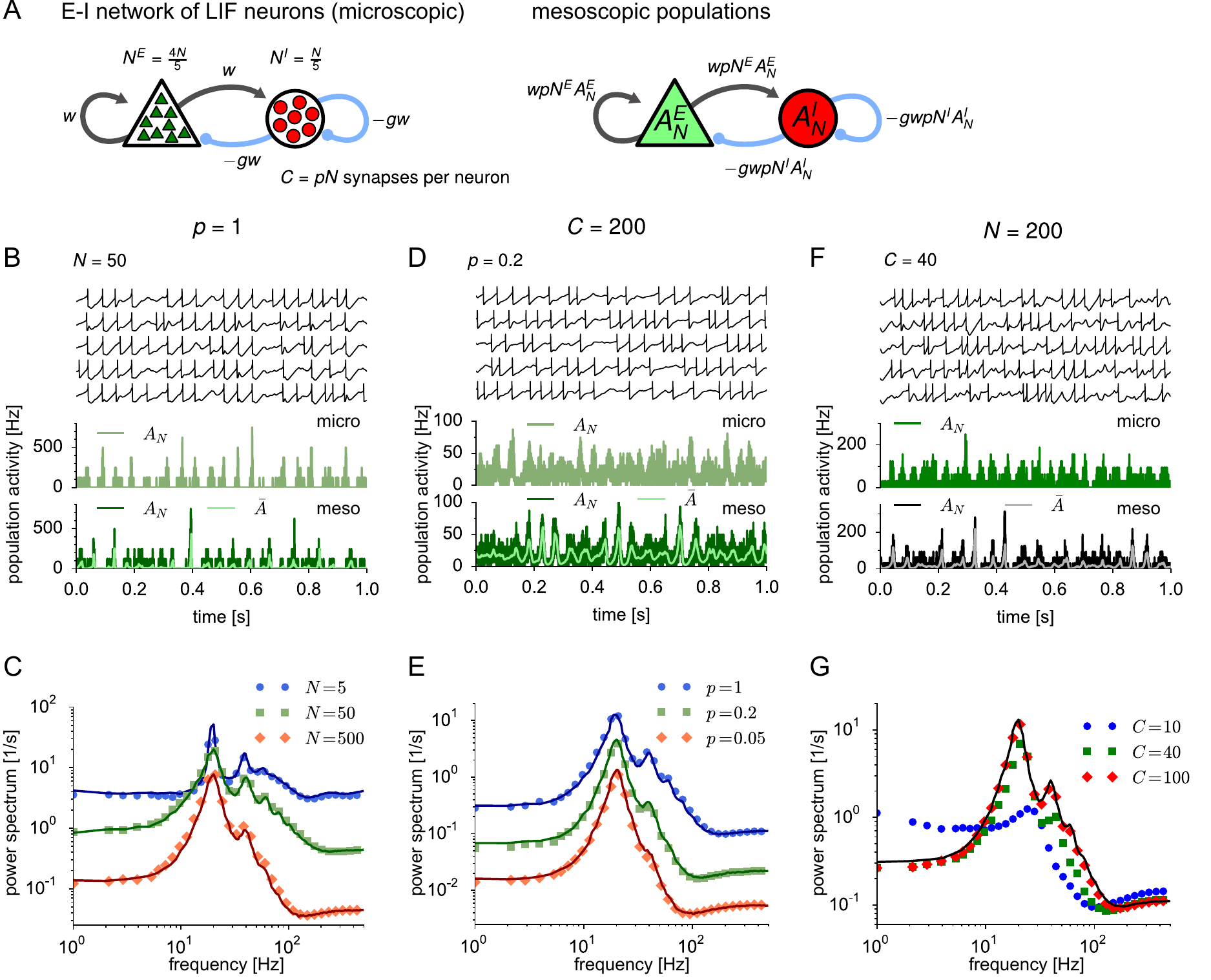}
  \caption{{\bf Mesoscopic dynamics of E-I network for varying network size $N$, connection probability $p$ and number of synapses per neuron $C$.}\\
    (A) Left: Schematic of the network of $N^E$ excitatory and
    $N^I=N^E/4$ inhibitory leaky integrate-and-fire (LIF) neurons,
    each receiving $C^E=pN^E$ ($C^I=pN^I$) connections from a random
    subset of excitatory (inhibitory) neurons. Total numbers are
    $N=N^E+N^I$ and $C=C^E+C^I$. At $C=200$, the synaptic strength is
    $w=0.3$~mV and $-gw=-1.5$~mV for excitatory and inhibitory
    connections, respectively. To preserve a constant mean synaptic
    input, the synaptic strength is scaled such that $Cw=const.$.
    Right: Schematic of a corresponding mesoscopic model of two
    interacting populations. (B) Trajectories of $u(t)$ for five
    example neurons (top) and of the excitatory population activity
    $A_N^E(t)$ obtained from the network simulation (middle) and the
    mesoscopic simulation (bottom, dark green) for $C=N=50$; time
    resolution $\Delta t=0.2$~ms. The light green trajectory (bottom
    panel) depicts the expected population activity $\bar{A}^E(t)$
    given the past activity. (C) Power spectra of $A_N^E(t)$ for
    different network sizes while keeping $p=1$ fixed (microscopic:
    symbols, mesoscopic: solid lines with corresponding dark
    colors). (D) Sample trajectories corresponding to the green curve
    in (E) ($N=1000$, $p=0.2$). (E) Analogously to (C) but varying the
    connection probabilities while keeping $C=pN=200$ fixed. (F)
    Sample trajectories corresponding to the green curve in (G)
    ($N=200$, $C=40$). (G) Analogously to (C) but varying the number
    of synapses $C$ while keeping $N=200$ fixed. Note that the
    mesoscopic theory (black solid line) is independent of $C$ because
    the product $Cw$, which determines the interaction strength in the
    mesoscopic model (see, left panel of (A)), is kept
    constant. Parameters: $\mu^{E/I}=24$~mV, $\Delta_u^{E/I}=2.5$~mV and $\theta^{E/I}(t)\equiv 0$ (no
    adaptation).}
  \label{fig:ei-net}
\end{adjustwidth}
\end{figure}

So far, we have studied populations of uncoupled neurons. This allowed
us to demonstrate that the mesoscopic dynamics captures effects of
single neuron dynamics on the fluctuations of the population
activity. Let us now suppose that each neuron in a population $\alpha$
is randomly connected to presynaptic neurons in population $\beta$
with probability $p^{\alpha\beta}$ such that the in-degree is fixed to
$p^{\alpha\beta}N^\beta$ connections. In the presence of synaptic
coupling, the fluctuations at time $t$ are propagated through the
recurrent connectivity and may significantly influence the population
activity at time $t'>t$. For instance, in a fully-connected network
\blau{($p^{\alpha\beta}=p=1$ for all $\alpha$ and $\beta$)} of excitatory and inhibitory neurons (E-I network,
Fig.~\ref{fig:ei-net}B,C), all neurons within a population receive
identical inputs given by the population activities $A_N^\alpha(t)$
(cf. Eq.~\eqref{eq:I_syn-fully}). Finite-size fluctuations of
$A_N^\alpha(t)$ generate common input to all neurons and tend to
synchronize neurons. This effect manifests itself in large
fluctuations of the population activity
(Fig.~\ref{fig:ei-net}B). Since the mean-field approximation of the
synaptic input becomes exact in the limit $p\rightarrow 1$, we expect
a good match between the microscopic and mesoscopic simulation in this
case. Interestingly, the power spectra of the population activities
obtained from these simulations coincide well even for an extremely
small E-I network consisting of only one inhibitory and four
excitatory neurons (Fig.~\ref{fig:ei-net}C). \rot{This extreme case of
  $N=5$ neurons with strong synapses (here, $w^{EE}=w^{IE}=12$~mV,
  $w^{II}=w^{EI}=-60$~mV) highlights the non-perturbative character
  of our theory for fully-connected networks, which does not require the inverse system size or the synaptic strength \blau{to be small}.}  In general, the power spectra
reveal pronounced oscillations that are induced by finite size
fluctuations \cite{WalBen11}. The amplitude of these stochastic
oscillations decreases as the network size increases and vanishes in
the large-$N$ limit.

\begin{figure}[t]
\begin{adjustwidth}{\figureoffsetleft}{\figureoffsetright}
  \centering
  \includegraphics[width=7.5in]{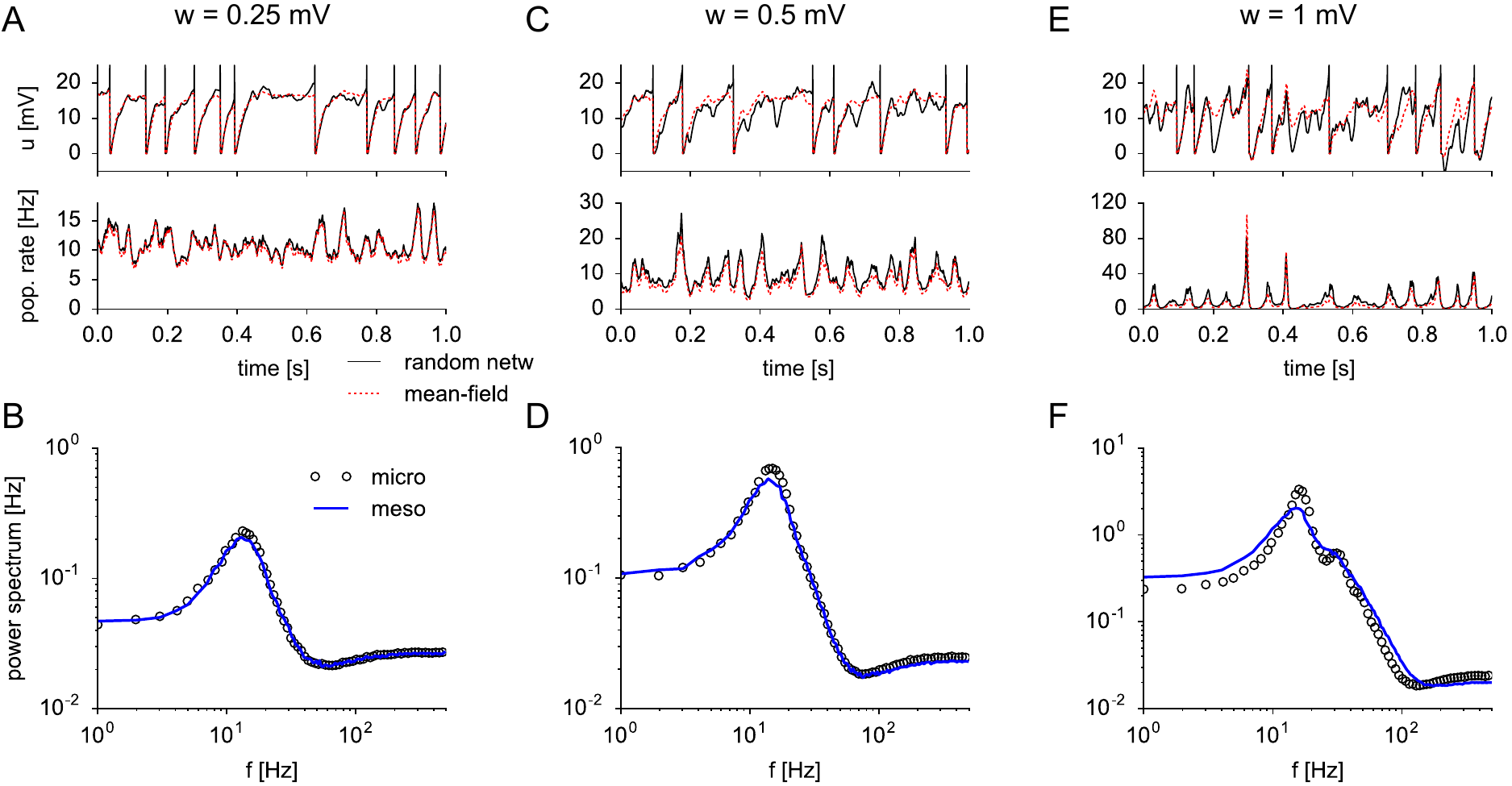}
  \caption{\rot{{\bf Mean-field approximation of synaptic input for randomly connected networks.}\\
The same E-I network as in Fig.~\ref{fig:ei-net} with $N=500$ neurons and connection probability $p=0.2$ was simulated for increasing synaptic strength $w^{EE}=w^{IE}=w$ ($w^{EI}=w^{II}=-5w$) of excitatory (inhibitory) connections: (A,B) $w=0.25$~mV,  (C,D) $w=0.5$~mV  (E,F) $w=1$~mV. (A,C,E) Top: Membrane potential of one example neuron shows fluctuations due to spike input from $C=100$ presynaptic neurons (black line), which represent a random subset of all 500 neurons. The mean-field approximation of the membrane potential (dashed red line) assumes that the neuron \blau{had the same firing times but was} driven by all neurons, i.e. by the population activities $A_N^{E}(t)$ and $A_N^{I}(t)$, with rescaled synaptic strength $w_\text{MF}^{E/I}=pw^{E/I}$. Although individual membrane potentials differ significantly from the mean-field approximation (top), the relevant population-averaged hazard rates $\bar A_\text{micro}^{E/I}(t)\equiv\frac{1}{N^{E/I}}\sum_{i=1}^{N^{E/I}}\lambda(t|\tl_i)$ (bottom) are well predicted by the mean-field approximation. (B,D,F) Corresponding power spectra of the (excitatory) population activity for microscopic (circles) and mesoscopic (blue solid line) simulation. Parameters as in Fig.~\ref{fig:ei-net} except $\mu^{E/I}=18$~mV.}}
  \label{fig:mf-approx}
\end{adjustwidth}
\end{figure}

If the network is not fully but randomly-connected ($0<p<1$), neurons
still share a part of the finite-size fluctuations of the population
activity. Earlier theoretical studies
\cite{BruHak99,MeyVre02,HelTet14} have pointed out that these common
fluctuations inevitably yield correlated and partially synchronized
neural activity, as observed in simulations
(Fig.~\ref{fig:ei-net}D,F). This genuine finite-size effect
\rot{decreases for larger networks approaching} an asynchronous state
\cite{RenRoc10} (Fig.~\ref{fig:ei-net}C,E). As argued in previous
studies \cite{BruHak99,Bru00,MatGiu02}, the fluctuations of the
synaptic input can be decomposed into two components, a coherent and
an incoherent one. The coherent fluctuations are given by the
fluctuations of the population activity and are thus common to all
neurons of a population. This component is exactly described by our
\rot{mean-field approximation,
  $u_i^\alpha(t)\approx u_A(t,\tl_i^\alpha)$ used in
  Eq.~\eqref{eq:lambda-short} (cf. {\sc Methods},
  Eq.~\eqref{eq:mf-appox})}. The incoherent fluctuations are caused by
the quenched randomness of the network (i.e. each neuron receives
input from a different subpopulation of the network) and have been
described as independent \rot{Poisson} input in earlier studies
\cite{BruHak99,Bru00,MatGiu02}. \rot{If we compare the membrane
  potential of a {\em single} neuron with the one expected from the
  mean-field approximation (Fig.~\ref{fig:mf-approx}A,C,E, top), we
  indeed observe a significant difference in fluctuations. This
  difference originates from the incoherent component. } \blau{Differences
in membrane potential will lead to differences in the instantaneous
spike emission probability for each individual neuron;
cf. Eq.~\eqref{eq:hazard-def-results}. However, in order to calculate the population activity we
need to average the conditional firing rate of Eq.~\eqref{eq:hazard-def-results} over all
neurons in the population (see Methods, Eq.~\eqref{eq:mf-lam} for details).
Despite the fact that each neuron is characterized by a different last
firing time $\tl$, the differences in firing rate caused by voltage
fluctuations will, for sufficiently large $N$ and not too small $p$,
average out whereas common fluctuations caused by past fluctuations in
the population activity will survive (Fig.~\ref{fig:mf-approx}A,C,E,
  bottom). In other words, the coherent
component is the one that dominates the finite-N activity whereas the
incoherent one is washed out.} \rot{Therefore, mesoscopic  population activities can be well
  described by our mean-field approximation even when the network is
  not fully connected (Fig.~\ref{fig:ei-net}E,G and
  Fig.~\ref{fig:mf-approx}B,D,F).}  Remarkably, even for $C=200$
synapses per neuron and $p=0.05$, the mesoscopic model agrees well
with the microscopic model. However, if both $N$ and $p$ are small,
the mesoscopic theory breaks down as expected (Fig.~\ref{fig:ei-net}G,
blue circles). \rot{Furthermore, strong synaptic weights imply strong
  incoherent noise, which is then passed through the exponential
  nonlinearity of the hazard function. This may lead to deviations of
  the population-averaged hazard rate from the corresponding
  mean-field approximation (Fig.~\ref{fig:mf-approx}E, bottom), and,
  consequently, to deviations between microscopic and mesoscopic
  population activities in networks with strong random connections
  (Fig.~\ref{fig:mf-approx}F). }

\paragraph{Finite-size induced switching in bistable networks.}

\begin{figure}[p]
\begin{adjustwidth}{\figureoffsetleft}{\figureoffsetright}
  \centering
  \includegraphics[width=7.5in]{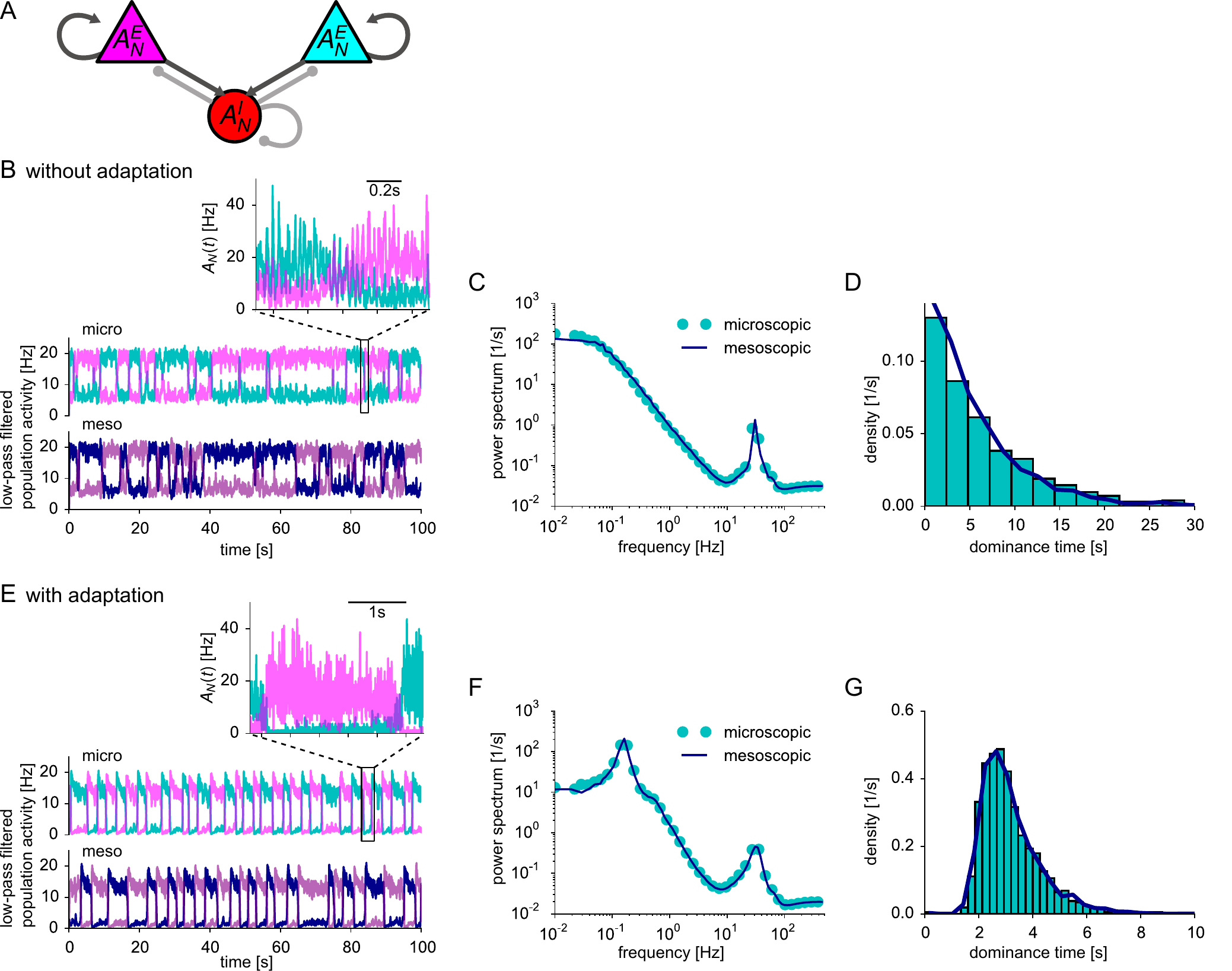}
  \caption{{\bf Finite-size induced switching in a bistable network.} \\
    (A) Schematic of a winner-take-all network architecture: Two
    competing excitatory populations ($N^{E1}=N^{E2}=400$) interact
    with a common inhibitory population ($N^I=200$). (B)-(D) In the
    absence of adaptation ($\theta^{E/I}(t)\equiv 0$), the excitatory
    populations switch between low and high activities in an irregular
    fashion (B). Activities in (B,E) are low-pass-filtered by a moving
    average of $100$~ms. Top: full network simulation. Inset:
    Magnified view of the activities for $1$~s (without moving
    average) showing fast large-amplitude oscillations. Bottom:
    mesoscopic simulation. (C) The power spectrum of the activity of
    the excitatory populations exhibits large low-frequency power and
    a high-frequency peak corresponding to the slow stochastic
    switching between high- and low-activity states and the fast
    oscillations, respectively. (D) The density of the dominance times
    (i.e. the residence time in the high-activity states) has an exponential
    form. (E-G) Like (B-D) but excitatory neurons exhibit weak and
    slow adaptation
    ($\theta^E(t)=(J_\theta/\tau_\theta)e^{-t/\tau_\theta}$
    with $J_\theta=0.1$~mV$\cdot$s, $\tau_\theta=1$~s for
    $t\ge\tref$). Switching between high- and low-activity states is
    more regular than in the non-adapting case as revealed by a
    low-frequency peak in the power spectrum (F) and a narrow,
    unimodal density of dominance times (G). In (C,D) and (F,G)
    microscopic and mesoscopic simulation correspond to cyan
    symbols/bars and dark blue solid lines, respectively. Parameters:
    $\mu^{E/I}=36$~mV except for $\mu^E=36.5$ in (E-G) to compensate
    adaptation. Time step $\Delta t=0.01$~ms (microscopic),
    $\Delta t=0.2$~ms (mesoscopic). Efficacies of excitatory and
    inhibitory connections: $w^{E}=0.0624$~mV and $w^I=-0.2496$~mV
    (B-D), and $w^{E}=0.096$~mV and $w^I=-0.384$~mV (E-G), $p=1$, $\Delta_u^{E/I}=2.5$~mV. }
  \label{fig:bistable}
\end{adjustwidth}
\end{figure}

In large but finite E-I networks, the main effect of weak finite-size
fluctuations is to distort the deterministic population dynamics of
the infinitely large network (stable asynchronous state or limit cycle
motion) leading to stochastic oscillations and phase diffusion that
can be understood analytically by linear response theory
\cite{MatGiu02,LinDoi05,WalBen11,DegSch14} and weakly nonlinear
analysis \cite{BruHak99}. This is qualitatively different in networks
with multiple stable states. In such networks, finite-size
fluctuations may have a drastic effect because they enable large switch-like transitions between metastable states that cannot be described by a
linear or weakly nonlinear theory. We will show now that our
mesoscopic population equation accurately captures strongly nonlinear
effects, such as large fluctuations in multistable networks.

Multistability in spiking neural networks can emerge as a collective
effect in balanced E-I networks with clustered connectivity
\cite{LitDoi12}, and, generically, in networks with a
winner-take-all architecture, where excitatory populations compete
through inhibitory interactions mediated by a common inhibitory
population (see,
e.g. \cite{HerPal91,Wan98,WonWan06,GerKis14,MazFon15,LagRot15} and
Fig.~\ref{fig:bistable}A). Jumps between metastable states have been
used to model switchings in bistable perception
\cite{MorRin07,ShpMor09,TheKov11}. 
To understand such finite-size induced switching in spiking neural
networks on a qualitative level, phenomenological rate models have been
usually employed \cite{MorRin07,LagRot15}. In these models, stochastic
switchings are enabled by noise added to the deterministic rate
equations in an {\it ad hoc} manner. Our mesoscopic mean-field
equations keep the spirit of such rate equations, however with the
important difference that the noisy dynamics is systematically derived
from the underlying spiking neural network without any free
parameter. Here, we show that the mesoscopic mean-field equations {\em
  quantitatively} reproduce finite-size induced transitions between
metastable states of spiking neural networks. We emphasize that the
switching statistics depends sensitively on the properties of the
noise that drive the transitions \cite{HanTal90}. Therefore, an
accurate account of finite-size fluctuations is expected to be
particularly important in this case.

We consider a simple bistable network of two excitatory populations
with activities $A^{E1}$ and $A^{E2}$, respectively, that are
reciprocally connected to a common inhibitory population with activity
$A^{I}$ (Fig.~\ref{fig:bistable}A). We choose the mean input and the
connection strength such that in the large-$N$ limit the population
activities exhibited two stable equilibrium states, one corresponding
to a situation, where $A^{E1}$ is high and $A^{E2}$ is low, the other
state corresponding to the inverse situation, where $A^{E1}$ is low
and $A^{E2}$ is high. We found that in smaller networks, finite-size
fluctuations are indeed sufficient to induce transitions between the
two states leading to repeated switches between high- and low-activity states
(Fig.~\ref{fig:bistable}B,E). The regularity of the switching appears
to depend crucially on the presence of adaptation, as has been
suggested previously \cite{MorRin07,TheKov11}. Remarkably, both in the
presence and absence of adaptation, the switching dynamics of the
spiking neural network appears to be well reproduced by the mesoscopic
mean-field model. 

For a more quantitative comparison, we use several statistical
measures that characterize the bistable activity. Let us first
consider the case without adaptation. As before, we compare the power
spectra of the population activity for both microscopic and mesoscopic
simulation and find a good agreement (Fig.~\ref{fig:bistable}C). The
peak in the power spectrum at relatively high-frequency reveals
strong, rapid oscillations that are visible in the population activity
after a switch to the high-activity state (inset of
Fig.~\ref{fig:bistable}B with magnified view). In contrast, the large
power at low frequencies corresponds to the slow fluctuations caused
by the switching of activity between the two excitatory populations,
as revealed by the low-pass filtered population activity
(Fig.~\ref{fig:bistable}B).  The Lorentzian shape of the power
spectrum caused by the slow switching dynamics is consistent with
stochastically independent, exponentially distributed residence times
in each of the two activity states (i.e., a homogeneous Poisson
process).  The residence time distribution shows indeed a monotonic,
exponential decay (Fig.~\ref{fig:bistable}D) both in the microscopic
and mesoscopic model.  Furthermore, \rott{we found that} residence times do not exhibit
significant serial correlations. Together, this
confirms the Poissonian nature of bistable switching in \blau{our
three-population model of neurons without adaptation.}

\rot{In models for perceptual bistability, residence times in the
  high-activity state are often called dominance times. The
  distribution of dominance times is usually not exponential but has
  been described by a more narrow, gamma-like distribution (see,
  e.g. \cite{CaoPas16}). 
\blau{Such a more narrow distribution emerges in a three-population network where excitatory neurons are weakly adaptive.} When the population enters a high-activity
  state, the initial strong increase of the population activity is now
  followed by a slow adaptation} to a somewhat smaller, stationary
activity (Fig.~\ref{fig:bistable}E). Eventually, the population jumps
back to the low-activity state. The switching dynamics is much more
regular with than without adaptation leading to slow stochastic
oscillations as highlighted by a second peak in the power spectrum at
low frequencies (Fig.~\ref{fig:bistable}F) and a narrow distribution
of dominance times (Fig.~\ref{fig:bistable}G), in line with
previous theoretical studies \cite{MorRin07,ShpMor09,TheKov11}. We
emphasize, however, that in contrast to these studies the underlying
deterministic dynamics for $N\rightarrow\infty$ is in our case not
oscillatory but bistable, because the adaptation level is below the
critical value necessary in the deterministic model to switch back to
the low-activity state.

The emergence of regular switching due to finite-size noise can be
understood by interpreting the residence time of a given population in
the high-activity state as arising from two stages: (i) the initial
transient of the activity to a decreased (but still large)
stationary value due to adaptation and (ii) the subsequent
noise-induced escape from the stationary adapted state. The first stage is
deterministic and hence does not contribute to the variability of the
residence times. The variability results mainly from the second
stage. The duration of the first stage is determined by the adaptation
time scale. If this time covers a considerable part of the total
residence time, we expect that the coefficient of variation (CV),
defined as the ratio of standard deviation and mean residence time, is
small. In the case without adaptation, a
deterministic relaxation stage can be neglected against the mean
noise-induced escape time so that the CV is larger.


\paragraph{Mesoscopic dynamics of cortical microcolumn.}

\begin{figure}[t]
\begin{adjustwidth}{\figureoffsetleft}{\figureoffsetright}
  \centering
  \includegraphics{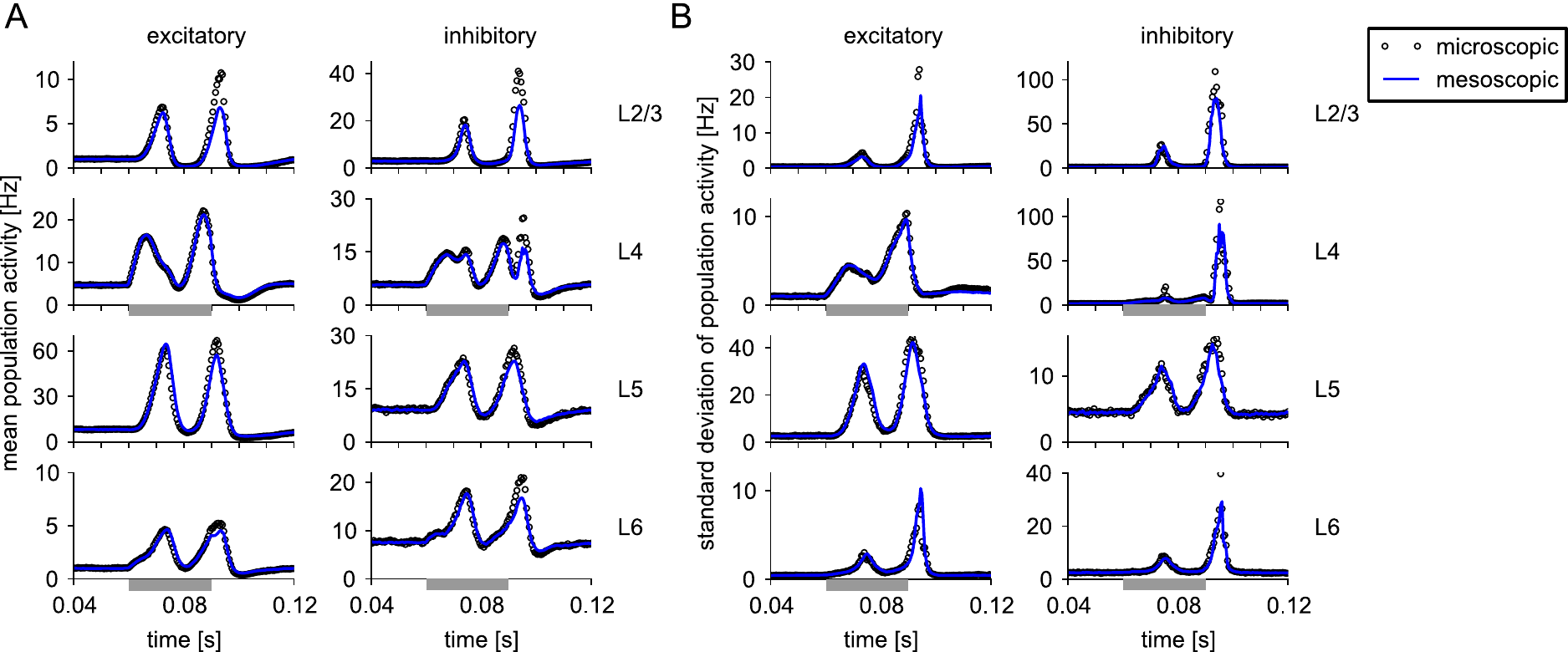}
  \caption{{\bf Time-dependent statistics of the population activities in a cortical column model.}\\
    (A) Trial-averaged population activities (peri-stimulus-time
    histogram, PSTH) in the modified Potjans-Diesmann model as
    illustrated for a single trial in
    Fig.~\ref{fig:potjans-scheme}. Circles and blue solid line show
    microscopic simulation (250 trials, simulation time step
    $\Delta t=0.01$~ms) and mesoscopic simulation (1000 trials,
    $\Delta t=0.5$~ms), respectively. A step current mimicking
    thalamic input is provided to neurons in layer 4 and 6 during a
    time window of 30 ms as indicated by the gray bar. Rows correspond
    to the layers L2/3, L4, L5 and L6, respectively, as
    indicated. Columns correspond to excitatory and inhibitory
    populations, respectively. (B) Corresponding, time-dependent
    standard deviation of $A_N(t)$ measured with temporal resolution
    $\Delta t=0.5$~ms.}
  \label{fig:potjans_psth}
\end{adjustwidth}
\end{figure}

\begin{figure}[t]
  \centering
  \includegraphics{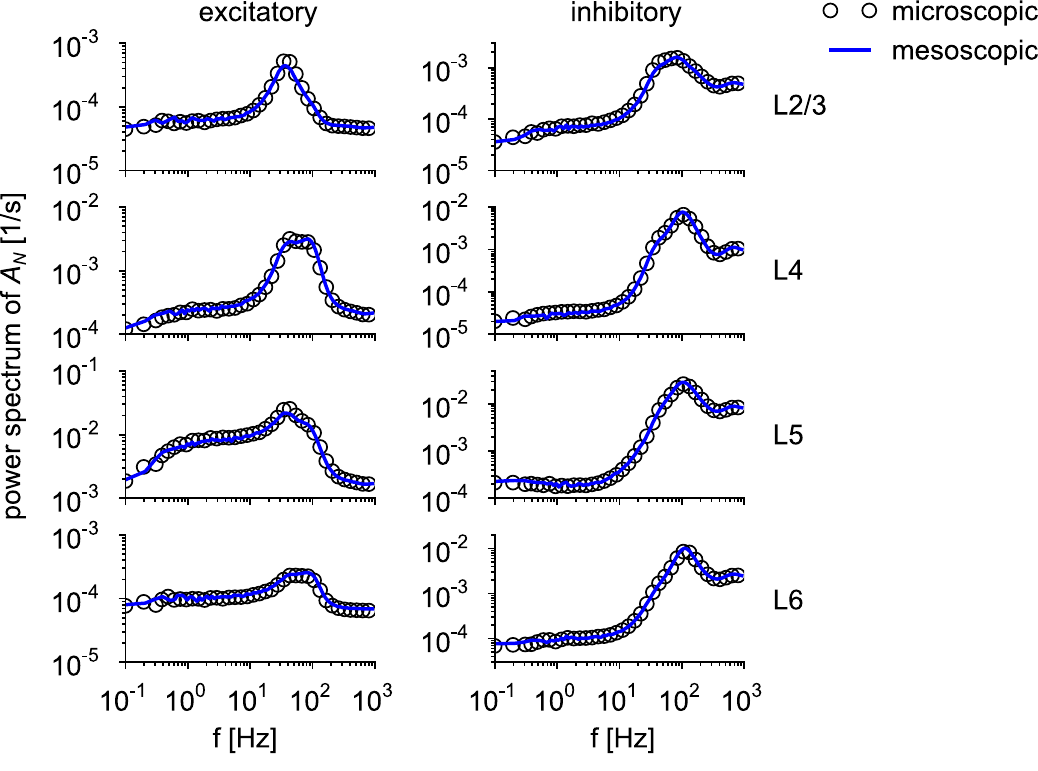}
  \caption{{\bf Stationary statistics of population activities in a cortical column model.}\\
    Power spectra of the spontaneous population activities $A_N(t)$ in
    the modified Potjans-Diesmann model in the absence of
    time-dependent thalamic input (corresponding to the activities shown
    in Fig.~\ref{fig:potjans-scheme}B (microscopic) and
    Fig.~\ref{fig:potjans-scheme}D (mesoscopic) outside of the
    stimulation window. Circles and blue solid lines represent
    microscopic and mesoscopic simulation, respectively. Rows
    correspond to the layers L2/3, L4, L5 and L6, respectively, as
    indicated. Columns correspond to excitatory and inhibitory
    populations, respectively.}
  \label{fig:potjans_spectra}
\end{figure}

As a final example, we applied the mesoscopic population equations to
a biologically more detailed model of a local cortical
microcircuit. Specifically, we used the multi-laminar column model of
V1 proposed by Potjans and Diesmann \cite{PotDie14} (see also \cite{BosDie16,CaiIye16} for an analysis of this model). It consists of
about $80'000$ non-adapting integrate-and-fire neurons organized into
four layers (L2/3, L4, L5 and L6), each accommodating an excitatory
and an inhibitory population (see schematic in
Fig.~\ref{fig:potjans-scheme}A). The neurons are randomly connected
within and across the eight populations. We slightly changed this
model to include spike-frequency adaptation of excitatory neurons, as
observed in experiments (see e.g. \cite{MenNau12}). Furthermore, we
replaced the Poissonian background noise in the original model by an
increase of mean current drive and escape noise (both in the
microscopic and mesoscopic model). The mean current drive was chosen
such that the firing rates of the spontaneous activity were matched to
the firing rates in the original model. We note that the fitting of
the mean current was made possible by the use of our population
equations, which allow for an efficient evaluation of the firing
rates.  The complete set of parameters is listed in \rott{{\sc Methods}, Sec.~\Newnameref{sec:potjans-model}.}

Sample trajectories of the population activities have already served
as an illustration of our approach in
Fig.~\ref{fig:potjans-scheme}, where neurons in layer 4
and 6 are stimulated by a step current of 30 ms duration, mimicking
input from the thalamus as in the original study
\cite{PotDie14}. Individual realizations obtained from the microscopic
and mesoscopic simulation differ due to the marked stochasticity of
the population activities (Fig.~\ref{fig:potjans-scheme}B,D). However,
trial-averaging reveals that the 
\blau{mean time-dependent activities that can be estimated from a 
peri-stimulus-time histogram (PSTH)
obtained from microscopic and mesoscopic simulations 
indeed agree well, except for a slight underestimation of the oscillatory peak during stimulus offset compared to the microscopic simulation (Fig.~\ref{fig:potjans_psth}A). However, during the short moments where the mean time-dependent activity (PSTH) of the mescoscopic and microscopic simulation do not match,  the time-dependent standard deviation across hundreds of trials (Fig.~\ref{fig:potjans_psth}B) is extremely high in both mesoscopic and microscopic simulation, indicating that fluctuations of the activity between one trial and the next are high after stimulus offset at $0.09$s.  The standard deviation as a function of time (Fig.~\ref{fig:potjans_psth}B) agrees overall nicely between microscopic and mesoscopic simulation, suggesting a good match of second-order statistics.} A closer
look at the second-order statistics, as provided by the power spectra
of spontaneous activities (``ground state'' of cortical activity),
also reveals a good agreement at all frequencies
(Fig.~\ref{fig:potjans_spectra}). This agreement is remarkable in
view of the low connection probabilities ($p<0.14$, see table 5 in
\cite{PotDie14}) that violate the assumption of dense random
connectivity used in the derivation of the mesoscopic mean-field
equations.  More generally, this example demonstrates that the range
of validity of our mesoscopic theory covers relevant cortical circuit
models.

Finally, we mention that the numerical integration of the mesoscopic
population equations yields a significant speed-up compared to the
microscopic simulation. While a systematic and fair comparison of the
efficiencies depends on many details and is thus beyond the scope of
this paper, we note that a simulation on a single core of 10s of
biological time took 811.2s using the microscopic model, whereas \blau{that of} the
mesoscopic model only took 6.6s. This corresponds to a speed-up factor
of around 120 achieved by using the mesoscopic population model. In
the simulation, we employed the same integration time step of
$\Delta t=0.5$~ms for both models for a first naive assessment of the
performance. However, a more detailed comparison of the performance
should be based on simulation parameters that achieve a given
accuracy. In this case, we expect an even larger speed-up of the
mesoscopic simulation because for the same accuracy the temporally
coarse-grained population equations allow for a significantly larger
time step than the microscopic simulation of spiking neurons.

\section*{Discussion}

In the present study we have derived stochastic population equations
that govern the evolution of mesoscopic neural activity arising from a
finite number of \blau{stochastic} neurons.  To our knowledge, this is the first time
that such a mesoscopic dynamics has been derived from an underlying
microscopic model of spiking neurons with pronounced spike-history
effects. The microscopic model consists of interacting homogeneous
populations of generalized integrate-and-fire (GIF) neuron models
\cite{MenNau12,GerKis14,PozNau13,PozMen15}, or alternatively,
spike-response (SRM) \cite{GerKis14} or generalized linear models
(GLMs) \cite{TruEde05,PilShl08,DegSch14}. These classes of neuron
models account for various spike-history effects like refractoriness
and adaptation \cite{GerKis14,WebPil16}. Importantly, parameters of
these models can be efficiently extracted from single cell experiments
\cite{PozMen15} providing faithful representations of real cortical
cells under somatic current injection. The resulting population
equations on the mesoscopic level yield the expected activity of each
population at the present time as a functional of population
activities in the past. 
Given the expected activities at the present time, the actual
mesoscopic activities can be obtained by drawing independent random
numbers. The derived mesoscopic dynamics captures nonlinear emergent
dynamics as well as finite-size effects, such as noisy oscillations
and stochastic transitions in multistable networks. Realizations
generated by the mesoscopic model have the same statistics as the
original microscopic model to a high degree of accuracy \blau{(as
  quantified by power spectra and residence time distributions)}.  The
equivalence of the population dynamics (mesoscopic model) and the
network of spiking neurons (microscopic model) holds for a wide range
of population sizes and coupling strengths, for time-dependent
external stimulation, random connectivity within and between
populations, and even if the single neurons are bursty or have
spike-frequency adaptation.

\subsection*{Quantitative modeling of mesoscopic neural data: applications and experimental predictions}
\label{sec:modell}

Our theory provides a general framework to replace spiking neural
networks that are organized into homogeneous populations by a network
of interacting mesoscopic populations. For example, the excitatory and
inhibitory neurons of a layer of a cortical column \cite{PotDie14} may
be represented by one population each, as in
Fig.~\ref{fig:potjans-scheme}. Weak heterogeneity in the neuronal
parameters are allowed in our theory because the mesoscopic equations
describe the population-averaged behavior. Further subdivisions of the
populations are possible, however, such as a subdivision of the
inhibitory neurons into fast-spiking and non fast-spiking types
\cite{MenNau12}. Populations that show initially a large degree of
heterogeneity can be further subdivided into smaller populations. In
this case, a correct description of finite-size fluctuations, as
provided by our theory, will be particularly important. However, as
with any mean-field theory, we expect that our theory breaks down if
neural activity and information processing is driven by a few
``outlier'' neurons such that a mean-field description becomes
meaningless. Further limitations may result from the mean-field and
quasi-renewal approximation, Eq.~\eqref{eq:lambda-short}. \rot{\blau{Formally,} the mean-field approximation of the synaptic input} requires dense connectivity \rot{and the heterogeneity in synaptic efficacies and in synapse numbers to be weak. \blau{Moreover,} the quasi-renewal approximation assumes} slow threshold dynamics. However, as
we have demonstrated here, our mesoscopic population equations \blau{can 
provide in concrete applications} excellent predictions even for sparse connectivity
(Fig.~\ref{fig:ei-net}D-G,~\ref{fig:potjans_psth} and
~\ref{fig:potjans_spectra}) and may qualitatively reproduce the
mesoscopic statistics in the presence of fast threshold dynamics
(Fig.~\ref{fig:bursting}D,E).

Using our mesoscopic population equations it is possible to make
  specific predictions about the response properties of local cortical
  circuits. For instance, recent progress in genetic methods now
enables experimentalists to selectively label and record from
genetically identified cell types, such as intratelencephalic (IT),
pyramidal tract (PT) and corticothalamic (CT) neurons among the
excitatory neurons, and vasoactive intestinal peptide (VIP),
somatostatin (Sst) and parvalbumin (Pvalb) expressing neurons among
the interneurons \cite{HarShe15}. These cell types have received much
attention recently as it has been proposed that they may form a basic
functional module of cortex, the canonical circuit
\cite{Car12,HarShe15}. The genetic classification of cells defines
subpopulations of the cortical network. A model of the canonical
circuits of the cortex in terms of interacting mesoscopic populations
can be particularly useful if used to describe experiments that use
optogenetic stimulation of genetically-defined populations by light,
which in our framework can be represented as a transient external
input current. To build a mesoscopic population model based on our
theory demands some assumptions about microscopic parameters such as
(i) typical neuron parameters for each subpopulation, (ii) structural
parameters as characterized by average synaptic efficacies and time
scales of connections between and within populations, and (iii)
estimates of neuron numbers per subpopulation. Parameters for a
typical neuron of each population could be extracted by the efficient
fitting procedures presented in \cite{MenNau12,PozMen15}. Structural
parameters and neuron numbers have been estimated, for instance, for
barrel columns in rodents somato-sensory cortex
\cite{LefTom09,AveTom12} and other studies (see e.g.,
\cite{PotDie14}). Our population equations could then be used to make
predictions about circuit responses to light stimuli, e.g. by imaging
the activity of a genetically-defined subpopulation in one column in
response to optogenetic stimulation of another cell class in another
column. 

As a first step in this direction, we have demonstrated here that our
population equations correctly predict the mesoscopic activities
(means and fluctuations) of a simulation of a detailed, microscopic
network model of a cortical microcircuit \cite{PotDie14} under
thalamic stimulation of layer 4 and 6 neurons. \rot{Using a population
  density method, mean activities of this model have also been
  predicted in a recent study to analyze its computational properties
  \cite{CaiIye16} \blau{with a special focus on predictive coding}. Our population density approach goes beyond that
  study by also predicting finite-size fluctuations of the activities
  and their effects on the mesoscopic network dynamics such as
  finite-size induced stochastic oscillations.} Predicting activities
in real experiments is, however, complicated by the fact that the
parameters of a microscopic network model are typically
underconstrained given the lack or uncertainty of measured or
estimated parameters \cite{SchSch15}. Here, our population equations
provide an efficient means to constrain unknown microscopic parameters
by requiring consistence with mesoscopic experimental data.


While the canonical circuit represents a model of interacting
populations on the mesoscopic level, recent interest in macroscopic
models of entire brain areas or even of whole brains has risen
\cite{IzhEde08,MarMul15}. Population dynamics can be used in this
context as a means to reduce large parts of the macroscopic neuronal
network to a system of interacting populations that is numerically
manageable, and requires less detailed knowledge of synaptic
connectivity (mean synaptic coupling of populations as opposed to
individual synapses). However, even this information about
  mesoscopic network structure might not be available given that it
  corresponds to an $M\times M$ matrix of mean synaptic efficacies,
  where the number $M$ of populations, or respectively cell types,
  might be large. In this case, our population equations can be
  utilized to efficiently constrain unknown structural parameters, such
  as synaptic weights, such that the resulting mesoscopic activities
  are consistent with experimental data. This leads in turn to
  experimentally testable predictions for synaptic
  connectivities. Such an approach \cite{SchSch15} has been recently
  applied to a network model of primate visual cortex demonstrating
  the usefulness of mean-field theories for predicting structural
  properties of large-scale cortical networks.

An interesting complementary route for further studies is a multiscale
model, in which a large-scale model is simulated in terms of reduced,
mesoscopic populations but with one or several areas in focus that are
simulated in full microscopic detail. As knowledge of anatomy and
computational capacity increases, more and more mesoscopic populations
can be replaced by a microscopic simulation, while at any time in this
process the full system is represented in the model. We therefore
expect our population dynamics model to be a useful tool to
continuously integrate experimental data into multiscale models of
whole mammalian brains.

Simplified whole brain models of interacting neuronal areas have
recently been proposed \cite{DecPon14,GilMor16}. Furthermore,
large-scale neuro imaging data are routinely modeled by
phenomenological population models such as neural mass, Wilson-Cowan,
or neural field models \cite{DecJir08,DavFri03}. Our new population
dynamics theory could be used in such approaches as an accurate
representation of the fluctuations of neural activity in the reduced
areas. For example, in macroscopic data such as resting
state fMRI, EEG or MEG, the endogenously generated fluctuations of
brain activity are of major interest \cite{GilMor16}. A fortiori the
same applies to mesoscopic data such as local field potentials (LFP)
or voltage-sensitive dye (VSD), in which finite-size fluctuation are
expected to be large. Our theory paves the way for relating macroscale
fluctuations to the underlying networks of spiking neurons and their
activity, and so to the neuronal circuits that underlie the
computations of the brain.

Another interesting application of our population model is to predict
the activity of neural networks grown in cultures. This model system
is much more accessible and controllable (e.g., by optogenetic
stimulations) than cortical networks in-vivo but may still provide
valuable insights into the complex network activity of excitatory and
inhibitory neurons as proposed in a recent study \cite{PulMus16}. In
particular, in that study the authors propose a critical role for
short-term plasticity \cite{TsoPaw98}. Although we have here used
static synapses, an extension of our mesoscopic mean-field theory to
synaptic short-term plasticity is feasible. Furthermore, finite-size
fluctuations appear to be particularly important in cell cultures as
suggested by a previous theoretical study \cite{GigDec15}. Our
mesoscopic population theory thus represents a framework to predict
spontaneous as well as evoked activity in neuronal cell cultures.
 
\subsection*{Theoretical aspects}
\label{sec:other-theo}

From a theoretical point of view, \blau{our study represents} a
generalization of {\em deterministic}, macroscopic population
equations \rot{for an infinite number of spiking neurons with
  refractoriness} \cite{Ger00,NauGer12,GerKis14} to
{\em stochastic}, \rot{mesoscopic population equations
for a finite number of neurons}. The resulting
dynamics can be directly used to generate \rot{single} stochastic
\rot{realizations} of mesoscopic activities, in analogy to a Langevin
dynamics. \rot{Our work is thus conceptually different from earlier
  studies of finite-size effects \cite{MeyVre02,ToyRad09,BuiCho13},
  who also considered finite networks of spiking neurons and
  refractoriness but derived deterministic evolution equations for
  moment and cross-correlation functions and hence characterized the
  ensemble dynamics. Furthermore, in contrast to these studies, our
  theory is not based on a perturbation expansion around the
  $N\rightarrow\infty$ limit, and thus captures large and non-Gaussian
  fluctuations in strongly nonlinear population dynamics such as
  bistable networks. }

Outside the low-rate Poisson firing regime, spiking neurons exhibit
history dependencies in their spike trains, the most prominent of
which is neuronal refractoriness, i.e. the strongly reduced firing
probability depending on the time since the last spike. On the
population level this means that a positive (negative) fluctuation of
the population rate affects the underlying refractory state of the
population because more (less) neurons than expected become
refractory. This altered refractory state in turn tends to decrease
(increase) the mean and variance of the population activity shortly
after the fluctuation. More generally, fluctuations of the population
activity influence the population density of state variables, which in
turn influences fluctuations. In this study, we have worked out how to
incorporate this interplay between fluctuations \rot{of the population
  activity and fluctuations of the refractory density} into a
mesoscopic population dynamics. The key insight to achieve this was
(i) to exploit the normalization \blau{condition for} the density of
microscopic states (in our case, the density of last spike times)
\blau{and (ii) to associate density fluctuations with a time-dependent
  but state-independent average rate that emphasizes the
  microscopic rates of those states that exhibit the largest
  finite-size fluctuations (in our case, the weighted average rate
  with respect to} the variance $v(t,\tl)$).

Our work is thus in marked contrast to previous stochastic rate models
for finite-size systems in the form of stochastic Wilson-Cowan
equations \cite{LagRot15,GigDec15}, or stochastic neural field
equations \cite{Bre09,FauIng15}. In these models, finite-size
fluctuations of the rate may feed back through the recurrent
connections but \blau{the strong negative self-feedback due to} refractoriness is neglected.
 This is the case even if the stationary or dynamic transfer
function employed in the rate dynamics corresponds to a spiking neuron
model \cite{BosDie16,GigDec15}. Furthermore, fluctuations of the
population rate have often been implemented {\it ad hoc} by a
phenomenological white-noise source, which was added to the
macroscopic (i.e. deterministic) rate dynamics
\cite{WonWan06,MorRin07,DecPon14}. The intensity of the noise is a free
parameter in these cases. Our mesoscopic equations are also driven by
a noise source, but two differences \rot{to these studies} are noteworthy: First, it is
derived from a microscopic model and does not contain any free
parameter; and second, the noise is white {\em given} the predicted
mean activity but since the activity predicted in one time step
depends on fluctuations in all earlier time steps, the effective noise
leads to a colored noise spectrum -- even if coupling is removed (see
Fig.~\ref{fig:lif}). This observation is consistent with \rot{previous
studies \cite{BruHak99,MatGiu02,LinDoi05,DegSch14}}, in which the power spectrum of the fluctuations
about a steady-state has been calculated analytically.

On the population level, refractoriness can be taken into account by
population density equations such as the Fokker-Planck equation  for
the membrane potential density (see
e.g. \cite{BruHak99,Bru00,NykTra00}, \blau{or \cite{IyeMen13,CaiIye16,LaiKam16_arxiv} for related master equations}), or the population integral
equation for the refractory density
\cite{Ger00,ChiGra07,ChiGra08,GerKis14}. These studies were mainly
concerned with macroscopic populations, which formally correspond to
the limit $N\rightarrow\infty$. \rot{For the refractory density
  formalism,} we have shown here how to \rot{extend} the population
integral equation \rot{to the case of finite population size. To this
  end, we} corrected for the missing normalization of the mesoscopic
density (e.g. $Q(t,\tl)=S(t|\tl)A_N(\tl)$ in
Eq.~\eqref{eq:master-cont-a-inf-main} or $q(t,\tl)$ in
Eq.~\eqref{eq:quasi-lin}), and \rot{thereby accounted} for the
interplay between fluctuations and refractoriness. Finite-size
fluctuations of the population rate have also been used in the
Fokker-Planck formalism \cite{BruHak99,Bru00,MatGiu02} but the
\rot{immediate} effect of these fluctuations on the membrane potential
density \rot{at threshold, and hence the refractoriness,} has been
neglected: \rot{in fact, a positive (negative) fluctuation of the
  population rate increases (decreases) the number of neurons at the
  reset potential while the number of neurons close to the threshold
  has to decrease (increase) such that the microscopic density remains
  normalized. The finite-$N$ membrane potential density used by Mattia
  and Del Giudice \cite{MatGiu02} does not account for this
  normalization effect. Whereas the numerical integration of their
  equation may still give} a satisfying solution in the low-rate,
Poissonian-firing regime, where refractory effects can be neglected,
it becomes unstable at higher rates unless the density is renormalized
manually at every time step \cite{BauAug16}.  How to correct for the
missing normalization in the Fokker-Planck approach is still an
unsolved theoretical question. In this respect, using analogies to and
insights from our approach might be promising.

The quasi-renewal approximation \cite{NauGer12,DegSch14} allowed us to
develop a finite-size theory for an effectively one-dimensional
population density equation even in the presence of adaptation. Here,
the only microscopic state variable is the last spike time $\tl$, or
equivalently the age of the neuron $\tau=t-\tl$. Longer lasting spike
history effects such as adaptation are captured by the dependence of
the conditional intensity on the population activity $A_N$, which as a
mesoscopic mean-field variable does not need to be treated as a state
variable. \rot{Furthermore, Chizhov and Graham have shown that the
  one-dimensional population density method in terms of the age $\tau$
  can also capture multiple gating variables in conductance-based
  neuron models with adaptation \cite{ChiGra07}. }  Such
one-dimensional descriptions have great advantages compared to
population density equations that include \rot{adaptation by
  additional state} variables and \rot{which} thus require a
multi-dimensional state-space
\cite{MulBue07,AugLad13,HerDur14,SchLin15}: Firstly, the numerical
solution of the density equations grows exponentially with the number
of dimensions \blau{and} becomes quickly infeasible if multiple adaptation
variables are needed as e.g. in the case of multi-timescale adaptation
\cite{PozNau13} or if an adiabatic approximation of slow variables
\cite{GigMat07,Ric09,AugLad13} is not possible. Secondly, it is
unclear how to treat finite-size fluctuations in the multi-dimensional
case.

Our theory is based on an effective fully-connected network, in which
neurons are coupled by the actual realization of the stochastic
population activity (the ``mean field''), both in the microscopic and
mesoscopic model. Thus, in the limit of a fully-connected network, the
problem of self-consistently matching the input and output statistics,
which arises in mean-field theories, is automatically satisfied to any
order by our finite-size theory. This is in marked contrast to the
opposite limit of a sparsely-connected network \cite{Bru00}. In that
case, the mean-field variables correspond to the {\em statistics} of
the spike trains (e.g. rate and auto-correlation function) rather than
to the actual realization of the population activity.  These
statistics must be matched self-consistently for input and output,
which is a hard theoretical problem
\cite{LerUrs06,RenMor07,SchDro15,WieBer15}. Between these two limit cases,
where the network is randomly connected with some finite connection
probability $0<p<1$, our examples (Fig.~\ref{fig:ei-net},
\ref{fig:potjans_psth} and \ref{fig:potjans_spectra}) indicate that
the approximation by an effective fully-connected network can still
yield reasonable results even for relatively sparse networks with
$p=5\%$. We emphasize that in our microscopic network model we used a
fixed in-degree in order to avoid additional variability due to the
quenched randomness in the number of synapses. This allowed us to
focus on dynamic finite-size noise in homogeneous populations and its
interactions with refractoriness. In contrast, the heterogeneity
caused by the quenched randomness is a further finite-size effect
\cite{BruHak99} that needs to be examined in a future study.

As an integral equation, the mesoscopic population model is formally
infinitely dimensional and represents a non-Markovian dynamics for the
population activity $A_N$. Such complexity is expected given that the
derived population equations are general and not limited to a specific
dynamical regime.  Loosely speaking, the equations must be rich
enough, and hence sufficiently complex, in order to reproduce the rich
repertoire of dynamical regimes that fully connected networks of
spiking neurons are able to exhibit (e.g. limit cycles,
multi-stability, cluster states \cite{Ger00}). For a mathematical
analysis, however, it is often desirable to have a low-dimensional
representation of the population dynamics in terms of a few
differential equations, at least for a certain parameter range. Apart
from the dynamics  \blau{in the neighborhood of} an equilibrium point (see
e.g. \cite{SchOst13}) or in the limit of slow synapses
\cite{ErmTer10}, such ``firing rate models'' are difficult to link to
the microscopic model already in the deterministic (macroscopic) case
(for notable \rott{exceptions see \cite{MonPaz15,AugLad16_arXiv}}), let alone the
stochastic, finite-size case. Here, our mesoscopic population rate
equations can serve as a suitable starting point for deriving
low-dimensional dynamics that links microscopic models to mesoscopic
rate equations with realistic finite-size noise.

\rot{
\subsection*{Extensions of the model}

There are several ways to extend our mesoscopic population model
towards more biological realism. We already mentioned the possibility
to include short-term synaptic plasticity in our mean-field
framework. Furthermore, the hazard function could be generalized to
capture Gaussian current noise as arising from background spiking
activity \cite{BruHak99,Bru00,PotDie14,SchDro15,MatGiu02,RenRoc10}.
Approximate mappings of white and colored current noise to an
effective hazard function in the escape noise formalism are available
\cite{ChiGra07,ChiGra08} and might be combined with our mesoscopic
population model. Yet another extension concerns the synaptic input
model. Here we only looked at current input but, as shown by Chizhov
and Graham \cite{ChiGra07}, it is straightforward to extend population
theories of the renewal type to the case of conductance inputs.  In
the simplest case, the synaptic current of neuron $i$ \blau{embedded} in
population $\alpha$ \blau{and driven by populations $\beta$} can be modelled
by a linear voltage-dependence:
\begin{equation}
  \label{eq:Isyn-conduct}
  I_{\text{syn},i}^\alpha(t)=-\sum_{\beta=1}^M\left(u_i^\alpha(t)-E^{\alpha\beta}\right)\sum_{j\in\Gamma_{i}^\beta}\bigl(g^{\alpha\beta}*s_{j}^\beta\bigr)(t)
\end{equation}
\blau{(cf. corresponding expression Eq.~\eqref{eq:I_syn} in {\sc
    Methods} for current-based synapses).} Here, $E^{\alpha\beta}$ is
the reversal potential of a synapse from population $\beta$, and
$g^{\alpha\beta}(t)$ is the conductance response (in $\text{nS}$)
elicited by a spike of a presynaptic neuron in population $\beta$. The
same mean-field arguments as for the current-based model carry over to
the case of conductance-based synapses. For example, the membrane
potential $u_A^\alpha(t,\tl)$ of a current-based leaky integrate-and-fire
neuron with a last spike time at time $\tl$ follows the equation
\begin{equation}
  \label{eq:V-mean-field-discuss}
  \taum^\alpha\pd{u_A^\alpha}{t}=-u_A^\alpha+\mu^\alpha(t)+\taum^\alpha \sum_{\beta=1}^M p^{\alpha\beta}N^\beta w^{\alpha\beta}(\epsilon^{\alpha\beta}*A_N^\beta)(t),
\end{equation}
where at $t=\tl$ and during an absolute refractory period $u_A(t,\tl)=\vreset$ is at the reset potential (see {\sc Methods}, Eq.~\eqref{eq:V-mean-field} for details). In the case of conductance-based input, Eq.~\eqref{eq:Isyn-conduct}, we only need to replace Eq.~\eqref{eq:V-mean-field-discuss} by
\begin{equation}
  \label{eq:V-mean-field-conduct}
  \taum^\alpha\pd{u_A^\alpha}{t}=-u_A^\alpha+\mu^\alpha(t)-R^\alpha\sum_{\beta=1}^M p^{\alpha\beta}N^\beta\lrrund{u_A^\alpha-E^{\alpha\beta}} (g^{\alpha\beta}*A_N^\beta)(t).
\end{equation}
where $R^\alpha$ is the membrane resistance.   How to model nonlinear voltage-dependence of synaptic
currents such as N-methyl-D-aspartate (NMDA) currents within a
mean-field approximation is less obvious but approximations also exist
for this case \cite{BruWan01}. It will be an interesting question for
the future how well these approaches work with the finite-N theory
developed in the present study. Alternatively, effective current
models \cite{Ric04,RicGer05} with activity-dependent, effective time
constant $\taum(t)$ and effective resting potential $\vrest(t)$ could
be another solution to treat conductance inputs.  }

Here, we have used a discrete set of populations. In large-scale
models of the brain, one often regards the spatial continuum limit,
resulting in so-called stochastic neural field equations
\cite{FauIng15}. These equations represent a compact description of
neural activity and do not depend on a specific discretization of
space. Just as discrete firing rate models, these field equations must
be considered phenomenological because the link to neuronal parameters
is not clear (note however that such equations have been derived from
non-spiking, two-state neuron models for $N<\infty$ \cite{Bre09}, and
from spiking models for $N\rightarrow\infty$
\cite{SpiGer01,GerKis02}). By taking the spatial continuum limit, our
mesoscopic population equations can be formulated as a stochastic
neural field equation that is directly derived from a finite-size,
spiking neural network. It would be interesting to employ this
continuous extension of our mesoscopic equations to study the effect of
spike-history effects on the stochastic behavior of bumps and waves in
neural fields.

A first simple comparison of the computational performance in {\sc
  Results, ``Mesoscopic dynamics of cortical microcolumn''},
demonstrated already that the mesoscopic population dynamics
outperformed the microscopic simulation by a speed-up factor of around
120. In this example, the numerical integration of the population
dynamics has not been particularly optimized with respect to time step
$\Delta t$ and history length $T$. A systematic comparison under the
condition of some given accuracy, has the potential for an even larger
speed-up because the population equations can be integrated with a
larger time step than the spiking neural network. \rot{In addition to that,
we have also compared the mesoscopic model to the full microscopic
simulation of the refractory density (cf. {\sc Results, ``Finite-size
  mean-field theory''}) and found a moderate enhancement in
performance for sufficiently large networks ($N\gtrsim 100$).} These computational aspects will be investigated in a separate
study.

\section*{Methods}

\subsection*{Model}
\label{sec:microsc-model}

\paragraph{Network setup.}

We consider a network of $M$ populations each consisting of $N^\alpha$
interconnected neurons of the same type (the superscript
$\alpha=1,\dotsc,M$ labels the populations). Neuron $i$ in population
$\alpha$ receives $p^{\alpha\beta}N^\beta$ connections (synapses) from
a random subset $\Gamma_i^\beta$ of presynaptic neurons in population
$\beta$. Here, $p^{\alpha\beta}$ denotes the probability for a
connection from a neuron in population $\beta$ to a neuron in
population $\alpha$. That is, the connections between any two
populations are random with fixed in-degree.

Let the spike train of neuron $i$ in population $\alpha$ be denoted by
\begin{equation}
  \label{eq:spike-train}
  s_i^\alpha(t)=\sum_{k}\delta(t-t_{i,k}^\alpha),
\end{equation}
where $t_{i,k}^\alpha$ is its $k$-th spike time and $\delta$ denotes
the Dirac $\delta$-function. The neuron receives spike train input
from its presynaptic partners in population $\beta$ with a
transmission delay $\Delta^{\alpha\beta}$ and synaptic weight
$w^{\alpha\beta}$. More precisely, the synaptic input current
$I_{\text{syn},i}^\alpha(t)$ is modeled as a sum of post-synaptic
currents caused by each spike of presynaptic neurons:
\begin{equation}
  \label{eq:I_syn}
 R^\alpha I_{\text{syn},i}^\alpha(t)=\tau_{\text{m}}^\alpha \sum_{\beta=1}^Mw^{\alpha\beta}\sum_{j\in\Gamma_{i}^\beta}\bigl(\epsilon^{\alpha\beta}*s_{j}^\beta\bigr)(t),
\end{equation}
where $R^\alpha$ and $\taum^\alpha$ are the membrane resistance and
membrane time constant of a neuron in population $\alpha$,
respectively, and $w^{\alpha\beta}$ sets the synaptic weights in units
of mV. The synaptic kernel $\epsilon^{\alpha\beta}(t)$ is defined as
the postsynaptic current (PSC) normalized by its charge that is
induced by one input spike from a neuron of population $\beta$. More
precisely, $\epsilon^{\alpha\beta}$ is the PSC divided by its
integral, and therefore it has units of $1/\text{sec}$.  In
Eq.~\eqref{eq:I_syn}, the first sum runs over all populations $\beta$,
whereas the second sum runs over the set $\Gamma_i^\beta$ of all
neurons in population $\beta$ that project onto neuron $i$ in
population $\alpha$.

In general, the filtered total synaptic input from population $\beta$,
$\sum_{j\in\Gamma_i^\beta}(\epsilon^{\alpha\beta}*s_j^\beta)(t)$,
may be modeled by a set of differential equations for a finite number
of synaptic variables $y_{i,\ell}^{\alpha\beta}$, $\ell=1,\dotsc,L$.
In simulations, we model the synaptic kernel by a single exponential
with constant delay $\Delta^{\alpha\beta}=\Delta$,
$\epsilon^{\alpha\beta}(t)=\Theta(t-\Delta)e^{-(t-\Delta)/\taus^\beta}/\taus^{\beta}$,
where $\Theta(t)$ denotes the Heaviside step function. The synaptic
time constants are $\taus^{E}=3$~ms and $\taus^{I}=6$~ms for
excitatory and inhibitory synapses, respectively. This kernel can be
realized by a single synaptic variable $y_i^{\alpha\beta}(t)$, which
obeys the first-order kinetics
$\taus^{\beta}\dot{y}_i^{\alpha\beta}=-y_i^{\alpha\beta}+\sum_{j\in\Gamma_i^\beta}s_j^\beta(t-\Delta)$
with $\beta\in\{E,I\}$.

\paragraph{Generalized integrate-and-fire model.}
\label{sec:glm-model}

Neurons are modeled by a leaky integrate-and-fire model with a dynamic
threshold \cite{GeiGol66,ChaLon00,LiuWan01,GerKis14} and an escape
noise mechanism \cite{MenNau12,GerKis14,PozNau13,PozMen15}. Following
\cite{PozMen15}, we refer to this model as the generalized
integrate-and-fire (GIF) model. The crucial variables of this model
are the membrane potential $u_i^\alpha(t)$ and the dynamic threshold
$\vartheta_i^\alpha(t)$. The membrane potential obeys the \gruen{subthreshold} dynamics
\begin{equation}
  \label{eq:glif}
  \taum^\alpha\od{u_i^\alpha}{t}=-u_i^\alpha+\mu^\alpha(t)+R^\alpha I_{\text{syn},i}^\alpha(t),
\end{equation}
where $\taum^\alpha$ is the membrane time constant and
$\mu^\alpha(t)=\vrest+R^\alpha I_{\text{ext}}^\alpha(t)$ is the
drive in the absence of synaptic input consisting of a constant resting potential $\vrest$ and an external
stimulus $I_{\text{ext}}^\alpha(t)$. The synaptic current
$I_{\text{syn},i}^\alpha(t)$ has been defined in Eq.~\eqref{eq:I_syn}. 

After each spike the voltage is reset to the potential $\vreset$,
where it is clamped for an absolute refractory period
$\tref=4$~ms. Furthermore, each spike
$t_{i,k}^\alpha$ adds a contribution $\theta^\alpha(t-t_{i,k}^\alpha)$ to the
dynamic threshold:
\begin{align}
  \label{eq:thresh}
  \vartheta_i^\alpha(t)&=\vth^\alpha+\sum_{t_{i,j}^\alpha<t}\theta^\alpha\left(t-t_{i,k}^\alpha\right),\nonumber\\
&=\vth^\alpha+\int_{-\infty}^t\theta^\alpha(t-t')s_i^\alpha(t')\,\mathrm{d}t'
\end{align}
where $\vth^\alpha$ is a baseline threshold and $\theta^\alpha(t)$ is
called the spike-triggered kernel \cite{MenNau12,PozMen15}.  Since the
increases in spike threshold accumulate over several spikes, the
spike-triggered kernel causes spike-frequency adaptation. We set
$\theta^\alpha(t)=\infty$ for $t\in(0,\tref)$ so as to ensure absolute
refractoriness. \rot{For the sake of simplicity, we assumed here that
  all spike-triggered accumulation effects can be lumped into the
  threshold variable (cf. Sec.~\Newnameref{sec:mapp-onto-gener}
  below). However, if realistic membrane potentials are needed (e.g.,
  for fitting membrane potential data \cite{PozMen15} or in a
  conductance-based extension of the model (see {\sc Discussion}) and
  \cite{ChiGra07}), adaptation mechanisms affecting the voltage should
  be kept in the voltage dynamics.}

Spikes are elicited stochastically by a conditional intensity (also
called hazard rate, escape rate or conditional rate)
\begin{equation}
  \label{eq:hazard-def}
  \lambda_i^\alpha(t)=f^\alpha\left(u_i^\alpha(t)-\vartheta_i^\alpha(t)\right),
\end{equation}
which depends on the momentary distance between the membrane potential
and the threshold via the exponential link function
$f^\alpha(x)=c^\alpha \exp(x/\Delta_u^\alpha)$. The parameter
$c^\alpha$ is the escape rate at threshold and the parameter
$\Delta_u^\alpha>0$ characterizes the softness of the threshold
(Fig.~\ref{fig:potjans-scheme}A, inset). Intuitively, a neuron fires
immediately if its membrane potential is $2\cdot\Delta_u^\alpha$
millivolts above the threshold and is unlikely to fire if its membrane
potential is $2\cdot\Delta_u^\alpha$ millivolts below the threshold
\cite{GerKis14}. In the limit $\Delta_u^\alpha\rightarrow 0$, the
model turns into a deterministic (but adaptive) leaky
integrate-and-fire model with a hard threshold. We emphasize that our
standard choice of $\Delta_u^\alpha\sim2$~mV is consistent with the
intrinsic stochasticity of neurons in cortical slices
\cite{JolRau06,MenNau12}. Alternatively, the softness of the threshold
$\Delta_u^\alpha$ may also be regarded as a phenomenological parameter
that accounts for all incoherent noise sources that are individual to
each neuron. This includes, e.g., any intrinsic noise but also
fluctuations of external background input from other neural
populations that are not modeled explicitly. For instance, to account
for the external Poisson input used in the original cortical column
model by Potjans an Diesmann \cite{PotDie14}, we increase in
Figs.~\ref{fig:potjans-scheme}, \ref{fig:potjans_psth} and
\ref{fig:potjans_spectra} the softness to
$\Delta_u^\alpha=5$~mV. \rot{We note that for more detailed
  comparisons with the original model, more elaborate approximations
  of the escape rate for the case of colored noise exist
  \cite{ChiGra08}, which in principle could be used to approximate
  external Poisson noise without a free parameter. However, because
  such a mapping is not the focus of the current study, \blau{we stick here
  for the sake of simplicity to} the phenomenological escape rate,
  Eq.~\eqref{eq:hazard-def}, of the exponential form.}

The parameters of the model used in simulations (unless specified
differently) are summarized in Table~\ref{table1}.

\paragraph{Mapping onto a generalized linear model.}
\label{sec:mapp-onto-gener}

We also considered a slightly different variant of the model, called
{\em spike-response model} \cite{GerKis14} or {\em generalized linear
  model (GLM)} \cite{TruEde05,PilShl08,ToyRad09,DegSch14,WebPil16}. This model does not
reqire the reset rule of the integrate-and-fire model but instead
relies on spike-triggered kernels to implement refractoriness and
other spike-history effects. Specifically, the membrane potential is
given by
\begin{equation}
  \label{eq:membran-pot}
  u_i^\alpha(t)=h_i^\alpha(t)+\sum_{\tl_{i,j}^\alpha<t}\eta^\alpha(t-\tl_{i,j}^\alpha),
\end{equation}
where $h_i^\alpha(t)$ is the free membrane potential given by
\begin{equation}
  \label{eq:h-eta}
  h_i^\alpha(t)=\left[\kappa^\alpha *\left(\mu^\alpha+R^\alpha I_{syn,i}^\alpha\right)(t)\right].
\end{equation}
For a membrane filter kernel
$\kappa^\alpha(t)=\Theta(t)e^{t/\taum^\alpha}/\taum^\alpha$, where
$\Theta(t)$ denotes the Heaviside step function, the dynamics of
$h_i^\alpha$ is equivalent to the dynamics of $u_i^\alpha$
(Eq.~\eqref{eq:glif}), except that $h_i^\alpha$ is not reset upon
spiking. Spike-history effects on the level of the membrane potential
are captured by the second term in Eq.~\eqref{eq:membran-pot}. This
term represents the convolution $(\eta^\alpha*s_i^\alpha)(t)$ of the
output spike train with a spike-triggered kernel $\eta(t)$ and
generates a spike-after-potential that accumulates over spikes.  As
before, the threshold $\vartheta_i^\alpha(t)$ obeys
Eq.~\eqref{eq:thresh}. Given the membrane potential $u_i^\alpha(t)$
and the dynamic threshold $\vartheta_i^\alpha(t)$, spikes are
generated by the same hazard rate $\lambda_i^\alpha(t)$ given by
Eq.~\eqref{eq:hazard-def}.

At low firing rates, the spike-triggered kernel $\eta$ can be used
to approximate the integrate-and-fire dynamics by choosing
$\eta(t)=(\vreset-\vth)e^{-(t-\tref^\alpha)/\taum^\alpha}\Theta(t)$. However,
this is not an exact mapping because the value of the membrane
potential is not reset to a fixed value $\vreset$ after spiking, in
contrast to the GIF model. This is due to the accumulation \blau{of the threshold and the} variability in the
voltage at the moment of firing.

We also mention that the kernel $\eta$ can be transformed into the
kernel $\theta$ of the threshold dynamics \cite{GerKis14}.  This is
possible because we are only interested in the spike emissions of the
neurons and not the membrane potentials. In fact, the conditional
firing rate, Eq.~\eqref{eq:hazard-def}, is invariant under the
transformation $\theta\rightarrow\theta-\eta$, $\eta\rightarrow 0$.

\begin{table}[!ht]
\centering
\caption{
  {\bf Default values of parameters used in simulations \gruen{unless stated otherwise}.}}
\begin{tabular}{lll}
$\taum$ & 20 ms & membrane time constant \\
$\tref$& 4 ms& absolute refractory period\\
$\vth$&15 mV& threshold (non-adapting part)\\
$\vreset$& 0 mV& reset potential\\
$c$ & 10 Hz & escape rate at threshold \\
$\Delta_u$& 2 mV& noise level\\
$\Delta$&1 ms& transmission delay\\
$\taus^{E}$&3 ms& decay time constant of excitatory synapses\\
$\taus^{I}$&6 ms& decay time constant of inhibitory synapses\\
\end{tabular}
\label{table1}
\end{table}

\subsection*{Mean-field approximation}
\label{sec:mf}

An important variable that characterizes the internal state of a
neuron is the time of its last spike, or, equivalently, the time
elapsed since the last spike (``age'' of the neuron)
\cite{ChiGra07}. \blau{The} time since the last spike is a
good predictor of the refractory state of a neuron at time $t$. Our
approach is to use a population density description for this
refractory state \cite{Ger00,MeyVre02,ChiGra07,ChiGra08}, in which the
coupling of neurons as well as the adaptation of single neurons are
mediated by the mesoscopic population activities $A_N^\alpha(t)$
defined by Eq.~\eqref{eq:pop-activ-def}.  \rot{To this end, we replace the conditional firing rate \blau{$\lambda_i^\alpha(t)$} of a neuron $i$ in
  population $\alpha$ by an effective rate
  $\lambda^\alpha_A(t|\tl_i^\alpha)$ that only depends on its} last
spike time $\tl_i^\alpha$ and the history of the population activity
$\{A_N^\alpha(t')\}_{t'<t}$ \cite{NauGer12}. Here and in the
following, the subscript $A$ indicates the dependence on the history
of $A_N^\alpha(t)$. \blau{We note that the expected total activity $\bar A^\alpha(t)$ of population $\alpha$ at time $t$ is the average of all the conditional firing rates summed over all neurons in this population: $\bar A^\alpha(t) = (1/N^\alpha) \sum_i \lambda_i^\alpha(t)$. The effective rate $\lambda_A^\alpha(t|\tl_i^\alpha)$} \rot{ is determined such that it approximates the conditional intensity on average:
\begin{equation}
  \label{eq:mf-lam}
\frac{1}{N^\alpha}\sum_{i=1}^{N^\alpha}\lambda_i^\alpha(t)\approx \frac{1}{N^\alpha}\sum_{i=1}^{N^\alpha} \lambda^\alpha_A(t|\tl_i^\alpha).
\end{equation}
}To find such an approximation, we
proceed in two steps \cite{DegSch14}: first, the membrane potential
$u_i^\alpha(t)$ is approximated by a function
$u_A^\alpha(t,\tl_i^\alpha)$ using a mean-field approximation of the
synaptic input. For fully connected populations, this first
approximation turns into an exact statement. Second, the dynamic
threshold $\vartheta_i^\alpha(t)$ is approximated by a function
$\vartheta_A^\alpha(t,\tl_i^\alpha)$ using the quasi-renewal
approximation \cite{NauGer12}. For renewal neurons, the second
approximation becomes exact. Once we have found an expression for the
mean-field approximation Eq.~\eqref{eq:mf-lam}, we are in a position
to use a population density description with respect to the last spike
times $\tl_i^\alpha$. \blau{In the following two paragraphs we explain the above two steps in detail.}

\paragraph{Mean-field approximation of synaptic input.}

In the special case of a fully connected network
($p^{\alpha\beta}=1$), the membrane potential can be completely
inferred from the last spike time $\tl_i^\alpha$ and the mean field
$A_N^\alpha$. In this case, the synaptic input Eq.~\eqref{eq:I_syn}
can be rewritten as
\begin{equation}
  \label{eq:I_syn-fully}
  R^\alpha I_{\text{syn},i}^\alpha(t)=\tau_{\text{m}}^\alpha \sum_{\beta=1}^M p^{\alpha\beta}N^\beta w^{\alpha\beta}(\epsilon^{\alpha\beta}*A_N^\beta)(t).
\end{equation}
Thus, in a fully connected network all neurons in population $\alpha$
``see'' the same synaptic input $R^\alpha I_\text{syn}^\alpha$ given by
the ``mean field'' $A_N(t)$. From Eq.~\eqref{eq:glif} follows that GIF
neurons with the same last spike time $\tl$ all have the same membrane
potential $u_A^\alpha(t,\tl)$ that obeys the differential equation
\begin{equation}
  \label{eq:V-mean-field}
  \taum^\alpha\pd{u_A^\alpha}{t}=-u_A^\alpha+\mu^\alpha(t)+\taum^\alpha \sum_{\beta=1}^M J^{\alpha\beta} (\epsilon^{\alpha\beta}*A_N^\beta)(t).
\end{equation}
with $J^{\alpha\beta}=p^{\alpha\beta}N^\beta w^{\alpha\beta}$. The
initial condition is $u_A^\alpha(\tl,\tl)=\vreset$ corresponding to
the reset of the membrane potential after the last spike. If we use
this insight for the conditional intensity
$f^\alpha(u_i^\alpha(t)-\vartheta_i^\alpha(t))$ we see that the
explicit dependence upon $u_i^\alpha(t)$ can be dropped as long as we
keep track of the last spike time $\tl_i^\alpha$,
cf. Eq.~\eqref{eq:mf-lam}; hence
$u_i^\alpha(t)=u_A^\alpha(t,\tl_i^\alpha)$.

In a randomly connected network ($p^{\alpha\beta}<1$), the synaptic
input is different for each neuron. \rot{On the population level,
  however, this heterogeneity is averaged allowing us to still use the
  mean-field approximation, Eq.~\eqref{eq:V-mean-field}.}  To see
this, we note that in our network with fixed in-degree, each neuron
$i$ in population $\alpha$ has $p^{\alpha\beta}N^\beta$ presynaptic
neurons in population $\beta$ ($\beta=\alpha$ is possible). This means
that in Eq.~\eqref{eq:I_syn} we can approximate the sum
$\sum_{j\in\Gamma_{i}^\beta}s_{j}^\beta(t)$ over the
$p^{\alpha\beta}N^\beta$ presynapic neurons by
\begin{equation}
  \label{eq:mf-appox}
p^{\alpha\beta}N^\beta\left(\frac{1}{p^{\alpha\beta}N^\beta}\sum_{j\in\Gamma_{i}^\beta}s_{j}^\beta(t)\right)\approx
  p^{\alpha\beta}N^\beta A_N^\beta(t)  
\end{equation}
(cf. definition of $A_N^\beta(t)$ in Eq.~\eqref{eq:pop-activ-def}). \rot{The mean-field approximation, Eq.~\eqref{eq:mf-appox}, only depends on the population activity and is thus identical for all neurons.  Therefore, fluctuations of the population activity can be regarded as common input fluctuations that are coherent across neurons. On the other hand, the deviation from the mean-field approximation (i.e. the difference between the left- and right-hand side of Eq.~\eqref{eq:mf-appox}) is different for each neuron and can be regarded as incoherent noise. For low connection probabilities, this incoherent part of the fluctuations may lead to a significant deviation of the membrane potential $u_i^\alpha(t)$ from the mean-field approximation $u_A^\alpha(t,\tl_i^\alpha)$ (Fig.~\ref{fig:mf-approx}A,C,E, top). On the mesoscopic scale, however, the total number of spikes in a small time step $\Delta t$ is determined by the population-averaged conditional firing rate, Eq.~\eqref{eq:mf-lam},  (cf. Eq.~\eqref{eq:n-poisson} below). Hence, for sufficiently large $N^\alpha$,} incoherent noise average out, whereas common finite-size fluctuations survive \rot{(Fig.~\ref{fig:mf-approx}A,C,E, bottom). Note, however, that incoherent noise may cause a small bias because we average a nonlinear function of the noisy membrane potential on the l.h.s. of Eq.~\eqref{eq:mf-lam}.} Effectively, the incoherent noise softens the threshold
of the escape noise mechanism.

\paragraph{Quasi-renewal approximation.}
\label{sec:qr}

So far we have reduced the conditional intensity to
$\lambda_i^\alpha(t)\approx
f\bigl(u_A(t,\tl_i^\alpha)-\vartheta_i^\alpha(t)\bigr)$.
This expression still involves the individual threshold
$\vartheta_i^\alpha(t)$ of neuron $i$ in population $\alpha$, which
depends on the full spike history of that neuron. This means that the
spike-train is generally not a time-dependent renewal process. Here,
we employ the quasi-renewal approximation \cite{NauGer12}
and average over the spikes before the last spike time assuming that
they occurred according to an inhomogeneous Poisson process with rate
$A_N^\alpha(t')$, $t'<\tl_i^\alpha$.  Averaging the conditional
intensity, Eq.~\eqref{eq:hazard-def}, in this way, conditioned on a
given last spike time $\tl_i^\alpha$ and a given history
$A_N^\alpha(t')$, $t'<\tl_i^\alpha$, yields
\cite{Str67I,van92,NauGer12}
\begin{equation}
  \label{eq:pop-avg-haz}
  \lambda_i^\alpha(t)\approx f^\alpha\bigl(u_A^{\alpha}(t,\tl_i^\alpha)-\vartheta_A^\alpha(t,\tl_i^\alpha)\bigr)\equiv \haz_A^{\alpha}(t|\tl_i^\alpha),
\end{equation}
where $\vartheta_A^\alpha(t,\tl_i^\alpha)$ is an effective dynamic
threshold given by
\begin{equation}
  \label{eq:qr-g}
  \vartheta_A^{\alpha}(t,\tl)=\vth+\theta^{\alpha}(t-\tl)+\int_{-\infty}^{\tl}\tilde{\theta}^\alpha(t-t')A_N^\alpha(t')\mathrm{d}t'.
\end{equation}
Here,
$
\tilde{\theta}^\alpha(t)=\Delta_u^\alpha\bigl[1-e^{-\theta^{\alpha}(t)/\Delta_u^{\alpha}}\bigr]$
is the so-called quasi-renewal kernel \cite{NauGer12}, while $\theta^\alpha(t-\tl)$
describes the increase of the threshold induced by the last
spike. Note that as a result of the two approximations, the
conditional firing rate no longer depends on the precise spiking
history of a given neuron and its presynaptic neurons, but only on its
last firing time, cf. Eq.~\eqref{eq:hazard-def} and
Eq.~\eqref{eq:pop-avg-haz}. This ends our explanation of
Eq.~\eqref{eq:mf-lam}.

\subsection*{Discretized population density equations.}
\label{mean-field}
Using the mean-field approximation Eq.~\eqref{eq:pop-avg-haz}, we have
reduced the model to a population of time-dependent renewal processes
\rot{\cite{Ger00,MeyVre02}}, where the conditional intensity of neuron $i$ is
$\lambda_A^\alpha(t|\tl_i^\alpha)$. Neurons are effectively coupled
through the dependence of $\lambda_A^\alpha(t|\tl_i^\alpha)$ upon the
membrane potential $u_A^\alpha(t|\tl_i)$, which in turn depends on the
activities $A_N^\beta$ of all populations $\beta$ that are connected
to population $\alpha$. This is the only place where population labels
different from $\alpha$ appear. For the sake of notational simplicity,
we will omit the population label $\alpha$ and the subscript $A$ in
this section, keeping in mind that all quantities refer to population
$\alpha$ and that the coupling with other populations is implicitly
contained in $u_A^\alpha(t|\tl_i)$.

\paragraph{Microscopic dynamics of the refractory density.}

\begin{figure}[t]
\begin{adjustwidth}{\figureoffsetleft}{\figureoffsetright}
  \centering
  \includegraphics[]{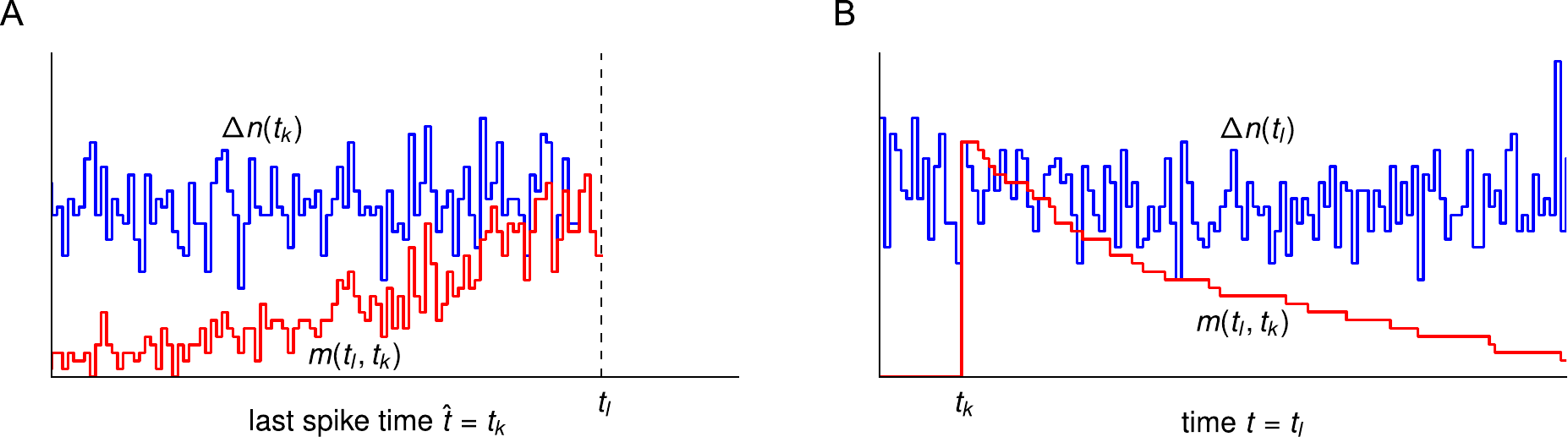}  
  \caption{{\bf Different interpretations of the function $m(t,\tl)$ (red line).}\\
    (A) As a function of $\tl$ (or as a function of $k$ in discrete
    time), $m(t_l,t_{k})$ represents the distribution of last spike
    times $\tl_i$ across the population at time $t=t_l$. (B) As a
    function of time $t$ (or as a function of the index $l$ in
    discrete time), $m(t_l,t_{k})$ represents the survival number,
    i.e. the number of neurons which fired in the interval
    $[t_k,t_k+\Delta t)$ which survived (did not fire) until time
    $t=t_l$. The activity $\Delta n(t_k)$, i.e. the number of neurons that
    fired in the $k$-th time bin, is depicted by a blue line. The
    population size is $N=1000$.  }
  \label{fig:m-n}
\end{adjustwidth}
\end{figure}

Because the firing probability of a neuron only depends on its last
spike time and the mesoscopic population activity in the past, we can
use a population density description of all last spike times $\tl_i$
in the population. To derive such representation it is useful to
discretize time by introducing the discrete time points
$t_{k}=t_0+k\Delta t$, ${k}\in\mathbb{Z}$, and the corresponding
intervals $\mathcal{I}_k=[t_k,t_k+\Delta t)$. Time is measured
relative to a reference time $t_0$, which, however, is irrelevant for
the following arguments. We require that the size of the intervals
$\Delta t$ is sufficiently small so that each neuron fires at most
once during any interval. Specifically, we require that
$\Delta t\le \tref$. \blau{We also require that the sum of axonal and synaptic delays is not smaller than $\Delta t$.} Furthermore, we identify the discrete
time point $t_l$ as the current time, whereas indices $k$ with $k<l$
correspond to the past. In the population density approach, we do not
keep track of the last spike time of each individual neuron but for
each past time interval $\mathcal{I}_k$ we only track the number of those
neurons that have their last spike time in this interval.  This
number is denoted by $m(t_l,t_{k})$. The collection
$\{m(t_l,t_k)\}_{k\in\mathbb{Z},k<l}$ of these numbers for all
intervals $\mathcal{I}_k$, $k<l$, represents the current distribution
of last spike times $\tl_i$ in the population at time $t_l$
(Fig.~\ref{fig:m-n}A). Because each neuron has exactly one last spike
time, the distribution $m(t_l,t_k)$ is normalized to the total number
of neurons:
\begin{equation}
  \label{eq:normalization-gen}
  \sum_{{k}=-\infty}^{l-1} m(t_l,t_{k})=N.
\end{equation}
\rot{Since the last spike time determines the refractory state of a neuron,
the distribution $m(t_l,t_k)$ will be also called refractory
distribution and the function $Q_N(t_l,t_k)\equiv m(t_l,t_k)/(N\Delta t)$ can be
regarded as the corresponding refractory density. The refractory
distribution fully} characterizes the microscopic state of the
population.

We now introduce the number of neurons that fired a spike in the
interval $\mathcal{I}_k$ (not necessarily the last spike). This number
is denoted by $\Delta n(t_{k})$ (Fig.~\ref{fig:m-n}A) \rot{and is related to the population activity by
  $A_N(t_k)=\Delta n(t_k)/(N\Delta t)$. Therefore, $\Delta n(t_{k})$}
will be often referred to as simply the ``activity'' at time $t_k$.
Knowing the past activities $\Delta n(t_{k'})$ for $k'<l$ and the last
spike time $t_k$ fully determines the membrane potentials $u(t_l,t_k)$
and thresholds $\vartheta(t_l,t_k)$, and hence the escape rate
$\lambda(t_l|t_k)=f\bigl(u(t_l,t_k)-\vartheta(t_l,t_k)\bigr)$
associated with the interval $\mathcal{I}_k$. Thus, the knowledge of
the past activities and the distribution of last spike times at time
$t_l$ is sufficient to statistically determine these quantities at
time $t_{l+1}$. In other words, the evolution of the system can be
described by a Markov process if we define the microscopic state
$\Xmicro(t_l)$ of the population at time $t_l$ by the sequence of
pairs
\begin{equation}
  \label{eq:Xmicro}
  \Xmicro(t_l)=\bigl\{\bigl(\Delta n(t_k),m(t_l,t_k)\bigr)\bigr\}_{k\in\mathbb{Z},k<l}\,.
\end{equation}
In the following, the main task will be to derive the statistics of
the number of spikes $\Delta n(t_l)$ in the next time interval
$[t_l,t_l+\Delta t)$ and the distribution $m(t_l+\Delta t,t_k)$ of
last spike times at time $t_{l+1}$ given the state $\Xmicro(t_l)$ at
time $t_l$. We mention that what we have lost in this population
density description is only the information about the identity of the
neurons, which, however, is irrelevant for the mesoscopic description
of homogeneous populations.

There is a second interpretation of $m(t_l,t_{k})$: let us consider
the group of neurons that have fired in the interval $\mathcal{I}_k$,
$k<l$. The number of neurons from this group that have ``survived''
(i.e. that have not fired again) until time $t_l$ is exactly given by
$m(t_l,t_{k})$. \rot{We will therefore also call it number of survived
  neurons or survival number for the refractory state
  $k$. Correspondingly, the ratio
  $S_N(t_l|t_k)=m(t_l,t_k)/\Delta n(t_k)$ is the fraction of survived
  neurons.}  As time $t_l$ evolves, the \rot{number of survived
  neurons} diminishes whenever there is a spike in that group
(Fig.~\ref{fig:m-n}B). Thus, if the group fires $X_{lk}$ spikes in the
time step $[t_l,t_l+\Delta t)$, then $m(t_l,t_{k})$ decreases by
$X_{lk}$. For $l>k$, this gives rise to the evolution equation
\begin{align}
  \label{eq:m-poiss}
  m (t_l+\Delta t,t_{k})&=m (t_l,t_{k})-X_{lk}.
\end{align}
The initial condition is given by $m (t_k+\Delta t,t_k)=\Delta n(t_k)$,
which follows from the absolute refractoriness during the first time
step after a spike. Absolute refractoriness also entails that each
neuron can fire only one spike per time step ($\Delta t\le \tref$)
with a firing probability
\begin{equation}
  \label{eq:firing-prob}
  P_{\lambda}(t_l|t_k)=1-\exp\lrrund{-\int_{t_l}^{t_l+\Delta t}\lambda(s|t_k)\,\mathrm{d}s}\approx 1-e^{-\bar\lambda(t_l|t_k)\Delta t}.
\end{equation}
In the last step, we introduced the average hazard rate
$\bar\lambda(t_l|t_k)=[\lambda(t_l|t_k)+\lambda(t_{l+1}|t_k)]/2$. Because
the past activities $\Delta n(t_k)$, $k<l$, completely determine the
probability to fire $P_\lambda(t_l|t_k)$, each neuron decides
independently from the others whether it fires in the next time step.
Furthermore, there is a total number of $m (t_l,t_{k})$ neurons from
the considered group that could potentially fire in the interval
$[t_l,t_l+\Delta t)$.  Therefore, the number of spikes $X_{lk}$ is the
result of $m (t_l,t_{k})$ independent Bernoulli trials with success
probability $P_\lambda (t_l|t_{k})$. This implies that $X_{lk}$ \rot{follows a binomial distribution:
\begin{equation}
  \label{eq:X-bino}
  X_{lk}\sim B\lrrund{m(t_l,t_k),P_\lambda(t_l|t_k)}.
\end{equation}}
Moreover, the random numbers $X_{lk}$ associated with different past
time intervals $\mathcal{I}_k$ are conditionally independent given the
current state of the system $\Xmicro(t_l)$
(cf. Eq.~\eqref{eq:Xmicro}).

The total number of spikes emitted in the current interval $[t_l,t_l+\Delta t)$ is equal
to the total reduction of survivals in that interval, hence
\begin{equation}
  \label{eq:n-t-micro}
  \Delta n(t_l)=\sum_{{k}=-\infty}^{l-1} X_{lk}.  
\end{equation}
Equations \eqref{eq:m-poiss} - \eqref{eq:n-t-micro} define the
microscopic kinetics in discrete time. In a simulation, for each past
time interval $\mathcal{I}_k$ one independent random number
$X_{lk}$ needs to be drawn per time step and population. These random
numbers determine the current spike count via Eq.~\eqref{eq:n-t-micro}
and the update of the distribution of last spike times $m(t_l,t_k)$,
via Eq.~\eqref{eq:m-poiss}. We call this description microscopic
because for small time steps, there will be many (order of $N$)
intervals $\mathcal{I}_k$ that contain survived neurons, i.e. for
which $m(t_l,t_k)>0$ and for each of which one needs to draw a random
number $X_{lk}$ in a simulation.  In the limit
$\Delta t\rightarrow 0$, such a simulation would be as complex as the
original microscopic simulation of $N$ neurons. 

\blau{
The microscopic population density description can be summarized in a particularly compact form by performing the continuum limit $\Delta t\rightarrow 0$ and by assuming large $N$. For large $N$, the statistics of $X(t_l,t_k)$ becomes Gaussian with mean and variance $P_\lambda(t_l|t_k)m(t_l,t_k)$. Thus, the dynamics of $m(t,\tl)$, Eq.~\eqref{eq:m-poiss}, can be rewritten as
\begin{equation}
  \label{eq:micro-Gauss}
  m (t_l+\Delta t,t_{k})-m (t_l,t_{k})=-P_\lambda(t_l|t_k)m(t_l,t_k)+\sqrt{P_\lambda(t_l|t_k)[m(t_l,t_k)]_+}\mathcal{N}(t_l,t_k),
\end{equation}
where $\{\mathcal{N}(t_l,t_k)\}_{k,l\in\mathbb{Z}}$ are independent,
standard normal random numbers and the ramp function $[x]_+=x\Theta(x)$ ensures non-negativity of $m$. Using the density of last spike times
$Q_N(t_l,t_k)\equiv S_N(t_l|t_k)A_N(t_k)\equiv m(t_l,t_k)/(N\Delta t)$, setting
$t_l=t$ and $t_k=\tl$, and expanding
$P_\lambda(t|\tl)\approx \lambda(t|\tl)\Delta t$ for small $\Delta t$
we arrive at the following dynamics in the limit $\Delta t\rightarrow 0$:
\begin{equation}
  \label{eq:QN-microdyn}
  \pd{Q_N(t,\tl)}{t}=-\lambda(t|\tl)Q_N(t,\tl)+\sqrt{\frac{\lambda(t|\tl)[Q_N(t,\tl)]_+}{N}}\xi(t,\tl).
\end{equation}
Here, $\xi(t,\tl)\equiv \lim_{\Delta t\rightarrow 0}\mathcal{N}(t,\tl)/\Delta t$ is a Gaussian random field with zero mean and correlation function $\langle\xi(t,\tl)\xi(t',\tl')\rangle=\delta(t-t')\delta(\tl-\tl')$. For a given last spike time $\tl$, Eq.~\eqref{eq:QN-microdyn} has the form of a Langevin equation. Its initial condition is $Q_N(\tl,\tl)=A_N(\tl)$. The population activity results from Eq.~\eqref{eq:n-t-micro} as the integral of changes of the refractory density $Q_N(t,\tl)$ over all \rott{refractory states, i.e. $A_N(t)=-\int_{-\infty}^t\partial_tQ_N(t,\tl)\,\mathrm{d}\tl$, or using Eq.~\eqref{eq:QN-microdyn}:
\begin{equation}
  \label{eq:popact-micro}
  A_N(t)=\int_{-\infty}^t \lambda(t|\tl)Q_N(t,\tl)\,\mathrm{d}\tl-\int_{-\infty}^t \sqrt{\frac{\lambda(t|\tl)[Q_N(t,\tl)]_+}{N}}\xi(t,\tl)\,\mathrm{d}\tl.
\end{equation}}
Equations \eqref{eq:QN-microdyn} and \eqref{eq:popact-micro} represent
the microscopic population density equations in continuous time  under the Gaussian and
quasi-renewal approximations.
}

\paragraph{Mesoscopic description.}

At the mesoscopic level, we want to describe the state of the
population at time $t_l$ only by the mesoscopic variables $\Delta n(t_k)$,
$k<l$, that have been observed so far. Therefore, we define the
history  of $n$ at time $t_l$ by 
\begin{equation}
  \label{eq:Xmeso}
  \mathcal{H}(t_l)=\bigl\{\Delta n(t_k)\bigr\}_{k\in\mathbb{Z},k<l},
\end{equation}
which completely determines the mesoscopic state. In contrast to the
microscopic state $\Xmicro(t_l)$ defined in Eq.~\eqref{eq:Xmicro}, the
mesoscopic state does not require the knowledge of the detailed
distribution of last spike times $m(t_l,t_k)$. We call a variable
mesoscopic if it only depends on the history
$\mathcal{H}(t_l)$. Likewise, an equation is called mesoscopic if it
only involves mesoscopic variables. \rot{In the following sections, all
averages at a given time $t_l$ have to be understood as conditional
averages given the history $\mathcal{H}(t_l)$. We will therefore
often omit an explicit notation of this condition.}

To derive a mesoscopic equation, we want to find an approximate
dynamics with only one effective, mesoscopic noise term that
summarizes the effect of all microscopic random variables $X_{lk}$. In
a simulation, this would imply to draw only one random number per time
step and per population.  Towards that end, \rot{we assume that
  $\Delta t$ can be chosen sufficiently small such that
  $P_\lambda(t_l|t_{k})\ll 1$, which is always possible if neurons are
  stochastic und hence do not perfectly synchronize. Under this
  assumption, the binomially-distributed random numbers $X_{lk}$ are
  approximately Poisson-distributed, i.e.
\begin{equation}
  \label{eq:X-pois}
  X_{lk}\sim \text{Pois}\left(\expect{X_{lk}\middle|m(t_l,t_k)}\right),
\end{equation}
where
\begin{equation}
  \label{eq:X-mean-var}
  \expect{X_{lk}|m(t_l|t_k)}=P_\lambda(t_l|t_k)m(t_l,t_k).
\end{equation}
is the conditional mean of $X_{lk}$ given the current survival number $m(t_l|t_k)$.  Given the conditional independence of $X_{lk}$ for different $k$, the
Poisson property} implies that the global activity $\Delta n(t_l)$ in
Eq.~\eqref{eq:n-t-micro} is also Poisson-distributed given the current refractory distribution $\{m(t_l,t_k)\}_{k<l}$, i.e.
\begin{equation}
  \label{eq:n-poisson}
  \Delta n(t_l)\sim \text{Pois}\lrrund{\nbar (t_l)},
\end{equation}
with mean
\begin{equation}
  \label{eq:mean-n-t}
  \nbar (t_l)\equiv\expect{\Delta n(t_l)|\{m(t_l,t_k)\}_{k<l},\mathcal{H}(t_l)}=\sum_{{k}=1}^\infty P_\lambda (t_l|t_{k})m(t_l,t_{k}).
\end{equation}
Because of the definition of refractory densities and
$P_\lambda\le 1$, we find that $\nbar(t_l)\le N$ is automatically
satisfied at any moment in time. \gruen{However, for the numerical
implementation with finite $\Delta t$ later on we need to keep in mind
that the Poisson number $\Delta n(t)$ could become larger than $N$, if
$\nbar(t_l)$ is close to $N$. In this case, using a binomial
statistics will be more appropriate, as explained in
Sec.~\Newnameref{sec:sim-algo}.}

Equations \eqref{eq:n-poisson} and \eqref{eq:mean-n-t} suggest the
possibility to generate $\Delta n(t_l)$ by a {\em single} Poisson-distributed
random number. However, Eq.~\eqref{eq:mean-n-t} is not a mesoscopic
equation yet because it still depends on the dynamics of $m(t_l,t_k)$,
Eq.~\eqref{eq:m-poiss}, which contains the microscopic random
variables $X_{lk}$. There is another, more subtle problem if we want
to use Eqs.~\eqref{eq:n-poisson} and \eqref{eq:mean-n-t} as a
mesoscopic dynamics that generates the activities $\Delta n(t_l)$. If we
regard $\Delta n(t_l)$ as an {\em independent} random variable, the
conservation of neurons, Eq.~\eqref{eq:n-t-micro}, imposes a
constraint on the microscopic random numbers $\{X_{lk}\}_{k<l}$, which
will therefore not be independent anymore. Conversely, if we consider
both $\{X_{lk}\}_{k<l}$ and $\Delta n(t_l)$ as independent variables, we
almost certainly violate the conservation of neurons,
Eq.~\eqref{eq:n-t-micro}, or equivalently, the normalization condition
Eq.~\eqref{eq:normalization-gen}. This problem does not occur in the
microscopic dynamics, where $\Delta n(t_l)$ is a {\em dependent} variable
generated from the independent random variables $\{X_{lk}\}_{k<l}$ via
Eq.~\eqref{eq:n-t-micro}, and hence the correct normalization is
guaranteed at any time. Nevertheless, the ``non-normalized'' or
``unconstrained'' process, in which $\{X_{lk}\}_{k<l}$ and $\Delta n(t_l)$
are drawn independently, will be useful for deriving mesoscopic
equations because it allows us to calculate the moments of the
survival numbers $m(t_l,t_k)$. Our main strategy is to use these
moments in conjunction with the normalization condition to express the
expected spike count $\nbar(t_l)$, Eq.~\eqref{eq:mean-n-t}, as a
deterministic functional of the past activities. In this way,
$\nbar(t_l)$ will not depend anymore on the actual microscopic
realizations of the constrained noise
$\{X_{l',k}\}_{k,l'\in\mathbb{Z},k<l'<l}$ (constrained by a given
history $\{\Delta n(t_k)\}_{k<l}$ via Eq.~\eqref{eq:n-t-micro}) and can thus
be used to generate $\Delta n(t_l)$ as a Poisson random number from the
knowledge of the past activities.

\paragraph{Moment equations.}

To achieve such deterministic relationship, we first derive mesoscopic
equations for the mean and variance of $m(t_l,t_k)$ given
the history $\mathcal{H}(t_l)$ in the so-called {\em non-normalized
  ensemble} or {\em unconstrained} ensemble. This means that the
history determines the initial conditions of the dynamics of
$m(t_l,t_k)$, Eq.~\eqref{eq:m-poiss}, as well as the conditional
intensities $\lambda(t_{l}|t_k)$, but it does not impose the
constraint Eq.~\eqref{eq:n-t-micro} on the random numbers
$\{X_{l',k}\}_{l'\le l}$. Although this unconstrained noise leads to a
non-normalized distribution $\hat m(t_l,t_k)$, it still yields a very good
approximation of its mean and variance in the actual
constrained ensemble.  Taking the average of Eq.~\eqref{eq:m-poiss},
and using \eqref{eq:X-pois} yields the evolution of the mean:
\begin{equation}
  \label{eq:mean-m-map}
\langle \hat m_{l+1,k}\rangle=[1-P_\lambda(t_l|t_k)]\langle
\hat m_{l,k}\rangle
\end{equation}
with initial condition $\langle \hat m_{k+1,k}\rangle =\Delta n(t_k)$.  Here and
in the following, $\hat m_{l,k}$ is short-hand for $\hat m(t_l,t_k)$ to simplify
the notation, and $\langle\cdot\rangle$ denotes the ensemble average
of the unconstrained process for a given history
$\mathcal{H}(t_l)$. Actually, the condition for the average
$\langle \cdot\rangle$ can be extended to the history
$\mathcal{H}(t_{l+1})$ (and to any future activities) because in the
unconstrained ensemble neither $\hat m_{l,k}$ nor $\hat m_{l+1,k}$ depend on the
most recent activity $\Delta n(t_l)$ (clearly, this also holds for any future
activity). Importantly, Eq.~\eqref{eq:mean-m-map} is a mesoscopic
equation because it is fully determined by the past activities.

As a next step we derive an equation for the variance of
$\hat m$. To this end, let
\begin{equation}
  \label{eq:Delta-m}
  \Delta \hat m_{l,k}=\hat m_{l,k}-\langle \hat m_{l,k}\rangle
\end{equation}
denote the deviation from the mean.  \rot{Using the law of total variance, we find for the variance in the next time step
\begin{equation}
  \label{eq:total-variance}
  \langle\Delta \hat m_{l+1,k}^2\rangle=\mathrm{Var}\lreckig{\mathrm{E}\lreckig{m_{l+1,k}|m_{l,k},\mathcal{H}_l}}+\lrk{\mathrm{Var}\lreckig{m_{l+1,k}|m_{l,k},\mathcal{H}_l}}.
\end{equation}
The conditional mean of $m_{l+1,k}$ given the current value
$m_{l,k}$, denoted by $\mathrm{E}\lreckig{m_{l+1,k}|m_{l,k}}$, follows from the evolution equation \eqref{eq:m-poiss} and
Eq.~\eqref{eq:X-mean-var} as $[1-P_\lambda(t_l|t_k)]m_{l,k}$. Therefore, its
variance is $[1-P_\lambda(t_l|t_k)]^2\langle\Delta m_{l,k}^2\rangle$. For the second term in Eq.~\eqref{eq:total-variance}, we note that the conditional variance $\mathrm{Var}\lreckig{m_{l+1,k}|m_{l,k}}$ is equal to the variance $\mathrm{Var}\lreckig{X_{lk}|m_{l,k}}$ of the decrement $X_{lk}$. Because $X_{lk}$ is a Poisson variable, this variance is equal to the mean given by Eq.~\eqref{eq:X-mean-var}. Taken together, we find the following update rule for the total variance
\begin{equation}
 \label{eq:dvar}
 \langle \Delta \hat m^2_{l+1,k}\rangle=\bigl[1-P_\lambda(t_l|t_k)\bigr]^2\langle\Delta \hat m^2_{l,k}\rangle+P_\lambda(t_l|t_k)\langle \hat m_{l,k}\rangle
\end{equation}
with initial condition
$\langle\Delta \hat m^2_{k+1,k}\rangle=0$.}  As a function of $t_l$
(Fig.~\ref{fig:scheme-theory}C bottom), the variance
$\langle\Delta \hat m^2(t_l,t_k)\rangle$ is initially zero because all
neurons have still survived immediately after firing at time $t_k$. On
the other hand, at long times $t_l\gg t_k$, the variance also vanishes
because according to Eq.~\eqref{eq:mean-m-map}, the mean number of
survived neurons $\langle m(t_l,t_k)\rangle$ appearing in
Eq.~\eqref{eq:dvar} goes to zero. As a consequence, the variance
obtains a maximum at an intermediate time. Similarly, the dependence
of the variance at time $t_l$ for different last spike times $\tl=t_k$
shows the same limiting behavior which implies a maximum at an
intermediate last spike time $\tl$ (Fig.~\ref{fig:scheme-theory}B
bottom). However, the rugged shape of this function with many local
maxima reflects the discontinuity of the driving force
$\langle m(t_l,t_k)\rangle$ as a function of $t_k$ that arises from
the stochastic initial condition
$\langle m(t_{k+1},t_k)\rangle=\Delta n(t_k)$.

\paragraph{Mesoscopic population equations.}

Let us return to Eq.~\eqref{eq:mean-n-t} for the expected activity
$\nbar(t_l)$, which is used to draw the activity $\Delta n(t_l)$ as a
Poisson random number (cf. Eq.~\eqref{eq:n-poisson}). Because
  we condition on the history $\{\Delta n(t_{l'})\}_{l'<l}$, the
  processes $m_{l,k}$ in this equation belong to the ``constrained''
  ensemble, in which the normalization condition,
  Eq.~\eqref{eq:normalization-gen}, is obeyed. We note that these
  constrained processes could in principle be generated
  microscopically by Eq.~\eqref{eq:m-poiss} if at each time $t_{l'}$
  in the past, the microscopic noise $X_{l'k}$ was sampled from a
  joint distribution that ensures the conservation of neurons,
  Eq.~\eqref{eq:n-t-micro}, i.e. $\sum_kX_{l'k}=\Delta n(t_{l'})$.
  However, as we will show in the following, such a construction is
  not needed because the dependence of the expected activity
  $\nbar(t_l)$ on a specific realization of $m_{l,k}$ can be
  eliminated by exploiting the normalization condition,
  Eq.~\eqref{eq:normalization-gen}. To this end, we take advantage of
the fact that the mean $\langle \hat m_{l,k}\rangle$ of the
unconstrained process is a mesoscopic variable.  This suggests to
split the constrained processes $m_{l,k}$ into the mean of the
unconstrained process and a fluctuation part:
\begin{equation}
  \label{eq:split}
  m_{l,k}=\langle \hat m_{l,k}\rangle+\delta m_{l,k}.
\end{equation}
The first contribution is deterministic given the past activities
$\Delta n(t_k)$ while the second contribution represents the microscopic
fluctuations.  We note that the fluctuation $\delta m_{l,k}$ is
not equivalent to the deviation $\Delta \hat m_{l,k}$ of the
unconstrained process because $\langle \hat m_{l,k}\rangle+\Delta \hat m_{l,k}$
does not obey the normalization condition,
Eq.~\eqref{eq:normalization-gen}, whereas $\langle \hat m_{l,k}\rangle+\delta m_{l,k}$ does.

To remove the microscopic fluctuations $\delta m_{l,k}$, we require that both
Eq.~\eqref{eq:normalization-gen} and Eq.~\eqref{eq:mean-n-t} are
simultaneously satisfied. Substituting 
Eq.~\eqref{eq:split} into these equations leads to
\begin{align}
  \label{eq:bla}
  N&=\sum_{{k}=-\infty}^{l-1} \langle \hat m_{l,k}\rangle+\sum_{{k}=-\infty}^{l-1} \delta m_{l,k},\\\label{eq:mean-n-2}
\Delta\bar{n} (t_l)&=\sum_{{k}=-\infty}^{l-1} P_\lambda (t_l|t_{k})\langle \hat m_{l,k}\rangle+\sum_{{k}=-\infty}^{l-1} P_\lambda (t_l|t_{k})\delta m_{l,k}.
\end{align}
The microscopic fluctuations $\delta m_{l,k}$ enter the dynamics only
in the form of two sums.  First, the normalization condition
Eq.~\eqref{eq:bla} imposes a strict relation between the summed
deviation $\sum_k\delta m_{l,k}$ and the means of the
unconstrained processes, $\langle \hat m_{l,k}\rangle$, irrespective of the
specific, underlying microscopic dynamics of $m_{l,k}$. In particular,
we can solve for $\sum_k\delta m_{l,k}=N-\sum_k\langle \hat m_{l,k}\rangle$
with terms on the r.h.s. that are completely determined given the past
activities.  Second, the total effect of the deviations on the
expected activity $\nbar(t_l)$ is given by the weighted sum
$\sum_kP_\lambda(t_l|t_k)\delta m_{l,k}$ in Eq.~\eqref{eq:mean-n-2} with $P_\lambda(t_l|t_k)\le 1$ for all $k<l$. 
The weighted sum
$\sum_kP_\lambda(t_l|t_k)\delta m_{l,k}$ is therefore tightly constrained by the
value of the summed fluctuation $\sum_k\delta m_{l,k}$.  These considerations
suggest to make the following decoupling approximation:
\begin{equation}
  \label{eq:approx-free}
  \sum_{{k}=-\infty}^{l-1} P_\lambda (t_l|t_{k})\delta m_{l,k}\approx P_\Lambda (t_l)\sum_{{k}=-\infty}^{l-1} \delta m_{l,k}
\end{equation}
with a still unknown factor $P_\Lambda(t_l)$, that we call effective firing probability. To determine the effective firing probability, we require
that \rot{in the unconstrained ensemble the corresponding approximation
\begin{equation}
  \label{eq:approx-free-unconstr}
  \sum_{{k}=-\infty}^{l-1} P_\lambda (t_l|t_{k})\Delta \hat m_{l,k}=P_\Lambda (t_l)\sum_{{k}=-\infty}^{l-1} \Delta \hat m_{l,k}+\varepsilon_l
\end{equation}
}minimizes the mean squared error
\rot{$\mathcal{E}(P_\Lambda)=\langle\varepsilon_l^2\rangle$,} where
$P_\Lambda$ is short-hand for $P_\Lambda(t_l)$. We use the
unconstrained deviations $\Delta \hat m_{lk}$ here because we are only
interested in the typical error. \rot{Note that if $\varepsilon_l$
  is Gaussian distributed, minimizing the mean squared error
  yields the optimal estimation of $P_\Lambda$ in the sense that it
  maximizes the log-likelihood of
  $\sum_kP_\lambda(t_l|t_k)\Delta \hat m_{l,k}$ given
  $\sum_k\Delta \hat m_{l,k}$ under the linear approximation
  Eq.~\eqref{eq:approx-free-unconstr}. The error can be rewritten as
  $\varepsilon_l=\sum_k[P_\lambda(t_l|t_k)-P_\Lambda(t_l)]\Delta
  m_{l,k}$,
  which for $N\gg 1$ is a sum of many independent variables that can
  indeed be considered to be Gaussian.}  The derivative of
$\mathcal{E}$ with respect to $P_\Lambda$ is
\begin{equation}
  \label{eq:dEdL}
  \od{\mathcal{E}}{P_\Lambda}=2P_\Lambda\sum_{k,k'<l}\langle\Delta \hat m_{l,k}\Delta \hat m_{l,k'}\rangle-2\sum_{k,k'<l}P_\lambda(t_l|t_k)\langle\Delta \hat m_{l,k}\Delta \hat m_{l,k'}\rangle,
\end{equation}
where we have exploited that $P_\lambda(t_l|t_k)$ is deterministic given
the past activities $\Delta n(t_k)$, $k<l$. Furthermore, under this
condition, the fluctuations $\Delta \hat m_{l,k}$ and $\Delta \hat m_{l,k'}$
with $k\neq k'$ are conditionally independent. Using this property and
setting $\mathrm{d}\mathcal{E}/\mathrm{d}P_\Lambda=0$ we find that the optimal effective firing probability is
\begin{equation}
  \label{eq:lambdabar}
  P_\Lambda(t_l)=\frac{\sum_{k=-\infty}^{l-1}P_\lambda(t_l|t_k)\langle \Delta \hat m^2_{l,k}\rangle}{\sum_{k=-\infty}^{l-1}\langle \Delta \hat m^2_{l,k}\rangle}.
\end{equation}
The variance $\langle \Delta \hat m^2_{l,k}\rangle$ in this formula obeys
the mesoscopic dynamics derived above in Eq.~\eqref{eq:dvar}.  Hence,
the effective firing probability is itself mesoscopic.

Using Eq.~\eqref{eq:bla} and \eqref{eq:approx-free},
$\sum_k\delta m_{l,k}$ can be eliminated in Eq.~\eqref{eq:mean-n-2}
resulting in
\begin{equation}
  \label{eq:master-discrete-inf}
  \nbar (t_l) =\sum_{k=-\infty}^{l-1} P_\lambda (t_l|t_k)\langle m _{l,k}\rangle +P_\Lambda (t_l)\left(N -\sum_{k=-\infty}^{l-1} \langle \hat m_{l,k}\rangle\right).
\end{equation}
Thus, we obtain an equation that yields the mean spike count
$\Delta\bar{n} (t_l)$ at the present time as a function of the past spike
counts $\{\Delta n(t_k)\}_{k<l}$. Equation \eqref{eq:master-discrete-inf} is
the desired mesoscopic equation in discrete time.  For sufficiently
small time steps $\Delta t$, the present spike count $\Delta n(t_l)$ can be
generated by drawing a Poisson random number with mean $\nbar(t_l)$
according to Eq.~\eqref{eq:n-poisson}.

\subsection*{Mesoscopic population density equations in continuous time}

In continuous time, we consider the rescaled variables
\begin{equation}
  \label{eq:a-def}
  A_N(t_l)=\frac{\Delta n(t_l)}{N\Delta t},\quad\bar A (t_l)=\frac{\nbar (t_l)}{N \Delta t}.
\end{equation}
Here, $\bar{A}(t)$ can be interpreted as the {\em expected population
  activity} given the past activity $A_N(t')$, $t'<t$. For $\Delta t$
small but positive, the spike count $\Delta n(t)$ is an independent Poisson
number with mean $\Delta\bar{n}(t)=N\bar{A}(t)\Delta t$. Thus, on a
coarse-grained time scale, the continuum limit of the population
activity may be written in the following suggestive way
\begin{equation}
  \label{eq:An-intuit}
  A_N(t)=\frac{\mathrm{d} n(t)}{N\mathrm{d}t},\qquad \mathrm{d} n(t)\sim \text{Pois}(N\bar A(t)\mathrm{d}t),
\end{equation}
where $\mathrm{d}t$ denotes an infinitesimal (but temporally
coarse-grained) time step and $\mathrm{d}n(t)$ is an independent
Poisson-distributed random number with mean $N\bar A(t)\mathrm{d}t$.
In the limit $\mathrm{d}t\rightarrow 0$, the spike count \rot{in an infinitesimal
time step is a Bernoulli random number, where}
$\mathrm{d} n(t)=1$ with probability $N\bar A(t)\mathrm{d}t$ and
$ n(t)=0$ with probability $1-N\bar A(t)\mathrm{d}t$.  Therefore, in
this limit the population activity $A_N(t)$ converges to a sequence of
Dirac $\delta$-functions occurring at random times $t_{\text{pop},k}$
with rate $N\bar A(t)$. Thus, $A_N(t)$ can be written more formally as
a population spike train or ``shot-noise''
\begin{equation}
  \label{eq:pop-act-pp}
  A_N(t)=\frac{1}{N}\sum_{k}\delta(t-t_{\text{pop},k}),
\end{equation}
where $(t_{\text{pop},k})_{k\in\mathbb{Z}}$ is a point process with
conditional intensity
$\lambda_\text{pop}(t|\mathcal{H}_t)=N\bar{A}(t)$. Here, the condition
$\mathcal{H}_t$ denotes the history of the population activity
$\{A_N(t')\}_{t'<t}$, or equivalently, the history of spike times
$\{t_{\text{pop},k}\}_{t_{\text{pop},k}<t}$, up to (but not including)
time $t$.

To obtain the dynamics for $\bar{A}(t)$, we also introduce the
rescaled variables
\begin{equation}
  \label{eq:S-v}
  S(t_l|t_k)=\frac{\langle \hat m_{l,k}\rangle}{\Delta n(t_k)},\quad v(t_l,t_k)=\frac{\langle \Delta \hat m^2_{l,k}\rangle}{N\Delta t}.  
\end{equation}
The function $S(t|\tl)$ can be interpreted as the survival probability
of neurons that have fired their last spike at time
$\tl$. Furthermore, for small $\Delta t$ the firing probability is
given by $P_\lambda(t_l|t_k)=\lambda(t_l|t_k)\Delta t+O(\Delta t^2)$.
Thus, the continuum limit of Eq.~\eqref{eq:master-discrete-inf} reads
\begin{multline}
  \label{eq:master-cont-a-inf-discr-def}
  \bar A (t)=\lim_{\Delta t\rightarrow 0}\left\{\sum_{k=-\infty}^{\frac{t}{\Delta t}-1} \lambda (t|k\Delta t)S (t|k\Delta t)\frac{\Delta n(k\Delta t)}{N}\right.
\\\left.+\frac{\sum_{k=-\infty}^{\frac{t}{\Delta t}-1} \lambda (t|k\Delta t)v(t,k\Delta t)\Delta t}{\sum_{k=-\infty}^{\frac{t}{\Delta t}-1}v(t,k\Delta t)\Delta t}\left(1-\sum_{k=-\infty}^{\frac{t}{\Delta t}-1}S (t|k\Delta t)\frac{\Delta n(k\Delta t)}{N}\right)\right\}.
\end{multline}
The sums in this equation can be regarded as the definition of stochastic integrals, which allows us to rewrite Eq.~\eqref{eq:master-cont-a-inf-discr-def}
as
\begin{equation}
  \label{eq:master-cont-a-inf}
  \bar A (t)=\int_{-\infty}^t \lambda (t|\tl)S (t|\tl)A_N (\tl)\,\mathrm{d}\tl+\Lambda (t)\left(1-\int_{-\infty}^t S (t|\tl)A_N (\tl)\,\mathrm{d}\tl\right).  
\end{equation}
Here,
\begin{equation}
\label{eq:Lambda-v}
\Lambda(t)=\frac{\int_{-\infty}^t\lambda(t|\tl)v(t,\tl)\,\mathrm{d}\tl}{\int_{-\infty}^tv(t,\tl)\,\mathrm{d}\tl}
\end{equation}
is an effective rate corresponding to the effective firing probability
$P_\Lambda(t)$. Note that according to
Eq.~\eqref{eq:master-cont-a-inf-discr-def}, the stochastic integrals
in Eq.~\eqref{eq:master-cont-a-inf} extend only over last spike times
$\tl<t$ not including time $\tl=t$.  Taking the continuum limit of
Eq.~\eqref{eq:mean-m-map} we find that the survival probability
satisfies the differential equation
\begin{equation}
\label{eq:S-dgl-meth}
\pd{S(t|\tl)}{t}=-\lambda(t|\tl)S(t|\tl),\qquad S(\tl|\tl)=1.  
\end{equation}
This equation has the simple solution
\begin{equation}
  \label{eq:Sexp}
  S(t|\tl)=\exp\left(-\int_{\tl}^t\lambda(t'|\tl)\,\mathrm{d}t'\right).
\end{equation}
Similarly, we find from Eq.~\eqref{eq:dvar} that the rescaled variance
obeys the differential equation
\begin{equation}
\label{eq:variance-dynamics}
\pd{v}{t}=-2\lambda(t|\tl)v+\lambda(t|\tl)S(t|\tl)A_N(\tl),\qquad v(\tl|\tl)=0.
\end{equation}
The set of coupled equations
\eqref{eq:pop-act-pp}~--~\eqref{eq:variance-dynamics} defines the
mesoscopic population dynamics. We emphasize that not only $A_N(t)$
depends on $\bar{A}(t)$ (cf. Eq.~\eqref{eq:pop-act-pp}) but that there
is also a feedback of $A_N(t)$ onto the dynamics of $\bar{A}(t)$,
cf. Eq.~\eqref{eq:master-cont-a-inf}. In fact, $\bar A(t)$ can be
regarded as a deterministic functional of the past activities up to
but not including time $t$. In particular, $A_N(t)$ is {\em not} an
inhomogeneous Poisson spike train because the specific realization of
the spike history of $A_N(t)$ determines the conditional intensity
function for the point process $(t_{\text{pop},k})$ via
Eq.~\eqref{eq:master-cont-a-inf}.  Furthermore, we note that, in the
case of synaptic coupling or adaptation, also the variables $S$ and
$v$ depend on the history of the population activity through the
dependence of $\lambda(t|\tl)$ on the membrane potential $u(t,\tl)$
and the threshold $\vartheta(t,\tl)$ (cf. Eqs.~\eqref{eq:V-mean-field}
and \eqref{eq:qr-g}).

For large $N$, the population activity can be approximated by a
Gaussian process. To this end, we note that in the discrete time
description the spike counts $\Delta n(t_l)$ are conditionally independent
random numbers with mean and variance $N\bar A(t_l)\Delta t$. Therefore,
in the large-$N$ limit, the variable
\begin{equation}
  \label{eq:dW}
  \Delta W(t_l)=\frac{\Delta n(t_l)-N\bar A(t_l)\Delta t}{\sqrt{N\bar A(t_l)}}
\end{equation}
is normally distributed  with mean zero and variance $\Delta t$, and hence corresponds to the increment of a Wiener process. Using Eq.~\eqref{eq:a-def} for the population activity and taking the continuum limit $\Delta t\rightarrow 0$, we obtain
\begin{equation}
  \label{eq:pop-act-zero-mean-noise}
  A_N(t)=\bar A(t) +\sqrt{\frac{\bar A (t)}{N }}\xi (t),
\end{equation}
where $\xi(t)=\lim_{\Delta t\rightarrow 0}\Delta W(t)/\Delta t$ is a
Gaussian white noise with correlation function
$\langle\xi(t)\xi(t')\rangle=\delta(t-t')$. This Gaussian
approximation has the advantage that the multiplicative character of
the noise in Eq.~\eqref{eq:pop-act-zero-mean-noise} becomes explicit
because $\xi(t)$ is independent of the state of the system. It also
explicitly reveals that the finite-size fluctuations scale like
$1/\sqrt{N}$. We stress again that $A_N(t)$ is not a white-noise
process with time-dependent mean, as
Eq.~\eqref{eq:pop-act-zero-mean-noise} might suggest at first glance,
but it is a sum of two mutually correlated processes, (i) a
white-noise term proportional to $\xi(t)$ that reflects the fact that
the population activity is a $\delta$-spike train and (ii) a colored
``noise'' $\bar{A}(t)$ that arises from the filtering of $\xi(t)$
through the dynamics in Eq.~\eqref{eq:master-cont-a-inf}. As a result,
the auto-correlation function of $A_N(t)$ contains a $\delta$-peak and
a continuous part, consistent with previous theoretical findings
\cite{DegSch14}.  In particular, at short lags the auto-correlation
function may be negative as a result of refractoriness: in this case,
$\xi$ and $\bar A$ are anti-correlated in line with the intuitive
picture discussed in the {\sc Results} section,
Fig.~\ref{fig:scheme-theory}, that a positive fluctuation $\xi(t)$ is
associated with the creation of a ``hole'' in the distribution of last
spike times leading to a reduced activity after time $t$. In the
frequency domain, refractoriness corresponds to a trough in the power
spectrum at low frequencies \cite{FraBai95} as visible, for instance,
in Fig.~\ref{fig:lif}. These considerations clearly highlight the
non-white character of the finite-size fluctuations in our theory.

\subsection*{Several populations}
\label{sec:several-pop}

It is straightforward to generalize the population equations to
several populations by adding a population label
$\alpha=1,\dotsc,M$. For the sake of completeness, we explicitly
state the full set of equations. The activity of population $\alpha$ is given by
\begin{equation}
  \label{eq:pop-act-pp-multi}
  A_N^\alpha(t)=\frac{1}{N^\alpha}\sum_{k}\delta(t-t_{\text{pop},k}^\alpha),
\end{equation}
where $(t_{\text{pop},k}^\alpha)_{k\in\mathbb{Z}}$ is a point process with
conditional intensity
$\lambda_\text{pop}^\alpha(t|\mathcal{H}_t)=N^\alpha\bar{A}^\alpha(t)$. Here,
the expected activity $\bar A(t)$ depends explicitly on the history
$\mathcal{H}_t=\{A_N^\beta(t')\}_{t'<t,\beta=1,\dotsc,M}$ by the following set of
equations
\begin{align}
  \label{eq:master-cont-a-inf-multi}
  \bar {A}^\alpha (t)&=\int_{-\infty}^t \lambda ^\alpha(t|\tl)S ^\alpha(t|\tl)A_N^\alpha (\tl)\,\mathrm{d}\tl+\Lambda ^\alpha(t)\left(1-\int_{-\infty}^t S ^\alpha(t|\tl)A_N^\alpha (\tl)\,\mathrm{d}\tl\right)\\
\label{eq:lam-multi}
  \haz^{\alpha}(t|\tl)&=c^{\alpha}\exp\left(\frac{u^{\alpha}(t,\tl)-\vartheta^\alpha(t,\tl)}{\Delta_u^{\alpha}}\right),\quad
\Lambda^\alpha(t)=\frac{\int_{-\infty}^t\lambda^\alpha(t|\tl)v^\alpha(t,\tl)\,\mathrm{d}\tl}{\int_{-\infty}^tv^\alpha(t,\tl)\,\mathrm{d}\tl},\\
\label{eq:S-multi}
\pd{S^\alpha}{t}&=-\lambda^\alpha(t|\tl)S^\alpha,\qquad S^\alpha(\tl|\tl)=1,\\
\label{eq:variance-dynamics-multi}
\pd{v^\alpha}{t}&=-2\lambda^\alpha(t|\tl)v^\alpha+\lambda^\alpha(t|\tl)S^\alpha(t|\tl)A_N^\alpha(\tl),\qquad v^\alpha(\tl,\tl)=0,\\
\label{eq:u-multi}
  \pd{u^\alpha}{t}&=-\frac{u^\alpha-\mu^\alpha(t)}{\taum^\alpha}+\sum_{\beta=1}^M w^{\alpha\beta}p^{\alpha\beta}N^\beta (\epsilon^{\alpha\beta}*A_N^\beta)(t),\qquad u^{\alpha}(\tl,\tl)=\vreset \\
\label{eq:vartheta-multi}
  \vartheta^{\alpha}(t,\tl)&=\vth+\theta^{\alpha}(t-\tl)+\int_{-\infty}^{\tl}\tilde{\theta}^\alpha(t-t')A_N^\alpha(t')\mathrm{d}t'.
\end{align}
For each population, the system of equations
\eqref{eq:master-cont-a-inf-multi} -- \eqref{eq:vartheta-multi}
contains \gruen{a family of ordinary differential equations for the variables $S$, $u$
and $v$ parametrized by
the continuous parameter $\tl$ with $-\infty<\tl<t$, and five
integrals over this parameter. In the next section, we show that the
family of ordinary differential equations is equivalent to three
first-order partial differential equations. Furthermore, in
Sec.~\Newnameref{sec:finite-hist}, we reduce the infinite integrals to
integrals over a finite range, which will be useful for the numerical
implementation of the population equations.}

\subsection*{Refractory density representation}

There is an equivalent formulation of the population equation in terms
of first-order partial differential equations for the density of ages
$\tau=t-\tl$ \cite{Ger00,MeyVre02,ChiGra07,ChiGra08}. The representation in
terms of age $\tau$ as a state variable is useful because it parallels
the Fokker-Planck formalism for the membrane potential density
\cite{NykTra00,BruHak99,Bru00,MatGiu02,GerKis14} or related density equations \cite{IyeMen13,LaiKam16_arxiv}, in which the state
variable is the membrane potential of a neuron. To keep the notation
simple we consider in the following population $\alpha$ but drop the index
$\alpha$ wherever confusion is not possible. Thus, we write e.g. $S$
for $S^\alpha$ and $A_N$ for $A_N^\alpha$ but we keep the index
$\beta$ as well as double indices $\alpha\beta$ occurring in
Eq.~\eqref{eq:u-multi}.

The density of ages at time $t$ is defined as
$q(t,\tau)=S(t|t-\tau)A_N(t-\tau)$. \gruen{We recall that because of finite-size fluctuations,} $q$ is not a normalized probability density. Furthermore, we regard the functions
$\lambda $, $u $ and $v $ as functions of $t$ and $\tau$.  With these
definitions the population equation, Eq.~\eqref{eq:master-cont-a-inf},
can be rewritten as
\begin{align}
  \label{eq:A-refract}
  \bar A (t)=\int_0^\infty\lambda (t,\tau)q(t,\tau)\,\mathrm{d}\tau+\Lambda (t)
\left(1-\int_{0}^\infty q(t,\tau)\,\mathrm{d}\tau\right).
\end{align}
\rot{The stochastic activity $A_N(t)$ then follows from Eq.~\eqref{eq:pop-act-coarse} or \eqref{eq:AN-Gaussian}.} Equation \eqref{eq:A-refract} yields the expected population rate at time $t$ for a
given density of ages.  In the Fokker-Planck formalism, this would
correspond to the calculation of the rate from the membrane potential
density as the probability flux across the threshold.

Noting that $\partial_tS(t|\tl)A_N(\tl)=(\partial_t+\partial_\tau)q(t,\tau)$, we find from Eq.~\eqref{eq:S-dgl-meth} the following first-order partial differential equation for the density of ages $q(t,\tau)$:
\begin{equation}
  \label{eq:quasi-lin}
  (\partial_t+\partial_\tau)q =-\lambda (t,\tau)q ,\qquad q (t,0)=A_N (t).
\end{equation}
Similarly, $u $ and $v $ obey from Eq.~\eqref{eq:u-multi} and
\eqref{eq:variance-dynamics-multi}, respectively,
\begin{align}
  \label{eq:uv-quasilin_u}
  (\partial_t+\partial_\tau)u &=-\frac{u -\mu }{\taum }+\sum_{\beta=1}^M w^{\alpha\beta}p^{\alpha\beta}N^\beta (\epsilon^{\alpha\beta}*A_N^\beta)(t),\\
  \label{eq:uv-quasilin_v}
  (\partial_t+\partial_\tau)v &=-\lambda (t,\tau)[2v -q ]
\end{align}
with boundary conditions $u (t,0)=\vreset$ and $v (t,0)=0$. These
functions, together with the threshold dynamics
\begin{equation}
  \label{eq:thresh-tau}
  \vartheta(t,\tau)=\vth+\theta(\tau)+\int_{\tau}^{\infty}\tilde{\theta}(\tau')A_N(t-\tau')\,\mathrm{d}\tau',
\end{equation}
determine $\lambda (t,\tau)$ and $\Lambda (t)$ via
Eq.~\eqref{eq:lam-multi}, i.e.
\begin{equation}
  \label{eq:lam-refr}
  \haz(t,\tau)=c\exp\left(\frac{u(t,\tau)-\vartheta(t,\tau)}{\Delta_u}\right),\quad
\Lambda(t)=\frac{\int_0^{\infty}\lambda(t,\tau)v(t,\tau)\,\mathrm{d}\tau}{\int_0^{\infty}v(t,\tau)\,\mathrm{d}\tau}.
\end{equation}
The equations \eqref{eq:S-multi} -- \eqref{eq:u-multi} of the previous
section can be regarded as the characteristic equations corresponding
to the partial differential equations \eqref{eq:quasi-lin} --
\eqref{eq:uv-quasilin_v} (``method of characteristics'').

\subsection*{Population equations for a finite history.}
\label{sec:finite-hist}

To simulate the population activity forward in time, the integrals in
Eq.~(\ref{eq:master-cont-a-inf}) over the past need to be evaluated,
starting at $\tl=-\infty$. For biological systems, however, it is
sufficient to limit the integrals to a finite history of length
$T$. This history corresponds to the range $t-T\le \tl<t$, where we
have to explicitly account for the dependence of the conditional
firing rate $\lambda(t|\tl)$ on the last spike time $\tl$. We will
call neurons with last spike time in this range ``refractory'' because
they still experience some degree of (relative) refractoriness caused by the last
spike. The remaining part of the integral corresponding to the range
$-\infty<\tl<t-T$ receives a separate, compact evaluation. We will
refer to neurons with their last spike time in this range as ``free''
because their conditional intensity is free of the influence of the
last spike.

How should we choose the length of the explicit history $T$? First of
all, this length can be different for different populations and is
mainly determined by the time scale of refractoriness, i.e. the time
it needs to forget the individual effect of a single spike in the
past. Furthermore, it depends on the properties of the spike-triggered
kernel, i.e. the dynamic threshold that is responsible for
adaptation. More precisely, we determine the length of the history by
the following conditions: first, the conditional intensity is
insensitive to the precise timing of the last spike at $\tl<t-T$ if
\begin{equation}
  \label{eq:T-relativ}
  T\gg\max [\tref,\taurel].
\end{equation}
Here, $\tref$ is the absolute refractory period and $\taurel$ is
the time scale of the relative refractory period. For the GIF model
$\taurel=\taum$. Second, we demand that $T$ is chosen such that for 
$t>T$, the quasi-renewal kernel
$\tilde{\theta}(t)=\Delta_u\bigl[1-e^{-\theta(t)/\Delta_u}\bigr]$ can
be well approximated by the original spike-triggered kernel
$\theta(t)$. Taylor expansion of the exponential yields the condition
\begin{equation}
  \label{eq:qr-appr-T}
  \theta(t)\ll\Delta_u,\qquad \forall t>T.
\end{equation}
The length of the history $T$ needs to be chosen such that both
conditions, Eq.~(\ref{eq:T-relativ}) and (\ref{eq:qr-appr-T}) are
fulfilled. It is important to note that condition
Eq.~(\ref{eq:qr-appr-T}) does not require the time window $T$ to be
larger than the largest time scale of the spike-triggered kernel. For
instance, consider the kernel
$\theta(t)=\frac{J_{\theta}}{\tau_{\theta}}e^{-t/\tau_{\theta}}$, where $J_{\theta}$ and $\tau_{\theta}$
are adaptation strength and time scale, respectively. In particular,
the adaptation strength $J_{\theta}$ sets the reduction in firing rate
compared to a non-adapting neuron in the limit of strong drive
irrespective of the time scale $\tau_{\theta}$ (see
e.g. \cite{LiuWan01,SchFis10}). Condition Eq.~(\ref{eq:qr-appr-T}) can
be fulfilled for a given $T$ if either $\tau_{\theta}$ is small enough such
that the exponential $e^{-t/\tau_{\theta}}$ is small, or, for a fixed
adaptation strength $J_{\theta}$, by increasing the adaptation time scale
$\tau_{\theta}$ such that $J_{\theta}/\tau_{\theta}\ll \Delta_u$.

\paragraph{Dynamic threshold of refractory and free neurons.}

For free neurons, i.e. for $-\infty<\tl<t-T$, we use the average
threshold under the assumption that spikes occurred in the range
$-\infty<\tl<t-T$ according to an inhomogeneous Poisson process with
rate $A_N(\tl)$. This average is given by \cite{Str67I,NauGer12}
\begin{align}
  \label{eq:qr-free-thresh}
  \vartheta_\text{free}(t)&=\vth+\int_{-\infty}^{t-T}\tilde{\theta}(t-t')A_N(t')\mathrm{d}t',\nonumber\\
  &\approx\vth+\int_{-\infty}^{t-T}\theta(t-t')A_N(t')\mathrm{d}t',
\end{align}
where in the last step we used Eq.~(\ref{eq:qr-appr-T}). We assume
that for $t>T$ the spike-triggered kernel can be sufficiently well approximated
by a sum of exponentials
\begin{equation*}
  \label{eq:gamma-exp}
  \theta(t)=\Theta(t)\sum_{\ell=1}^{N_\theta}\frac{J_{\theta,\ell}}{\tau_{\theta,\ell}} e^{-t/\tau_{\theta,\ell}}.
\end{equation*}
This allows us to express the threshold for free neurons as
  \begin{equation}
  \label{eq:thresh-nr}
    \vartheta_\text{free}(t)=\vth+\sum_{\ell=1}^{N_\theta}J_{\theta,\ell} e^{-T/\tau_{\theta,\ell}}g_\ell(t),
  \end{equation}
  where the variables $g_\ell(t)$ satisfy the differential equations
  \begin{equation}
    \label{eq:dgl-thresh-variables}
    \tau_{\theta,\ell}\od{g_\ell}{t}=-g_\ell+A_N(t-T).
  \end{equation}  

For refractory neurons, i.e. if $t-T\le \tl<t$, we need to evaluate in
the effective threshold, Eq.~\eqref{eq:qr-g}, an integral over the
exact quasi-renewal kernel $\tilde{\theta}(t)$. Splitting this
integral into the free and refractory part yields the threshold of
refractory neurons:
\begin{equation}
  \label{eq:vartheta}
  \vartheta(t,\tl)=\vartheta_\text{free}(t)+\theta(t-\tl)+\int_{t-T}^{\tl} \tilde{\theta}(t-t')A_N(t')\mathrm{d}t'.
\end{equation}
We can use the threshold for free and refractory neurons,
Eqs.~\eqref{eq:thresh-nr}, \eqref{eq:dgl-thresh-variables} and Eq.~\eqref{eq:vartheta},
respectively, to obtain the respective conditional intensities:
\begin{equation}
  \label{eq:cond-int-free}
  \freehaz(t)=f\bigl(h(t)-\freetheta(t)\bigr),\qquad \haz(t,\tl)=f\bigl(u(t,\tl)-\theta(t,\tl)\bigr),
\end{equation}
where $h(t)$ is the free membrane potential given by
Eq.~\eqref{eq:h-eta}. Let us remind the reader that $h(t)$ obeys the dynamics
Eq.~\eqref{eq:glif} but without resetting of the membrane potential
after a spike.

\paragraph{Population equations.}

We now apply the split of the history to the integrals that appear
in the population equations, specifically
Eq.~(\ref{eq:master-cont-a-inf}) and \eqref{eq:Lambda-v}.  By
definition, the conditional intensity of free neurons does not depend
explicitly on the last spike time $\tl$. That is,
$\lambda(t|\tl)=\lambda_{\text{free}}(t)$ for $-\infty<\tl<t-T$, where
the free hazard rate $\lambda_{\text{free}}(t)$ is given by
Eq.~\eqref{eq:cond-int-free}. In the free part of the integrals in
Eq.~(\ref{eq:master-cont-a-inf}) and \eqref{eq:Lambda-v}, the free
hazard rate can be pulled out of the integral, which yields
\begin{align}
\label{eq:Abar-splitting}
    \bar A (t)&=\int_{t-T}^t \lambda (t|\tl)S (t|\tl)A_N (\tl)\,\mathrm{d}\tl+\lambda_\text{free}(t)\frac{x(t)}{N}\\
&\qquad+\Lambda (t)\left(1-\int_{t-T}^t S (t|\tl)A_N (\tl)\,\mathrm{d}\tl-\frac{x(t)}{N}\right),\nonumber\\
\Lambda(t)&=\frac{\int_{t-T}^t\lambda(t|\tl)v(t,\tl)\,\mathrm{d}\tl+\freehaz(t)z(t)/N}{\int_{t-T}^tv(t,\tl)\,\mathrm{d}\tl+z(t)/N}.
\end{align}  
Here we have introduced the expected number of free neurons
$x(t)=N\int_{-\infty}^{t-T}S(t|\tl)A_N(\tl)\,\mathrm{d}\tl$ and the
partial integral over the variance function
$z(t)=N\int_{-\infty}^{t-T}v(t,\tl)\,\mathrm{d}\tl$. Differentiating
these new variables and employing Eqs.~\eqref{eq:S-v} and
\eqref{eq:variance-dynamics}, we find that they obey the differential
equations
\begin{align}
\label{eq:x-dynamics}
  \od{x}{t}&=-\freehaz(t)x+NS(t|t-T)A_N(t-T),\\
\label{eq:z-dynamics}
  \od{z}{t}&=-2\freehaz(t)z+\freehaz(t)x+Nv(t,t-T).
\end{align}
Thus, the integrals no longer run from $-\infty$ to $t$ but are now limited to the range $[t-T,t)$. The long tails over the past have been reduced to differential equations. 
 
\subsection*{Numerical implementation}
\label{sec:sim-algo}

\paragraph{Discretization of time.}

We discretize the time axis into a grid with step size
$\Delta t$ and grid points
\begin{equation}
  t_k=k\Delta t, \qquad k\in\mathbb{Z}.
\end{equation}
Because we keep track of a finite history with the oldest last spike
time $\tl=t-T$, the history consists of a finite number $K$ of bins
such that $T=K\Delta t$. If the index $k=l$ corresponds to the current
time, the oldest last spike time of the explicit history corresponds
to an index $k=l-K$ and the most recent one corresponds to the index
$k=l-1$.  Note that the numerical implementation requires the absolute
refractory period $\tref$ to be at least as large as the integration
time step $\Delta t$ (see below).

To facilitate the notation of the update rules, it is convenient to
introduce the following notations:
\begin{align}
  \bar m_k(t_l)&\equiv\langle m_{l,k}\rangle=N\Delta
tS(t_i|t_k)A_N(t_k),\\
  v_k(t_l)&\equiv\langle\Delta m^2_{l,k}\rangle=N\Delta tv(t_l,t_k),\\
  u_k(t_l)&\equiv u(t_l|t_k),\\
  \lambda_k(t_l)&\equiv\lambda(t_l|t_k).
\end{align}
In particular, $\bar m_k$ and $v_k$ correspond to the mean and
variance of the unconstrained survival numbers, respectively. \gruen{We also
recall that the population index
$\alpha$ is dropped wherever confusion is not possible, while the index $\beta$ as well as double indices $\alpha\beta$ will be kept.}

\paragraph{Choice of $\Delta t$.}

A crucial assumption of the derivation of the population equations in
discrete time was that the time step $\Delta t$ is small enough such
that each neuron fires at most one spike per time step. This can be achieved by the condition
\begin{equation}
  \label{eq:refract-cond}
  \Delta t\le\tref
\end{equation}
(cf. Sec.~\Newnameref{mean-field}). Clearly, this condition implies that the
total number of spikes per time step must be bounded by the number of
neurons, i.e. the population activity must obey $\Delta n(t_l)\le N$. The
equality sign corresponds to the case that all neurons fire in the
same time step. In addition to condition Eq.~\eqref{eq:refract-cond}, we also had to require that $\Delta t$
is not larger than the transmission delay $\Delta$, i.e.
\begin{equation}
  \label{eq:delay-cond}
  \Delta t\le\Delta.
\end{equation}

In order to justify the use of the Poisson statistics in the
  derivation of the population equations, we further assumed that
  $\Delta t$ is sufficiently small such that the expected number of
  spikes per time step, $\nbar(t)$, is much smaller than $N$, or
  equivalently $\bar A(t)\Delta t\ll 1$. While this does not pose a
  problem for the theory, which ultimately concerns with the continuum
  limit $\Delta t\rightarrow 0$, an efficient numerical integration of
  the population equations benefits from a time step that is as large
  as possible and should thus not be limited by such a condition. In
  particular, we should allow a large fraction of neurons to fire
  during one time step, either as a result of an external
  synchronization of many neurons (e.g. by a strong, sudden stimulus)
  or because of synchronous oscillations emerging from the network
  dynamics. In this case, a Poisson-distributed spike count
  $\Delta n(t)$ may exceed the number of neurons $N$. This problem can
  be remedied by drawing $\Delta n(t)$ from a binomial distribution
  with mean $\nbar(t)$ and maximal value $N$. For $\nbar(t)\ll N$,
  this binomial distribution agrees with the Poisson distribution used
  in our theory, whereas at large $\nbar(t)$ it ensures that the spike
  count is bounded by the total number of neurons $N$. Although the
  binomial distribution does not follow strictly from our theory, it
  is expected to yield a very good approximation even at large
  $\nbar(t)$.  The reason is as follows: \blau{A statement analogous
    to} our argument that the sum of Poisson numbers
  [Eq.~\eqref{eq:n-t-micro}] yields again a Poisson number
  [Eq.~\eqref{eq:n-poisson}]
  \blau{is, in general, not valid for
    binomial random numbers if the random numbers (i.e. the firing
    probabilities $P_\lambda(t_l|t_k)$ in our model) are very
    different.  However, if neurons are strongly synchronized, and
    hence $\nbar(t)\sim N$, they fire with a similar probability,
    which implies indeed a binomial distribution of the spike count
    $\Delta n(t)$.}

Besides Eq.~\eqref{eq:refract-cond} and \eqref{eq:delay-cond}, a third condition concerns the approximation of
the integral $\int_t^{t+\Delta t}\lambda(t')\mathrm{d}t'$ in
Eq.~\eqref{eq:firing-prob} by $\bar\lambda(t)\Delta t$ (trapezoidal
rule). This approximation is valid if the membrane potential $u$ and
threshold $\vartheta$ do not vary too strongly during a time
step. More precisely, the absolute error of the trapezoidal rule is
known to be $\Delta t^3|\lambda''|/12$, which we require to be much
smaller than $\lambda\Delta t$. Thus, the relative error is of order
$\Delta t^2$. Using the definition of $\lambda$,
Eq.~\eqref{eq:hazard-def}, this leads to the condition
\begin{equation}
  \label{eq:dlamda}
  \left|\lrrund{\pd{u}{t}-\pd{\vartheta}{t}}^2+\Delta_u\lrrund{\frac{\partial^2u}{\partial t^2}-\frac{\partial^2\vartheta}{\partial t^2}}\right|\Delta t^2\ll 12\Delta_u^2.
\end{equation}
In summary, $\Delta t$ should be chosen such that all three conditions,
Eqs.~\eqref{eq:refract-cond} -- \eqref{eq:dlamda} are satisfied for
all populations.

\paragraph{Update of the membrane potential.}

To compute the firing probabilities, we need to update both the
membrane potential and the threshold.  In the presence of an
exponential synaptic filter
$\epsilon^{\alpha\beta}(s)=\Theta(s-\Delta)e^{-(s-\Delta)/\taus^{\beta}}/\taus^\beta$, the
membrane potential of free neurons $u(t)=h(t)$ in population $\alpha$ obeys the differential
equation
 \begin{align}
   \label{eq:u-y}
   \taum\od{h}{t}&=-h+\vrest+RI_\text{ext}(t)+\taum \sum_{\beta=1}^Mp^{\alpha\beta}N^\beta w^{\alpha\beta}y^{\alpha\beta},\\
   \taus^\beta\od{y^{\alpha\beta}}{t}&=-y^{\alpha\beta}+A_N^{\alpha\beta}(t-\Delta),&\beta=1,\dotsc,M.
 \end{align}
Assuming that the external stimulus $I_\text{ext}(t)$ and the population activity $A_N(t)$ are constant during one time step, the solution over one time step is given by
\begin{align}
  \label{eq:u-y-sol}
  h(t_{l+1})&=\vrest+\lrrund{h(t_l)-\vrest}e^{-\Delta t/\taum}+h_\text{tot},\\
 \label{eq:y-sol}
  y^{\alpha\beta}(t_{l+1})&=A_N^\beta(t_l-\Delta)+\lreckig{y^{\alpha\beta}(t_l)-A_N^\beta(t_l-\Delta)}e^{-\Delta t/\taus^\beta},&\beta=1,\dotsc,M
\end{align}
where $h_\text{tot}$ is the total input of population $\alpha$ given by
\begin{multline}
  \label{eq:Itot}
  h_\text{tot}=RI_\text{ext}(t_l)\lrrund{1-e^{-\frac{\Delta t}{\taum}}}+\taum \sum_{\beta=1}^Mp^{\alpha\beta}N^\beta w^{\alpha\beta}\left\{A_N^\beta(t_l-\Delta)+\vphantom{\frac{\taus^\beta e^{-\frac{\Delta t}{\taus^\beta}}\lreckig{y^{\alpha\beta}(t_l)-A_N^\beta(t_l-\Delta)}}{\taus^\beta-\taum}}\right.\\
+\left.\frac{\taus^\beta e^{-\frac{\Delta t}{\taus^\beta}}\lreckig{y^{\alpha\beta}(t_l)-A_N^{\beta}(t_l-\Delta)}-e^{-\frac{\Delta t}{\taum}}\lreckig{\taus^\beta y^{\alpha\beta}(t_l)-\taum A_N^\beta(t_l-\Delta)}}{\taus^\beta-\taum}\right\}
\end{multline}

For refractory neurons, we obtain the membran potential in the GLM
model by the simple formula
$u_k(t_{l+1})=h(t_{l+1})+\eta(t_{l+1}-t_k)$. For the GIF model, 
the same update rule as for $h(t)$, Eq.~\eqref{eq:u-y-sol}, can be applied for  $k=l-K,\dotsc,l-k_{\text{ref}}$:
\begin{equation}
  \label{eq:u-iter}
  u_k(t_{l+1})=\vrest+\lrrund{u_k(t_l)-\vrest}e^{-\Delta t/\taum}+h_\text{tot}.
\end{equation}
For \rot{the absolute refractory period}, $l-k_\text{ref}<k<l$, \rot{the membrane potential remains at} $u_k(t_{l+1})=\vreset$. Note that the total integrated input $h_\text{tot}$ needs to
be computed only once per time step.

\paragraph{Update of the threshold.}
Let us first discuss, how to compute the threshold at time $t_l$ given
the values of $g_\ell(t_l)$ and $\Delta n(t_k)$ for $k=l-K,\dotsc,l-1$. For free
neurons, the threshold $\freetheta(t_l)$ is given by
Eq.~\eqref{eq:thresh-nr} evaluated at time $t=t_l$. For refractory neurons, we find from
Eq.~\eqref{eq:vartheta} that the threshold can be written in the
discretized form
\begin{equation}
  \label{eq:vartheta-discrete}
  \vartheta_k(t_l)=\vartheta_\text{free}(t_l)+\theta(t_l-t_k)+\frac{1}{N}\sum_{k'=l-K}^{k-1}\tilde{\theta}(t_l-t_{k'})\Delta n(t_{k'}),
\end{equation}
$k=l-K,\dotsc,l-k_\text{ref}$.  Equation \eqref{eq:vartheta-discrete} can be
rewritten as
\begin{equation}
  \label{eq:theta}
  \vartheta_k(t_l)=\hat\vartheta_k(t_l)+\theta\bigl(t_l-t_k\bigr),
\end{equation}
where the variables $\hat\vartheta_k(t)$ can be calculated iteratively starting at $k=l-K$:
\begin{equation}
  \label{eq:hattheta}
  \hat\vartheta_{k+1}(t_l)=\hat\vartheta_{k}(t_l)+N^{-1}\tilde{\theta}\bigl(t_l-t_{k}\bigr)\Delta n(t_{k}),\qquad k=l-K,\dotsc,l-1-\kref
\end{equation}
with initial condition
$\hat\vartheta_{l-K}(t_l)=\vartheta_\text{free}(t_l)$. Thus, at each
time step $t_l$, the threshold can be rapidly evaluated in one sweep
via Eq.~\eqref{eq:thresh-nr}, \eqref{eq:theta} and
\eqref{eq:hattheta}.

For the computation of the firing probabilities below, it is necessary
to compute the threshold one time step ahead, i.e. at time
$t_{l+1}$. To this end, we first update the variables $g_\ell$
according to Eq.~\eqref{eq:dgl-thresh-variables}:
\begin{equation}
  \label{eq:dgl-thresh-variables-discret}
  g_\ell(t_{l+1})=g_\ell(t_l)e^{-\Delta t/\tau_{\theta,\ell}}+\frac{\Delta n(t_{l-K})}{N\Delta t}\lrrund{1-e^{-\Delta t/\tau_{\theta,\ell}}}.
\end{equation}
This yields the threshold $\freetheta(t_{l+1})$ of free neurons via
the formula Eq.~\eqref{eq:thresh-nr}. For refractory neurons we find from Eqs.~\eqref{eq:theta} and  \eqref{eq:hattheta}
\begin{equation}
  \label{eq:theta-new}
\vartheta_k(t_{l+1})=\hat\vartheta_k(t_{l+1})+\theta\bigl(t_{l+1}-t_k\bigr),
\end{equation}
where $\hat\vartheta_k(t_{l+1})$ can be iterated by \rot{
\begin{equation}
  \label{eq:hattheta-methods}
  \hat\vartheta_{k+1}(t_{l+1})=\hat\vartheta_{k}(t_{l+1})+N^{-1}\tilde{\theta}\bigl(t_{l+1}-t_{k}\bigr)\Delta n(t_{k}), \qquad k=l-K,\dotsc,l-k_\text{ref}-1.
\end{equation}
with initial condition
\begin{equation}
  \label{eq:hattheta-initial}
  \hat\vartheta_{l-K}(t_{l+1})=\vartheta_\text{free}(t_{l+1})-N^{-1}\tilde{\theta}\bigl(t_{l+1}-t_{l-K}\bigr)\Delta n(t_{l-K}).
\end{equation}
}

\paragraph{Firing probabilities.}

The firing probabilities for free and refractory neurons
 are given by
\begin{equation}
  \label{eq:pfire}
  \pfree(t_l)=1-e^{-\bar\lambda_{\text{free}}(t_l)\Delta t},\qquad  P_\lambda(t_l|t_k)=1-e^{-\bar\lambda(t_l|t_k)\Delta t},
\end{equation}
respectively. Here,
\begin{align}
  \label{eq:lambda-bar-free}
  \bar\lambda_\text{free}(t_l)&=[\freehaz(t_l)+\freehaz(t_{l+1})]/2,\\
  \label{eq:lambda-bar}
  \bar\lambda(t_l|t_k)&=[\lambda_k(t_l)+\lambda_k(t_{l+1})]/2,
\end{align}
are the arithmetic mean of the respective intensities at the beginning
and end of the time interval (cf. Eq.~\eqref{eq:firing-prob}). \rot{The free intensity $\freehaz(t)$ is given by Eq.~\eqref{eq:cond-int-free}. For refractory neurons, the conditional intensities are given by
\begin{equation}
  \label{eq:intens-refarc}
  \lambda_k(t)=
  \begin{cases}
    f\bigl(u_k(t)-\vartheta_k(t)\bigr),&t_k< t-\kref\Delta t\\    
    0,&t-\kref\Delta t\le t_k<t,
  \end{cases}
\end{equation}
where the last case corresponds to the absolute refractory period.}

\paragraph{Population dynamics.} 

We can directly use the discretized form of the population equations given by
Eqs.~\eqref{eq:lambdabar} and \eqref{eq:master-discrete-inf}.  As in
Eq.~\eqref{eq:Abar-splitting} the infinite sums in
Eq.~\eqref{eq:master-discrete-inf} can be split into an explicit,
finite history of length $K$ and a remaining part corresponding to
$k<l-K$. This results in
\begin{multline}
  \label{eq:nbar}
  \nbar(t_l)=\sum_{k=l-K}^{l-1}P_{\lambda} (t_l|t_k)\bar{m}_k(t_l) +P_{\text{free}}(t_l)x(t_i)+\\
+P_\Lambda(t_l)\left(N-\sum_{k=l-K}^{l-1}\bar{m}_k(t_l) -x(t_l)\right),  
\end{multline}
where
\begin{equation}
  \label{eq:Lambda-discrete-finite-hist}
P_\Lambda(t_l)=\frac{\sum\limits_{k=l-K}^{l-1}P_\lambda(t_l|t_k)v_k(t_l) +\pfree(t_l)z(t_l)}{\sum\limits_{k=l-K}^{l-1}v_k(t_l) +z(t_l)}.
\end{equation}
is the firing probability of ``neurons'' belonging to the ``holes and overshoots'' $\delta m_{l,k}$ (cf. {\sc Results},
Sec.~\Newnameref{sec:mesosc-popul-equat}). The variables $x$ and $z$ have the
discrete time definition
\begin{equation}
  \label{eq:x-z-discrete-def}
  x(t_l)=\sum_{k=-\infty}^{l-K-1}\bar{m}_k(t_l) ,\qquad z(t_l)=\sum_{k=-\infty}^{l-K-1}v_k(t_l) ,
\end{equation}
corresponding to a discretization of their integral definition above.
Having calculated the expected spike count $\nbar(t_l)$, the actual
spike count $\Delta n(t_l)$ is obtained by drawing a binomially distributed random number
\begin{equation}
  \label{eq:binom}
  \Delta n(t_l)\sim B\bigl(N,p_B=\nbar(t_l)/N\bigr)
\end{equation}
as discussed above. In Eq.~\eqref{eq:binom}, $B(N,p_B)$ denotes the binomial distribution corresponding to $N$ Bernoulli trials with success probability $p_B$.

The discrete evolution equations for $\bar{m}_k(t_l) $ and $v_k(t_l)$ are given by Eq.~\eqref{eq:mean-m-map} and \eqref{eq:dvar}, respectively, which we repeat here for convenience:
\begin{align}
\label{eq:m-update}
  \bar m_k(t_{l+1})&=[1-P_{\lambda}(t_l|t_k)]\bar{m}_k(t_l) \\
\label{eq:v-update}
  v_k(t_{l+1})&=[1-P_{\lambda}(t_l|t_k)]^2v_k(t_l) + P_{\lambda}(t_l|t_k)\bar m_k(t_{l}).
\end{align}  
To find the update rule for $x$, we use the definition
Eq.~\eqref{eq:x-z-discrete-def} and the update rule for
$\bar{m}_k(t_l) $, Eq.~\eqref{eq:m-update}:
\begin{align}
  x(t_{l+1})&=\sum_{k=-\infty}^{l-K}\bar m_k(t_{l+1})=\sum_{k=-\infty}^{l-K}[1-P_{\lambda}(t_l|t_k)]\bar{m}_k(t_l) \nonumber\\
  \label{eq:x-upd-prelim}
            &=[1-P_{\text{free}}(t_l)]\sum_{k=-\infty}^{l-K-1}\bar{m}_k(t_l) +[1-P_{\lambda}(t_l|t_{l-K})]\bar m_{l-K}(t_l).
\end{align}
Here, we have exploited that $P_{\lambda}(t_l|t_k)=P_{\text{free}}(t_l)$
for $k<l-K$. Using again Eq.~\eqref{eq:m-update} we find
\begin{equation}
  \label{eq:x-update}
  x(t_{l+1})=[1-P_{\text{free}}(t_l)]x(t_l)+\bar m_{l-K}(t_{l+1}).
\end{equation}
This equation is the discrete analog of the continuous-time equation
\eqref{eq:x-dynamics}. Note that $\bar m_{l-K}(t_{l+1})$ is given by Eq.~\eqref{eq:m-update}.

An update rule for $z$ can be found from the definition
Eq.~\eqref{eq:x-z-discrete-def} and the update rule for
$\langle \Delta m_{l,k}^2\rangle$, Eq.~\eqref{eq:v-update}. A similar
calculation that led to Eq.~\eqref{eq:x-update} results in
\begin{align}
  \label{eq:z-update}
z(t_{l+1})=[1-P_{\text{free}}(t_l)]^2z(t_l)+P_{\text{free}}(t_l)x(t_l)+v_{l-K}(t_{l+1}).
\end{align}
This equation is the discrete analog of the continuous-time equation
\eqref{eq:z-dynamics}. Note that $v_{l-K}(t_{l+1})$ is given by Eq.~\eqref{eq:v-update}.

Finally, the initial conditions can be accounted for by setting 
\begin{equation}
  \label{eq:bc-res}
  \bar m_l(t_{l+1})=\Delta n(t_l),\qquad  v_l(t_{l+1})=0,\qquad  u_{l}(t_{l+1})=\vreset.    
\end{equation}
The last update step corresponding to the reset of the membrane potential only needs to be performed for the GIF model.

\paragraph{Initialization and storage of history.}
One simple way to initialize the system is to fully synchronize the
network at time $-\Delta t$ such that at time $t_0=0$ all neurons are
refractory. This gives rise to the sharp initial condition
$\Delta n(t_k)=\bar m_{k}(0)=N\delta_{k,-1}$ and $v_{k}(0)=u_{k}(0)=0$
for the refractory epoch ($k=-K,\dotsc,-1$). Here, $\delta_{k,l}$
denotes the Kronecker delta, which is unity for $k=l$ and zero
otherwise. After synchronization there are no free neurons, hence
$x(t_0)=z(t_0)=0$ and, if there were no further spikes in the past, $g_\ell(t_0)=0$ for $\ell=1,\dotsc,N_\theta$.
The initialization of $g_\ell$ corresponds to a zero adaptation level at the beginning
of the simulation.

For
the representation of the variables $\bar m_k(t_l)$, $v_k(t_l)$,
$u_k(t_l)$ and $\lambda_k(t_l)$, $k=l-K,\dotsc,l-1$, in memory, it is
convenient to employ circular buffers. That is, the ``running'' range
of the explicit history $k=l-K,\dotsc,l-1$ is mapped to a static range
$\hat k=0,\dotsc,K-1$ in memory by applying the modulo operation
\begin{equation}
  \label{eq:mod-map}
  \hat k= (k\text{ mod } K)
\end{equation}
to all temporal indices.  

\begin{figure}[p]
\begin{adjustwidth}{\figureoffsetleft}{\figureoffsetright}
  \centering
  \includegraphics[width=7.5in]{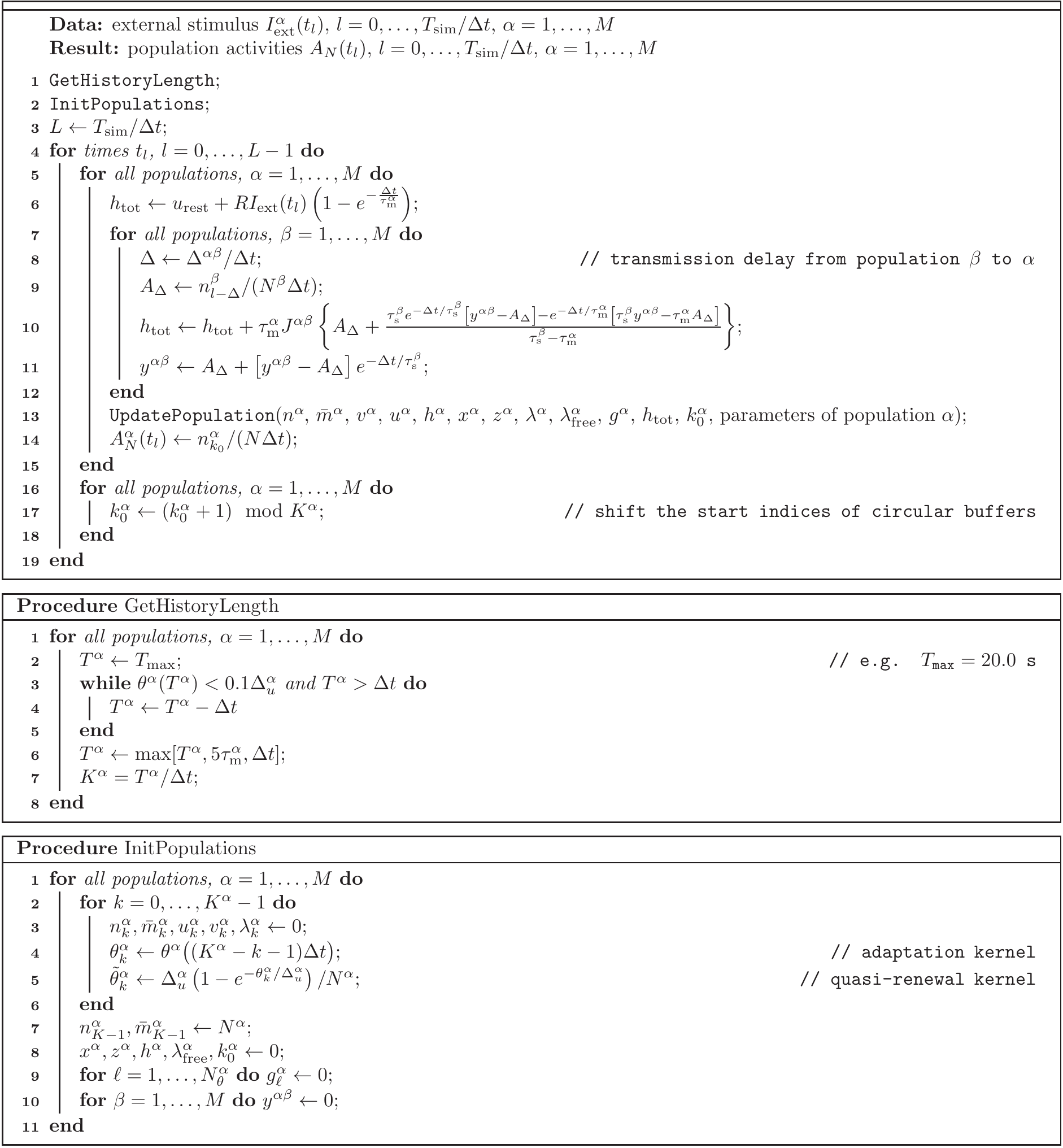}
  \caption{{\bf Pseudocode for the integration of the mesoscopic
      population equation.}\\
  Note that procedure {\tt UpdatePopulation} in line 12 is shown in Fig.~\ref{fig:algorithm}.}
  \label{fig:algorithm-main}
\end{adjustwidth}
\end{figure}

\begin{figure}[p]
\begin{adjustwidth}{\figureoffsetleft}{\figureoffsetright}
  \centering
  \includegraphics[width=7.5in]{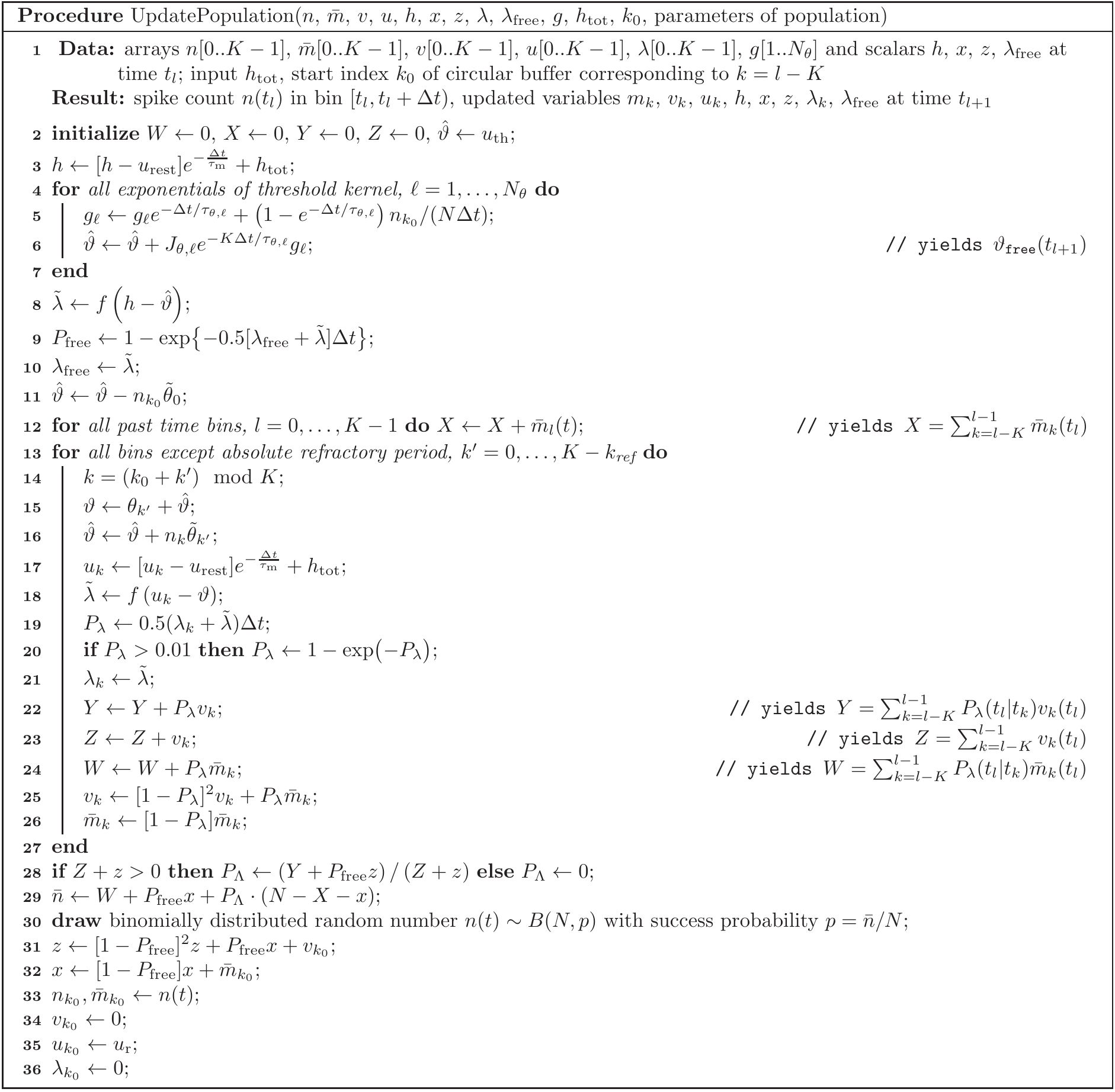}
    \caption{{\bf Pseudocode for the update of the variables of a given population.}\\
      Note that the adaptation kernel
      $\theta_{k'}\equiv\theta\bigl((K-k')\Delta t\bigr)$, the
      quasi-renewal kernel
      $\tilde{\theta}_{k'}\equiv\tilde{\theta}\bigl((K-k')\Delta
      t\bigr)/N$, Eq.~\eqref{eq:qr-g},
      as well as the exponentials $e^{-\frac{\Delta t}{\taum}}$ and
      $J_{\theta,\ell} e^{-K\Delta t/\tau_{\theta,\ell}}$ can be precomputed.}
  \label{fig:algorithm}
\end{adjustwidth}
\end{figure}

\rot{
\paragraph{Summary of the update step and pseudocode.}

Let us summarize the steps needed to evolve the population equation
from time $t_l$ to time $t_{l+1}$:
\begin{enumerate}
\item Calculate the total integrated input $h_\text{tot}$ using Eq.~\eqref{eq:Itot} and then update the synaptic variables
  $y^{\alpha\beta}(t_{l+1})$ according to Eq.~\eqref{eq:y-sol}.
\item Update the free membrane potential $h(t_{l+1})$ and threshold variable
  $g_\ell(t_{l+1})$ for free neurons using Eq.~\eqref{eq:u-y-sol} and Eq.~\eqref{eq:dgl-thresh-variables-discret} and use these values to compute the threshold $\vartheta_\text{free}(t_{l+1})$ and
  conditional intensity $\freehaz(t_{l+1})$ of free neurons by means of Eq.~\eqref{eq:thresh-nr} and Eq.~\eqref{eq:cond-int-free}. This yields the firing probability of free neurons $\pfree(t_l)$ via Eq.~\eqref{eq:pfire} and \eqref{eq:lambda-bar-free}.
\item \label{enum:u-update} For all refractory states  $k=l-K,\dotsc,l-k_{\text{ref}}$, compute the membrane potential $u_k(t_{l+1})$ from Eq.~\eqref{eq:u-iter}, the threshold $\vartheta_k(t_{l+1})$ from Eq.~\eqref{eq:hattheta-methods} and the conditional intensity $\lambda_k(t_{l+1})$ from Eq.~\eqref{eq:intens-refarc}. The firing probabilities  $P_{\lambda}(t_l|t_k)$ are then given by Eq.~\eqref{eq:pfire} and \eqref{eq:lambda-bar}.
\item \label{enum:PLam} Calculate the effective firing probability $P_\Lambda(t_l)$ from Eq.~\eqref{eq:Lambda-discrete-finite-hist}.
\item \label{enum:main} Calculate the expected activity $\nbar(t_l)$
  by Eq.~\eqref{eq:nbar}.  The empirical population activity $\Delta n(t_l)$ can be
  obtained by drawing a binomially distributed random number according to Eq.~\eqref{eq:binom}.
\item \label{enum:m-v-update} Update the mean and variance of the survival numbers $\bar m_k(t_{l+1})$, $v_k(t_{l+1})$, $x(t_{l+1})$ and $z(t_{l+1})$ using Eqs.~\eqref{eq:m-update},~\eqref{eq:v-update}, \eqref{eq:x-update} and \eqref{eq:z-update}.
\item Realize the boundary conditions at $\tl=t$ according to Eq.~\eqref{eq:bc-res}.

\end{enumerate}
These steps have to be performed for all populations
$\alpha=1,\dotsc,M$. A detailed implementation of the algorithm is
provided by the pseudocode shown in Figs.~\ref{fig:algorithm-main} and
\ref{fig:algorithm}.  }

Whereas the complexity of the microscopic simulation is of order
$O(N\Delta t^{-1})$, the integration of the population equation is of
order $O(\Delta t^{-2})$ because in each time step one has to update a
history of length $K=T/\Delta t$ (step \ref{enum:u-update},
\ref{enum:PLam} and \ref{enum:m-v-update}). Hence, at low neuron
numbers (e.g. $N<100$), the direct simulation of the microscopic
system may become more efficient.  We emphasize, however, that to
achieve a comparable accuracy, the integration of the mesoscopic
population equation can be performed on a coarser, millisecond time
scale (e.g., $\Delta t=1$~ms), whereas the microscopic simulation
requires precise spike times and hence a sub-millisecond simulation
(e.g., $\Delta t=0.1$~ms). If we take advantage of this fact, the
mesoscopic population model performs well even at low neuron numbers.

\subsection*{Power spectrum}
\label{sec:power-spectrum}

We characterize the fluctuations of the stationary population activity
by the power spectrum defined as
\begin{equation}
  \label{eq:psd}
  \tilde{C}(f)=\lim_{T\rightarrow\infty}\frac{\langle |\tilde{A}(f;T)|^2\rangle}{T},
\end{equation}
where
\begin{equation}
  \label{eq:A_fourier}
  \tilde{A}(f;T)=\int_0^TA_N(t)e^{2\pi i ft}\,\mathrm{d}t
\end{equation}
is the Fourier transform of the population activity on a
time window of length $T$. 

For a population of renewal neurons the power spectrum is known analytically. It is given by \cite{Str67I}
\begin{equation}
  \label{eq:renewal-psd}
  \tilde{C}(f)=\frac{r}{N}\frac{1-|\tilde{P}_\text{ISI}(f)|^2}{|1-\tilde{P}_\text{ISI}(f)|^2},
\end{equation}
where $\tilde{P}_\text{ISI}(f)$ is the Fourier transform of the interspike interval density
\begin{equation}
  \label{eq:renewal-isi}
  P_\text{ISI}(t)=\lambda(t|0)\exp\lrrund{-\int_0^t\lambda(s|0)\,\mathrm{d}s}
\end{equation}
and  $r$ is the stationary firing rate given by 
\begin{equation}
  \label{eq:rate-renewal}
  r=\lreckig{\int_0^\infty \exp\lrrund{-\int_0^t\lambda(s|0)\,\mathrm{d}s}\,\mathrm{d}t}^{-1}.
\end{equation}
Note that the power of the fluctuations in Eq.~\eqref{eq:renewal-psd}
scales like $1/N$, vanishing in the macroscopic limit
$N\rightarrow\infty$. For the LIF model with escape noise, the hazard rate $\lambda(t|0)$ is
given by
\begin{equation}
  \label{eq:hazard-gif}
  \lambda(t|0)=f\bigl(u(t,0)-\vth\bigr),\qquad u(t,0)=\mu+(\vreset-\mu)\exp\lrrund{-\frac{t-\tref}{\taum}}
\end{equation}
for $t>\tref$ and $\lambda(t|0)=0$ for $t\le \tref$.

\subsection*{Modified Potjans-Diesmann model}
\label{sec:potjans-model}
To model the cortical column of \cite{PotDie14} in our framework, we
used the parameters of the original publication and modified the
model in two ways: First, the background Poisson input was replaced by
a constant drive and an increased escape noise such that the
populations exhibited roughly the same stationary firing
rates. Specifically, we set $\vreset=0$~mV, $\vth=15$~mV and
$\Delta_u=5$~mV; and, using the mesoscopic dynamics, fitted the
resting potentials of the GIF model (here denoted by $\hat\mu^\alpha$)
without adaptation $J_{\theta}=0$ to obtain firing rates $\hat r$ that
roughly match the target firing rates. Second, we introduced
adaptation on excitatory cells with strength $J_{\theta}$ and time
scale $\tau_{\theta}$, and re-adjusted the resting potential as
follows $\vrest=\hat\mu+J_{\theta}\hat r$. This yielded correct
stationary firing rates in the presence of adaptation. The resulting
parameters of the modified model are summarized in
Table~\ref{parampotjans}.

\begin{table}[!h]
\begin{adjustwidth}{\figureoffsetleft}{\figureoffsetright} 
\centering
\caption{
  {\bf Parameters of the modified Potjans-Diesmann model.}  }
\begin{tabular}{|l|llllllll|}
\hline
\multicolumn{1}{|l|}{\bf population} & \multicolumn{1}{l}{\bf L2/3e}& \multicolumn{1}{l}{\bf L2/3i}& \multicolumn{1}{l}{\bf L4e}& \multicolumn{1}{l}{\bf L4i}& \multicolumn{1}{l}{\bf L5e}& \multicolumn{1}{l}{\bf L5i}& \multicolumn{1}{l}{\bf L6e}& \multicolumn{1}{l|}{\bf L6i}\\\hline
\multicolumn{1}{|l|}{$\taum$ [s]} & \multicolumn{8}{|c|}{0.01}\\
\multicolumn{1}{|l|}{$\taus$ [s]} & \multicolumn{8}{|c|}{0.0005}\\
\multicolumn{1}{|l|}{$\Delta$ [s]} & \multicolumn{8}{|c|}{0.0015}\\
\multicolumn{1}{|l|}{$\tref$ [s]} & \multicolumn{8}{|c|}{0.002}\\
\multicolumn{1}{|l|}{$\Delta_u$ [mV]} & \multicolumn{8}{|c|}{5.0}\\
\hline
$\hat\mu$ [mV]&19.149 & 20.362  &30.805&  28.069&  29.437&  29.33&   34.932&  32.081\\

$\hat r$ [Hz]&0.974&  2.861&  4.673&  5.65&   8.141&  9.013&  0.988&  7.53 \\\hline
\multicolumn{9}{|l|}{adaptation: $\theta(t)=(J_{\theta}/\tau_{\theta})e^{-t/\tau_{\theta}}$ for $t>\tref$}\\\hline
$J_{\theta}$ [mV s]&1.0&0.0&1.0&0.0&1.0&0.0&1.0&0.0\\
$\tau_{\theta}$ [s]&1.0&-&1.0&-&1.0&-&1.0&-\\
\multicolumn{1}{|l|}{$\vrest=\hat\mu+J_{\theta}\hat r$ [mV]} &20.123&  20.362&  35.478&  28.069&  37.578&  29.33&   35.92&   32.081\\\hline
\multicolumn{9}{|l|}{step stimulus (``thalamic input'')}\\\hline
$RI_\text{ext}$ [mV]&0.&      0.&     19.&     11.964&   0.&      0.&      9.896&   3.788\\
\multicolumn{1}{|l|}{$\mu(t)$ [mV]} & \multicolumn{8}{|c|}{$\vrest+RI_\text{ext}$ for $t\in [0.06s,0.09s]$, else $\vrest$}\\ \hline
\multicolumn{9}{|l|}{network parameters}\\\hline
$N$&20683&  5834& 21915&  5479&  4850&  1065& 14395&  2948\\\hline
\multicolumn{1}{|l|}{connection prob. $p^{\alpha\beta}$} & \multicolumn{8}{c|}{from population $\beta$}\\
$\alpha=$~L2/3e&0.1009& 0.1689& 0.0437& 0.0818& 0.0323& 0.0& 0.0076& 0.0\\
$\alpha=$~L2/3i&0.1346& 0.1371& 0.0316& 0.0515& 0.0755& 0.0& 0.0042& 0.0\\
$\alpha=$~L4e& 0.0077& 0.0059& 0.0497& 0.135& 0.0067& 0.0003& 0.0453& 0.0\\
$\alpha=$~L4i& 0.0691& 0.0029& 0.0794& 0.1597& 0.0033& 0.0& 0.1057& 0.0\\
$\alpha=$~L5e& 0.1004& 0.0622& 0.0505& 0.0057& 0.0831& 0.3726& 0.0204& 0.0\\
$\alpha=$~L5i& 0.0548& 0.0269& 0.0257& 0.0022& 0.06& 0.3158& 0.0086& 0.0\\
$\alpha=$~L6e& 0.0156& 0.0066& 0.0211& 0.0166& 0.0572& 0.0197& 0.0396& 0.2252\\
$\alpha=$~L6i& 0.0364& 0.001& 0.0034& 0.0005& 0.0277& 0.008& 0.0658& 0.1443\\
\hline
$w^{\alpha\beta}$ [mV], $\alpha=$~L4e&0.176& -0.702&  0.351& -0.702&  0.176& -0.702&  0.176& -0.702\\
$w^{\alpha\beta}$ [mV], $\alpha\neq$~L4e&0.176& -0.702&  0.176& -0.702&  0.176& -0.702&  0.176& -0.702\\\hline
\end{tabular}
\label{parampotjans}
\end{adjustwidth}
\end{table}

\section*{Acknowledgments}
We thank Hesam Setareh for useful discussions. This project received funding from the European Union’s Horizon 2020 research and innovation programme under grant agreement No. 720270 and from the European Research Council under grant agreement No. 268689, MultiRules.

\nolinenumbers



\begin{thebibliography}{100}

\bibitem{WanTol04}
Wang Y, Toledo-Rodriguez M, Gupta A, Wu C, Silberberg G, Luo J, et~al.
\newblock Anatomical, physiological and molecular properties of Martinotti
  cells in the somatosensory cortex of the juvenile rat.
\newblock J Physiol. 2004;561(1):65--90.

\bibitem{SugHem06}
Sugino K, Hempel CM, Miller MN, Hattox AM, Shapiro P, Wu C, et~al.
\newblock Molecular taxonomy of major neuronal classes in the adult mouse
  forebrain.
\newblock Nat Neurosci. 2006;9(1):99--107.

\bibitem{LefTom09}
Lefort S, Tomm C, Sarria JCF, Petersen CCH.
\newblock The excitatory neuronal network of the C2 barrel column in mouse
  primary somatosensory cortex.
\newblock Neuron. 2009;61(2):301--316.

\bibitem{HarShe15}
Harris KD, Shepherd GMG.
\newblock The neocortical circuit: themes and variations.
\newblock Nat Neurosci. 2015;18(2):170--181.

\bibitem{PotDie14}
Potjans TC, Diesmann M.
\newblock The cell-type specific cortical microcircuit: relating structure and
  activity in a full-scale spiking network model.
\newblock Cereb Cortex. 2014;24(3):785--806.

\bibitem{MarMul15}
Markram H, Muller E, Ramaswamy S, Reimann MW, Abdellah M, Sanchez CA, et~al.
\newblock Reconstruction and simulation of neocortical microcircuitry.
\newblock Cell. 2015;163(2):456--492.

\bibitem{IzhEde08}
Izhikevich EM, Edelman GM.
\newblock Large-scale model of mammalian thalamocortical systems.
\newblock Proc Natl Acad Sci U S A. 2008;105(9):3593--3598.

\bibitem{Fre75}
Freeman WJ.
\newblock Mass action in the nervous system.
\newblock New York, NY: Academic Press; 1975.

\bibitem{DavFri03}
David O, Friston KJ.
\newblock A neural mass model for MEG/EEG: coupling and neuronal dynamics.
\newblock Neuroimage. 2003;20(3):1743--1755.

\bibitem{MorPin13}
Moran R, Pinotsis D, Friston K.
\newblock Neural masses and fields in dynamic causal modeling.
\newblock Front Computat Neuroscie. 2013;7:57.

\bibitem{JirHak97}
Jirsa VK, Haken H.
\newblock A derivation of a macroscopic field theory of the brain from the
  quasi-microscopic neural dynamics.
\newblock Physica D. 1997;99(4):503--526.

\bibitem{Coo10}
Coombes S.
\newblock Large-scale neural dynamics: simple and complex.
\newblock NeuroImage. 2010;52(3):731--739.

\bibitem{BojOos10}
Bojak I, Oostendorp TF, Reid AT, K{\"o}tter R.
\newblock Connecting mean field models of neural activity to EEG and fMRI data.
\newblock Brain Topogr. 2010;23(2):139--149.

\bibitem{GerKis14}
Gerstner W, Kistler WM, Naud R, Paninski L.
\newblock Neuronal Dynamics: From Single Neurons to Networks and Models of
  Cognition.
\newblock Cambridge: Cambridge University Press; 2014.

\bibitem{DayAbb05}
Dayan P, Abbott LF.
\newblock Theoretical Neuroscience: Computational and Mathematical Modeling of
  Neural Systems.
\newblock 1st ed. The {MIT} Press; 2005.

\bibitem{WilCow72}
Wilson HR, Cowan JD.
\newblock Excitatory and inhibitory interactions in localized populations of
  model neurons.
\newblock Biophys J. 1972;12(1):1.

\bibitem{FriHar03}
Friston KJ, Harrison L, Penny W.
\newblock Dynamic causal modelling.
\newblock Neuroimage. 2003;19(4):1273--1302.

\bibitem{DecJir11}
Deco G, Jirsa VK, McIntosh AR.
\newblock Emerging concepts for the dynamical organization of resting-state
  activity in the brain.
\newblock Nat Rev Neurosci. 2011;12(1):43--56.

\bibitem{GerHem92}
Gerstner W, van Hemmen JL.
\newblock Universality in neural networks: the importance of the ‘mean firing
  rate’.
\newblock Biol Cybern. 1992;67(3):195--205.

\bibitem{BruWan01}
Brunel N, Wang XJ.
\newblock Effects of Neuromodulation in a Cortical Network Model of Object
  Working Memory Dominated by Recurrent Inhibition.
\newblock J Comput Neurosci. 2001;11:63--85.

\bibitem{DecRol05}
Deco G, Rolls ET.
\newblock Neurodynamics of Biased Competition and Cooperation for Attention: A
  Model With Spiking Neurons.
\newblock J Neurophysiol. 2005;94:295--313.

\bibitem{DecJir08}
Deco G, Jirsa VK, Robinson PA, Breakspear M, Friston K.
\newblock The dynamic brain: from spiking neurons to neural masses and cortical
  fields.
\newblock PLoS Comput Biol. 2008;4(8):e1000092.

\bibitem{Ger00}
Gerstner W.
\newblock Population Dynamics of Spiking Neurons: Fast Transients, Asynchronous
  States, and Locking.
\newblock Neural Comput. 2000;12:43.

\bibitem{EinKay13}
Einevoll GT, Kayser C, Logothetis NK, Panzeri S.
\newblock Modelling and analysis of local field potentials for studying the
  function of cortical circuits.
\newblock Nat Rev Neurosci. 2013;14(11):770--785.

\bibitem{GerNau09}
Gerstner W, Naud R.
\newblock Neuroscience. How good are neuron models?
\newblock Science. 2009 Oct;326(5951):379--380.

\bibitem{MenNau12}
Mensi S, Naud R, Pozzorini C, Avermann M, Petersen CCH, Gerstner W.
\newblock Parameter Extraction and Classification of Three Cortical Neuron
  Types Reveals Two Distinct Adaptation Mechanisms.
\newblock J Neurophysiol. 2012;.

\bibitem{PozMen15}
Pozzorini C, Mensi S, Hagens O, Naud R, Koch C, Gerstner W.
\newblock Automated High-Throughput Characterization of Single Neurons by Means
  of Simplified Spiking Models.
\newblock PLoS Comput Biol. 2015;11(6):e1004275.

\bibitem{PozNau13}
Pozzorini C, Naud R, Mensi S, Gerstner W.
\newblock Temporal whitening by power-law adaptation in neocortical neurons.
\newblock Nat Neurosci. 2013;16(7):942--948.

\bibitem{BoyZha05}
Boyden ES, Zhang F, Bamberg E, Nagel G, Deisseroth K.
\newblock Millisecond-timescale, genetically targeted optical control of neural
  activity.
\newblock Nat Neurosci. 2005;8(9):1263--1268.

\bibitem{Dei11}
Deisseroth K.
\newblock Optogenetics.
\newblock Nat Methods. 2011;8(1):26--29.

\bibitem{LiuRam12}
Liu X, Ramirez S, Pang PT, Puryear CB, Govindarajan A, Deisseroth K, et~al.
\newblock Optogenetic stimulation of a hippocampal engram activates fear memory
  recall.
\newblock Nature. 2012;484(7394):381--385.

\bibitem{MorRin07}
Moreno-Bote R, Rinzel J, Rubin N.
\newblock Noise-induced alternations in an attractor network model of
  perceptual bistability.
\newblock J Neurophysiol. 2007;98(3):1125--1139.

\bibitem{ShpMor09}
Shpiro A, Moreno-Bote R, Rubin N, Rinzel J.
\newblock Balance between noise and adaptation in competition models of
  perceptual bistability.
\newblock J Comput Neurosci. 2009;27(1):37--54.

\bibitem{TheKov11}
Theodoni P, Kov{\'a}cs G, Greenlee MW, Deco G.
\newblock Neuronal adaptation effects in decision making.
\newblock J Neurosci. 2011;31(1):234--246.

\bibitem{ELu11}
E W, Lu J.
\newblock {M}ultiscale modeling.
\newblock Scholarpedia. 2011;6(10):11527.

\bibitem{NykTra00}
Nykamp DQ, Tranchina D.
\newblock A population density approach that facilitates large-scale modeling
  of neural networks: Analysis and an application to orientation tuning.
\newblock J Comput Neurosci. 2000;8(1):19--50.

\bibitem{MulBue07}
Muller E, Buesing L, Schemmel J, Meier K.
\newblock Spike-Frequency Adapting Neural Ensembles: Beyond Mean Adaptation and
  Renewal Theories.
\newblock Neural Comp. 2007;19(11):2958--3110.

\bibitem{BalFas12}
Baladron J, Fasoli D, Faugeras O, Touboul J.
\newblock Mean-field description and propagation of chaos in networks of
  Hodgkin-Huxley and FitzHugh-Nagumo neurons.
\newblock J Math Neurosci. 2012;2(1):1--50.

\bibitem{Bre09}
Bressloff PC.
\newblock Stochastic neural field theory and the system-size expansion.
\newblock SIAM J Appl Math. 2009;70(5):1488--1521.

\bibitem{BuiCow07}
Buice MA, Cowan JD.
\newblock Field-theoretic approach to fluctuation effects in neural networks.
\newblock Phys Rev E. 2007;75(5):051919.

\bibitem{BuiCow10}
Buice MA, Cowan JD, Chow CC.
\newblock Systematic fluctuation expansion for neural network activity
  equations.
\newblock Neural Comput. 2010;22(2):377--426.

\bibitem{Bre10}
Bressloff PC.
\newblock Metastable states and quasicycles in a stochastic Wilson-Cowan model
  of neuronal population dynamics.
\newblock Phys Rev E. 2010;82(5):051903.

\bibitem{WalBen11}
Wallace E, Benayoun M, Van~Drongelen W, Cowan JD.
\newblock Emergent oscillations in networks of stochastic spiking neurons.
\newblock Plos one. 2011;6(5):e14804.

\bibitem{TouErm11}
Touboul JD, Ermentrout GB.
\newblock Finite-size and correlation-induced effects in mean-field dynamics.
\newblock J Comput Neurosci. 2011;31(3):453--484.

\bibitem{GoyGoy15}
Goychuk I, Goychuk A.
\newblock Stochastic Wilson--Cowan models of neuronal network dynamics with
  memory and delay.
\newblock New J Phys. 2015;17(4):045029.

\bibitem{BerMei98}
Berry MJ, Meister M.
\newblock Refractoriness and neural precision.
\newblock J Neurosci. 1998;18(6):2200--2211.

\bibitem{GeiGol66}
Geisler C, Goldberg JM.
\newblock A Stochastic Model of the Repetitive Activity of Neurons.
\newblock Biophys J. 1966;6(1):53--69.

\bibitem{RatNel00}
Ratnam R, Nelson ME.
\newblock Nonrenewal Statistics of Electrosensory Afferent Spike Trains:
  Implications for the Detection of Weak Sensory Signals.
\newblock J Neurosci. 2000;20:6672.

\bibitem{ChaLon00}
Chacron MJ, Longtin A, St-Hilaire M, Maler L.
\newblock Suprathreshold stochastic firing dynamics with memory in {P-type}
  electroreceptors.
\newblock Phys Rev Lett. 2000;85:1576.

\bibitem{NawBou07}
Nawrot MP, Boucsein C, Rodriguez-Molina V, Aertsen A, Grun S, Rotter S.
\newblock Serial interval statistics of spontaneous activity in cortical
  neurons in vivo and in vitro.
\newblock Neurocomp. 2007;70:1717.

\bibitem{FisSch12}
Fisch K, Schwalger T, Lindner B, Herz AVM, Benda J.
\newblock Channel noise from both slow adaptation currents and fast currents is
  required to explain spike-response variability in a sensory neuron.
\newblock J Neurosci. 2012;32(48):17332--17344.

\bibitem{Lin06}
Lindner B.
\newblock Superposition of many independent spike trains is generally not a
  {Poisson} process.
\newblock Phys Rev E. 2006;73:022901.

\bibitem{CatRey06}
C\^{a}teau H, Reyes AD.
\newblock Relation between Single Neuron and Population Spiking Statistics and
  Effects on Network Activity.
\newblock Phys Rev Lett. 2006;96:058101.

\bibitem{DegHel12}
Deger M, Helias M, Boucsein C, Rotter S.
\newblock Statistical properties of superimposed stationary spike trains.
\newblock J Comput Neurosci. 2012;32:443--463.

\bibitem{DegSch14}
Deger M, Schwalger T, Naud R, Gerstner W.
\newblock Fluctuations and information filtering in coupled populations of
  spiking neurons with adaptation.
\newblock Phys Rev E. 2014 Dec;90(6-1):062704.

\bibitem{WieBer15}
Wieland S, Bernardi D, Schwalger T, Lindner B.
\newblock Slow fluctuations in recurrent networks of spiking neurons.
\newblock Phys Rev E. 2015;92(4):040901.

\bibitem{SchDro15}
Schwalger T, Droste F, Lindner B.
\newblock Statistical structure of neural spiking under non-Poissonian or other
  non-white stimulation.
\newblock J Comput Neurosci. 2015;39(1):29--51.

\bibitem{BruHak99}
Brunel N, Hakim V.
\newblock Fast global oscillations in networks of integrate-and-fire neurons
  with low firing rates.
\newblock Neural Comput. 1999;11:1621.

\bibitem{Bru00}
Brunel N.
\newblock Sparsely Connected Networks of Spiking Neurons.
\newblock J Comput Neurosci. 2000;8:183.

\bibitem{MatGiu02}
Mattia M, {Del Giudice} P.
\newblock Population dynamics of interacting spiking neurons.
\newblock Phys Rev E. 2002;66:051917.

\bibitem{LagRot14}
Lagzi F, Rotter S.
\newblock A Markov model for the temporal dynamics of balanced random networks
  of finite size.
\newblock Front Comput Neurosci. 2014;8:142.

\bibitem{GigDec15}
Gigante G, Deco G, Marom S, Del~Giudice P.
\newblock Network events on multiple space and time scales in cultured neural
  networks and in a stochastic rate model.
\newblock PLoS Comput Biol. 2015;11(11):e1004547.

\bibitem{NauGer12}
Naud R, Gerstner W.
\newblock Coding and decoding with adapting neurons: a population approach to
  the peri-stimulus time histogram.
\newblock PLoS Comput Biol. 2012;8(10).

\bibitem{GigMat07}
Gigante G, Mattia M, Del~Giudice P.
\newblock Diverse Population-Bursting Modes of Adapting Spiking Neurons.
\newblock Phys Rev Lett. 2007;98(14):148101.

\bibitem{OckJos16_arxiv}
{Ocker} GK, {Josi{\'c}} K, {Shea-Brown} E, {Buice} MA.
\newblock {Linking structure and activity in nonlinear spiking networks}.
\newblock ArXiv e-prints. 2016;.

\bibitem{ToyRad09}
Toyoizumi T, Rad KR, Paninski L.
\newblock Mean-field approximations for coupled populations of generalized
  linear model spiking neurons with Markov refractoriness.
\newblock Neural Comput. 2009;21(5):1203--1243.

\bibitem{BuiCho13}
Buice MA, Chow CC.
\newblock Dynamic finite size effects in spiking neural networks.
\newblock PLoS Comput Biol. 2013;9(1):e1002872.

\bibitem{MeyVre02}
Meyer C, van Vreeswijk C.
\newblock Temporal correlations in stochastic networks of spiking neurons.
\newblock Neural Comput. 2002;14(2):369--404.

\bibitem{LinDoi05}
Lindner B, Doiron B, Longtin A.
\newblock Theory of oscillatory firing induced by spatially correlated noise
  and delayed inhibitory feedback.
\newblock Phys Rev E. 2005;72(6):061919--14.

\bibitem{TroHu12}
Trousdale J, Hu Y, Shea-Brown E, Josić K.
\newblock Impact of network structure and cellular response on spike time
  correlations.
\newblock PLoS Comput Biol. 2012;8(3).

\bibitem{BarMaz14}
Barbieri F, Mazzoni A, Logothetis NK, Panzeri S, Brunel N.
\newblock Stimulus dependence of local field potential spectra: experiment
  versus theory.
\newblock J Neurosci. 2014;34(44):14589--14605.

\bibitem{BosDie16}
Bos H, Diesmann M, Helias M.
\newblock Identifying Anatomical Origins of Coexisting Oscillations in the
  Cortical Microcircuit.
\newblock PLoS Comput Biol. 2016;12(10):e1005132.

\bibitem{AveTom12}
Avermann M, Tomm C, Mateo C, Gerstner W, Petersen CCH.
\newblock Microcircuits of excitatory and inhibitory neurons in layer 2/3 of
  mouse barrel cortex.
\newblock J Neurophysiol. 2012;107(11):3116--3134.

\bibitem{GenKre12}
Gentet LJ, Kremer Y, Taniguchi H, Huang ZJ, Staiger JF, Petersen CCH.
\newblock Unique functional properties of somatostatin-expressing GABAergic
  neurons in mouse barrel cortex.
\newblock Nat Neurosci. 2012;15(4):607--612.

\bibitem{CruLew07}
Cruikshank SJ, Lewis TJ, Connors BW.
\newblock Synaptic basis for intense thalamocortical activation of feedforward
  inhibitory cells in neocortex.
\newblock Nat Neurosci. 2007;10(4):462--468.

\bibitem{PacYus11}
Packer AM, Yuste R.
\newblock Dense, unspecific connectivity of neocortical parvalbumin-positive
  interneurons: a canonical microcircuit for inhibition?
\newblock J Neurosci. 2011;31(37):13260--13271.

\bibitem{PfeXue13}
Pfeffer CK, Xue M, He M, Huang ZJ, Scanziani M.
\newblock Inhibition of inhibition in visual cortex: the logic of connections
  between molecularly distinct interneurons.
\newblock Nat Neurosci. 2013;16(8):1068--1076.

\bibitem{LiJi14}
Li L, Ji X, Liang F, Li Y, Xiao Z, Tao HW, et~al.
\newblock A feedforward inhibitory circuit mediates lateral refinement of
  sensory representation in upper layer 2/3 of mouse primary auditory cortex.
\newblock J Neurosci. 2014;34(41):13670--13683.

\bibitem{KarJac16}
Karnani MM, Jackson J, Ayzenshtat I, Tucciarone J, Manoocheri K, Snider WG,
  et~al.
\newblock Cooperative subnetworks of molecularly similar interneurons in mouse
  neocortex.
\newblock Neuron. 2016;90(1):86--100.

\bibitem{LiuWan01}
Liu YH, Wang XJ.
\newblock Spike-frequency adaptation of a generalized leaky integrate-and-fire
  model neuron.
\newblock J Comp Neurosci. 2001;10:25.

\bibitem{BibIva85}
Bibikov NG, Ivanitskii GA.
\newblock Modelling spontaneous pulsation and short-term adaptation in the
  fibres of the auditory nerve.
\newblock Biophysics. 1985;30:152--156.

\bibitem{SchFis10}
Schwalger T, Fisch K, Benda J, Lindner B.
\newblock How Noisy Adaptation of Neurons Shapes Interspike Interval Histograms
  and Correlations.
\newblock PLoS Comput Biol. 2010;6(12):e1001026.

\bibitem{SchLin13}
Schwalger T, Lindner B.
\newblock Patterns of interval correlations in neural oscillators with
  adaptation.
\newblock Front Comput Neurosci. 2013;7(164):164.

\bibitem{WebPil16}
{Weber} AI, {Pillow} JW.
\newblock {Capturing the dynamical repertoire of single neurons with
  generalized linear models}.
\newblock ArXiv e-prints. 2016 Feb;.

\bibitem{JolRau06}
Jolivet R, Rauch A, L{\"u}scher HR, Gerstner W.
\newblock Predicting spike timing of neocortical pyramidal neurons by simple
  threshold models.
\newblock J Comput Neurosci. 2006;21(1):35--49.

\bibitem{PlaBre11}
Platkiewicz J, Brette R.
\newblock Impact of Fast Sodium Channel Inactivation on Spike Threshold
  Dynamics and Synaptic Integration.
\newblock PLoS Comput Biol. 2011;7(5):1--15.

\bibitem{MenHag16}
Mensi S, Hagens O, Gerstner W, Pozzorini C.
\newblock Enhanced sensitivity to rapid input fluctuations by nonlinear
  threshold dynamics in neocortical pyramidal neurons.
\newblock PLoS Comput Biol. 2016;12(2):e1004761.

\bibitem{ChiGra07}
Chizhov AV, Graham LJ.
\newblock Population model of hippocampal pyramidal neurons, linking a
  refractory density approach to conductance-based neurons.
\newblock Phys Rev E. 2007;75(1):011924.

\bibitem{ChiGra08}
Chizhov AV, Graham LJ.
\newblock Efficient evaluation of neuron populations receiving colored-noise
  current based on a refractory density method.
\newblock Phys Rev E. 2008;77(1):011910.

\bibitem{PerGer67}
Perkel DH, Gerstein GL, Moore GP.
\newblock Neuronal spike trains and stochastic point processes. I. The single
  spike train.
\newblock Biophys J. 1967;7(4):391--418.

\bibitem{GewDie07}
Gewaltig MO, Diesmann M.
\newblock NEST (NEural Simulation Tool).
\newblock Scholarpedia. 2007;2(4):1430.

\bibitem{DegHel10}
Deger M, Helias M, Cardanobile S, Atay FM, Rotter S.
\newblock Nonequilibrium dynamics of stochastic point processes with
  refractoriness.
\newblock Phys Rev E. 2010;82(2):021129.

\bibitem{Kni72}
Knight BW.
\newblock Dynamics of encoding in a population of neurons.
\newblock J Gen Physiol. 1972;59:734.

\bibitem{FraBai95}
Franklin J, Bair W.
\newblock The effect of a refractory period on the power spectrum of neuronal
  discharge.
\newblock SIAM J Appl Math. 1995;55:1074.

\bibitem{GerHem96}
Gerstner W, Van~Hemmen JL, Cowan JD.
\newblock What matters in neuronal locking?
\newblock Neural Comput. 1996;8(8):1653--1676.

\bibitem{GerDeg17}
Gerhard F, Deger M, Truccolo W.
\newblock On the stability and dynamics of stochastic spiking neuron models:
  nonlinear Hawkes process and point process GLMs.
\newblock PLOS Computat Biol. 2017;13(2):e1005390.

\bibitem{HelTet14}
Helias M, Tetzlaff T, Diesmann M.
\newblock The correlation structure of local neuronal networks intrinsically
  results from recurrent dynamics.
\newblock PLoS Comput Biol. 2014;10(1):e1003428.

\bibitem{RenRoc10}
Renart A, De~La~Rocha J, Bartho P, Hollender L, Parga N, Reyes A, et~al.
\newblock The asynchronous state in cortical circuits.
\newblock Science. 2010;327(5965):587--590.

\bibitem{LitDoi12}
Litwin-Kumar A, Doiron B.
\newblock Slow dynamics and high variability in balanced cortical networks with
  clustered connections.
\newblock Nature Neurosci. 2012;15(11):1498--1505.

\bibitem{HerPal91}
Hertz J, Krogh A, Palmer RG.
\newblock Introduction to the theory of neural computation.
\newblock Redwood City Calif: Addison-Wesley; 1991.

\bibitem{Wan98}
Wang XJ.
\newblock Calcium Coding and Adaptive Temporal Computation in Cortical
  Pyramidal Neurons.
\newblock J Neurophysiol. 1998;79(3):1549--1566.

\bibitem{WonWan06}
Wong KF, Wang XJ.
\newblock A recurrent network mechanism of time integration in perceptual
  decisions.
\newblock J Neurosci. 2006;26(4):1314--1328.

\bibitem{MazFon15}
Mazzucato L, Fontanini A, La~Camera G.
\newblock Dynamics of multistable states during ongoing and evoked cortical
  activity.
\newblock J Neurosci. 2015;35(21):8214--8231.

\bibitem{LagRot15}
Lagzi F, Rotter S.
\newblock Dynamics of Competition between Subnetworks of Spiking Neuronal
  Networks in the Balanced State.
\newblock PloS one. 2015;10(9):e0138947.

\bibitem{HanTal90}
H\"anggi P, Talkner P, Borkovec M.
\newblock Reaction Rate Theory: Fifty Years After Kramers.
\newblock Rev Mod Phys. 1990;62:251.

\bibitem{CaoPas16}
Cao R, Pastukhov A, Mattia M, Braun J.
\newblock Collective Activity of Many Bistable Assemblies Reproduces
  Characteristic Dynamics of Multistable Perception.
\newblock J Neurosci. 2016;36(26):6957--6972.

\bibitem{CaiIye16}
Cain N, Iyer R, Koch C, Mihalas S.
\newblock The Computational Properties of a Simplified Cortical Column Model.
\newblock PLoS Comput Biol. 2016;12(9):e1005045.

\bibitem{TruEde05}
Truccolo W, Eden UT, Fellows MR, Donoghue JP, Brown EN.
\newblock A point process framework for relating neural spiking activity to
  spiking history, neural ensemble, and extrinsic covariate effects.
\newblock J Neurophysiol. 2005;93(2):1074--1089.

\bibitem{PilShl08}
Pillow JW, Shlens J, Paninski L, Sher A, Litke AM, Chichilnisky EJ, et~al.
\newblock Spatio-temporal correlations and visual signalling in a complete
  neuronal population.
\newblock Nature. 2008;454(7207):995--999.

\bibitem{Car12}
Carandini M.
\newblock From circuits to behavior: a bridge too far?
\newblock Nat Neurosci. 2012;15(4):507--509.

\bibitem{SchSch15}
Schuecker J, Schmidt M, van Albada SJ, Diesmann M, Helias M.
\newblock Fundamental Activity Constraints Lead to Specific Interpretations of
  the Connectome.
\newblock PLOS Comput Biol. 2017;13(2):1--25.

\bibitem{DecPon14}
Deco G, Ponce-Alvarez A, Hagmann P, Romani GL, Mantini D, Corbetta M.
\newblock How local excitation--inhibition ratio impacts the whole brain
  dynamics.
\newblock J Neurosci. 2014;34(23):7886--7898.

\bibitem{GilMor16}
Gilson M, Moreno-Bote R, Ponce-Alvarez A, Ritter P, Deco G.
\newblock Estimation of Directed Effective Connectivity from fMRI Functional
  Connectivity Hints at Asymmetries of Cortical Connectome.
\newblock PLoS Comput Biol. 2016;12(3):e1004762.

\bibitem{PulMus16}
Pulizzi R, Musumeci G, Van~den Haute C, Van De~Vijver S, Baekelandt V,
  Giugliano M.
\newblock Brief wide-field photostimuli evoke and modulate oscillatory
  reverberating activity in cortical networks.
\newblock Sci Rep. 2016;6:24701.

\bibitem{TsoPaw98}
Tsodyks M, Pawelzik K, Markram H.
\newblock Neural networks with dynamic synapses.
\newblock Neural Comput. 1998;10(4):821--835.

\bibitem{FauIng15}
Faugeras O, Inglis J.
\newblock Stochastic neural field equations: a rigorous footing.
\newblock J Math Biol. 2015;71(2):259--300.

\bibitem{IyeMen13}
Iyer R, Menon V, Buice M, Koch C, Mihalas S.
\newblock The influence of synaptic weight distribution on neuronal population
  dynamics.
\newblock PLoS Comput Biol. 2013;9(10):e1003248.

\bibitem{LaiKam16_arxiv}
{Lai} YM, {de Kamps} M.
\newblock {Population Density Equations for Stochastic Processes with Memory
  Kernels}.
\newblock ArXiv e-prints. 2016;.

\bibitem{BauAug16}
Baumann F, Augustin M, Ladenbauer J, Obermayer K.
\newblock A stochastic Fokker-Planck equation for the dynamics of finite-sized
  neuronal populations.
\newblock In: Bernstein Conference 2016; 2016. .

\bibitem{AugLad13}
Augustin M, Ladenbauer J, Obermayer K.
\newblock How adaptation shapes spike rate oscillations in recurrent neuronal
  networks.
\newblock Front Comput Neurosci. 2013;7(9).

\bibitem{HerDur14}
Hert{\"a}g L, Durstewitz D, Brunel N.
\newblock Analytical approximations of the firing rate of an adaptive
  exponential integrate-and-fire neuron in the presence of synaptic noise.
\newblock Front Comput Neurosci. 2014;8:116.

\bibitem{SchLin15}
Schwalger T, Lindner B.
\newblock Analytical approach to an integrate-and-fire model with
  spike-triggered adaptation.
\newblock Phys Rev E. 2015 Dec;92:062703.

\bibitem{Ric09}
Richardson MJE.
\newblock Dynamics of populations and networks of neurons with
  voltage-activated and calcium-activated currents.
\newblock Phys Rev E. 2009;80(2):021928--16.

\bibitem{LerUrs06}
Lerchner A, Ursta C, Hertz J, Ahmadi M, Ruffiot P, Enemark S.
\newblock Response variability in balanced cortical networks.
\newblock Neural Comput. 2006;18(3):634.

\bibitem{RenMor07}
Renart A, Moreno-Bote R, Wang XJ, Parga N.
\newblock Mean-driven and fluctuation-driven persistent activity in recurrent
  networks.
\newblock Neural Computat. 2007;19(1):1--46.

\bibitem{SchOst13}
Schaffer ES, Ostojic S, Abbott LF.
\newblock A complex-valued firing-rate model that approximates the dynamics of
  spiking networks.
\newblock PLoS Comput Biol. 2013;9(10):e1003301.

\bibitem{ErmTer10}
Ermentrout GB, Terman DH.
\newblock Mathematical Foundations of Neuroscience.
\newblock Springer; 2010.

\bibitem{MonPaz15}
Montbri{\'o} E, Paz{\'o} D, Roxin A.
\newblock Macroscopic description for networks of spiking neurons.
\newblock Phys Rev X. 2015;5(2):021028.

\bibitem{AugLad16_arXiv}
{Augustin} M, {Ladenbauer} J, {Baumann} F, {Obermayer} K.
\newblock {Low-dimensional spike rate models derived from networks of adaptive
  integrate-and-fire neurons: comparison and implementation}.
\newblock ArXiv e-prints. 2016 Nov;.

\bibitem{Ric04}
Richardson MJE.
\newblock Effects of synaptic conductance on the voltage distribution and
  firing rate of spiking neurons.
\newblock Phys Rev E. 2004;69:051918.

\bibitem{RicGer05}
Richardson MJE, Gerstner W.
\newblock Synaptic shot noise and conductance fluctuations affect the membrane
  voltage with equal significance.
\newblock Neural Comput. 2005;17:923.

\bibitem{SpiGer01}
Spiridon M, Gerstner W.
\newblock Effect of lateral connections on the accuracy of the population code
  for a network of spiking neurons.
\newblock Network. 2001;12(4):409--421.

\bibitem{GerKis02}
Gerstner W, Kistler WM.
\newblock Spiking Neuron Models: Single Neurons, Populations, Plasticity.
\newblock Cambridge: Cambridge University Press; 2002.

\bibitem{Str67I}
Stratonovich RL.
\newblock Topics in the Theory of Random Noise. vol.~1.
\newblock New York: Gordon and Breach; 1967.

\bibitem{van92}
van Kampen NG.
\newblock Stochastic Processes in Physics and Chemistry.
\newblock Amsterdam: North-Holland; 1992.

\end{thebibliography}

\end{document}